\documentclass[%
reprint,
superscriptaddress,
 amsmath,amssymb,
 aps,
 onecolumn,
]{revtex4-2}

\usepackage[title]{appendix}% appendix titles
\usepackage{comment}% comments in latex
\usepackage[caption=false]{subfig}% multi-panel figures
\usepackage[normalem]{ulem}% underlining
\usepackage{url}% web links
\usepackage{xcolor}% colors for text
\usepackage{amsmath}% standard math formatting
\usepackage{mathrsfs}% script math fonts
\usepackage{newtxmath,newtxtext}% text and math fonts
\usepackage{siunitx}% SI units

\usepackage{graphicx}% include figure files
\usepackage{dcolumn}% align table columns on decimal point
\usepackage{bm}% bold math
\usepackage[colorlinks=true,linkcolor=blue,filecolor=blue,citecolor=blue,urlcolor=blue]{hyperref}% add hypertext capabilities
\newcommand{\rev}[1]{{{\color{black}#1}}}% ubiquitous revision command

\makeatletter% changes the catcode of @ to 11
\newcommand{\vast}{\bBigg@{2}}% vast parentheses
\newcommand{\Vast}{\bBigg@{3}}% really vast parentheses
\makeatother% changes the catcode of @ back to 12

\begin{document}
%%%%%%%%%%%%%%%%%%%%%%%%%%%%%%%%%%%%%%%%%%

%% PRELIMINARY INFORMATION

\preprint{APS/123-QED}% preprint style

% force line breaks with \\ if necessary
\title{Towards real-time reconstruction of velocity fluctuations in turbulent channel flow}
%\thanks{}% a footnote to the article title

% Rahul
\author{Rahul Arun}
\email{\textcolor{black}{rarun@caltech.edu}}
\affiliation{
Graduate Aerospace Laboratories,
California Institute of Technology,
Pasadena, California 91125, USA
}
% \altaffiliation[]{}

% Jane
\author{H. Jane Bae}
% \email{jbae@caltech.edu}
\affiliation{
Graduate Aerospace Laboratories,
California Institute of Technology,
Pasadena, California 91125, USA
}
% \altaffiliation[]{}

% Beverley
\author{Beverley J. McKeon}
\thanks{\textcolor{black}{Present address: Department of Mechanical Engineering, Stanford University, Stanford, California 94305, USA.}}
\affiliation{
Graduate Aerospace Laboratories,
California Institute of Technology,
Pasadena, California 91125, USA
}
%\affiliation{
%Department of Mechanical Engineering,
%Stanford University,
%Stanford, CA 94305, USA
%}

% use \today (automatic) or explicitly specify date
%\date{\today}

%%%%%%%%%%%%%%%%%%%%%%%%%%%%%%%%%%%%%%%%%%
\begin{abstract}
%%%%%%%%%%%%%%%%%%%%%%%%%%%%%%%%%%%%%%%%%%
We develop a framework for efficient streaming reconstructions of turbulent velocity fluctuations from limited sensor measurements with the goal of enabling real-time applications. The reconstruction process is simplified by computing linear estimators using flow statistics from an initial training period and evaluating their performance during a subsequent testing period with data obtained from direct numerical simulation. We address cases where (i) no, (ii) limited, and (iii) full-field training data are available using estimators based on (i) resolvent modes, (ii) resolvent-based estimation, and (iii) spectral proper orthogonal decomposition modes. During training, we introduce blockwise inversion to accurately and efficiently compute the resolvent operator in an interpretable manner. During testing, we enable efficient streaming reconstructions by using a temporal sliding discrete Fourier transform to recursively update Fourier coefficients using incoming measurements. We use this framework to reconstruct with minimal time delay the turbulent velocity fluctuations in a minimal channel at ${\rm Re}_\tau \approx 186$ from sparse planar measurements. We evaluate reconstruction accuracy in the context of the extent of data required and thereby identify potential use cases for each estimator. The reconstructions capture large portions of the dynamics from relatively few measurement planes when the linear estimators are computed with sufficient fidelity. We also evaluate the efficiency of our reconstructions and show that the present framework has the potential to \rev{help} enable real-time reconstructions of turbulent velocity fluctuations in an analogous experimental setting.
%%%%%%%%%%%%%%%%%%%%%%%%%%%%%%%%%%%%%%%%%%
\end{abstract}
%%%%%%%%%%%%%%%%%%%%%%%%%%%%%%%%%%%%%%%%%%

% use showkeys to display keywords, if defined
%\keywords{Suggested keywords}% define keywords
\maketitle% make the title
%\tableofcontents% make the table of contents

%%%%%%%%%%%%%%%%%%%%%%%%%%%%%%%%%%%%%%%%%%
\section{Introduction}\label{sec:intro}
%%%%%%%%%%%%%%%%%%%%%%%%%%%%%%%%%%%%%%%%%%

%%%%%%%%%%%%%%%%%%%%%%%%%%%%%%%%%%%%%%%%%%
\subsection{Real-time estimation and control}\label{sec:intro:estcont}
%%%%%%%%%%%%%%%%%%%%%%%%%%%%%%%%%%%%%%%%%%

\begin{figure}[!htp]
    \centering
    \includegraphics[width=0.55\textwidth]{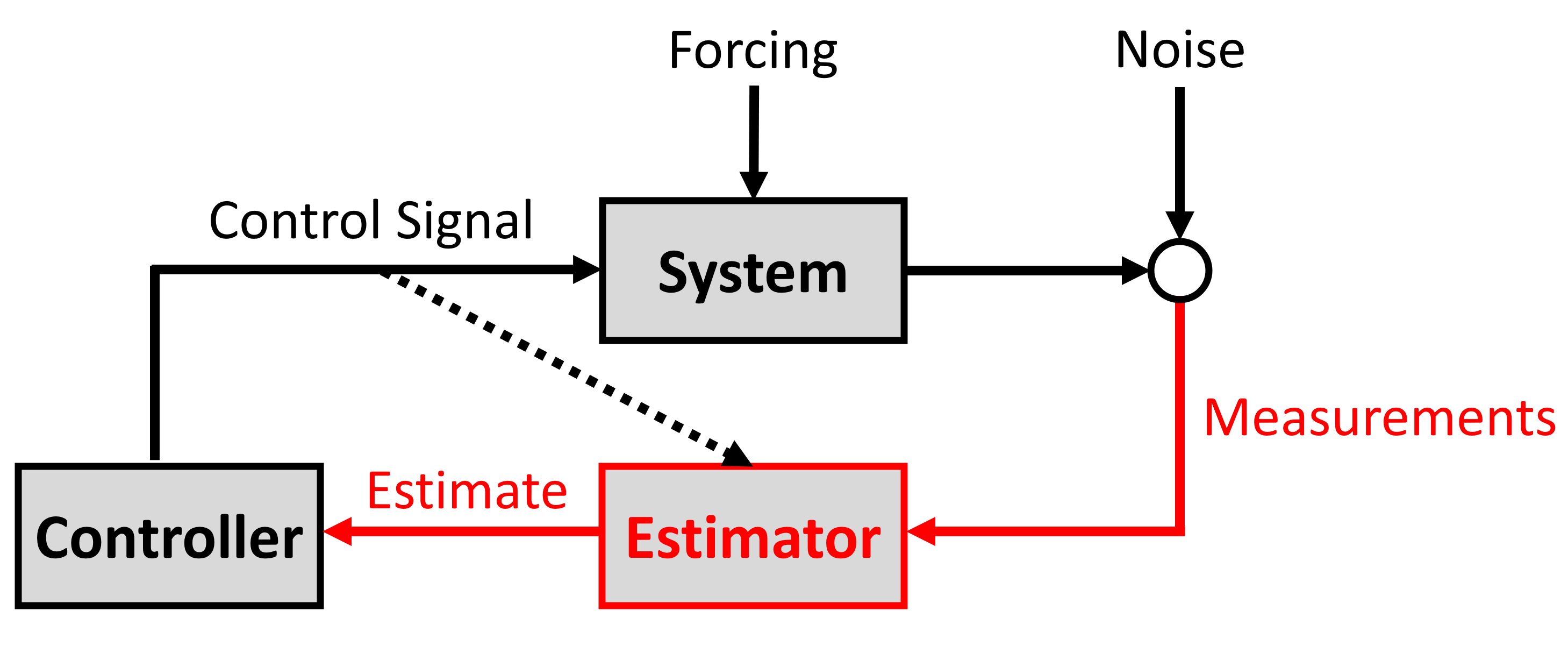}
    \caption{High-level block diagram depicting a framework in which a real-time control scheme is informed by real-time estimates generated by noisy measurements. We focus on the real-time estimation problem (red) in the context of flow reconstruction.}
    \label{fig:intro_fig}
\end{figure}

Processing measurements generated by turbulent flows in real time is an increasingly important problem in applications including flow estimation and control. As depicted in Fig. \ref{fig:intro_fig}, estimates can be used to inform control schemes in real-time applications. Flow estimation problems may be further subdivided into smoothing, filtering, and prediction problems. These problems aim to inform estimates of past, present, and future states (respectively) using incoming measurements that are usually incomplete and noisy. Real-time predictions can be useful for active control, but they are typically more challenging than (smoothing or filtering) reconstructions of turbulent flows due to multiscale dynamics and \rev{sensitivity to initial conditions}.

Real-time flow estimation tasks are often practically limited by low-fidelity measurements and task-related time constraints, e.g., in the context of aviation meteorology \citep{Gul2019}. Measurements can be improved by optimizing the design \citep{Fan2015} and configuration \citep{Man2018} of sensors. \rev{Computational processing can be expedited by leveraging data-driven and physics-based methods \citep{Bru2015} that often reduce the dimensionality of the problem or introduce a simplifying model.} In the context of turbulent jets, \citet{Sas2017} demonstrated that the parabolized stability equations can be used to estimate (in real time) the pressure at downstream sensors using upstream pressure measurements.

Beyond estimation, real-time control schemes require yet further empirically or heuristically prescribed simplifications to ensure in-time actuation. \citet{Mai2021} used a reactive control scheme to experimentally attenuate centerline velocity fluctuations in a forced jet (${\rm Re}_D =$ 50\,000) by generating wave packets that destructively interfere with stochastic disturbances generated at the nozzle. \citet{Abb2017} demonstrated that actuating an array of cross-flowing jets that penetrate into the log region of a turbulent boundary layer (${\rm Re}_\tau =$ 14\,400) can influence the resulting skin friction drag and turbulence intensity.

Previous real-time flow estimation and control schemes demonstrate that, by combining governing equations with measurements and simplifying models, hybrid methods allow for a tailored balance between the efficacy and the speed of the scheme. One commonality of these methods is that their simplifying assumptions are often not well justified from first principles. In the present investigation, we focus on the problem of real-time flow reconstruction in the context of wall-bounded turbulence. Even for this relatively simple estimation problem, the task of efficiently and accurately representing turbulent dynamics from first principles [i.e., the Navier-Stokes equations (NSE)] using minimal assumptions has remained challenging.

%%%%%%%%%%%%%%%%%%%%%%%%%%%%%%%%%%%%%%%%%%
\subsection{Reconstruction techniques}\label{sec:intro:rectech}
%%%%%%%%%%%%%%%%%%%%%%%%%%%%%%%%%%%%%%%%%%

Reduced-order models (ROMs) often provide efficient flow representations to reduce the complexity of high-dimensional turbulent flows, e.g., using proper orthogonal decomposition (POD) \citep{Ber1993} and dynamic mode decomposition (DMD) \citep{Sch2010}. For statistically stationary flows, spectral POD (SPOD) provides an efficient means of identifying coherent structures that retains a direct relationship to resolvent analysis and DMD \citep{Tow2018}. \rev{Unless otherwise stated, the SPOD modes we refer to are those of the velocity fluctuations.} Truncated SPOD mode estimation (TSME) often provides an efficient means of flow reconstruction since SPOD modes form an optimal orthogonal basis in terms of the variance captured by a given subset of modes \citep{Sch2020}. Data-driven methods like SPOD are often limited in that they neglect the governing equations and require extensive postprocessing. However, \citet{Gha2020} showed that supplementing SPOD-based truncations with physics-based enrichment using Gabor modes enables representations of flows over a broad range of scales. Moreover, SPOD is amenable to a streaming formulation \citep{Sch2019} and convolution-based strategies for time-domain analysis \citep{Nek2021}, thereby reducing its computational requirements.

Equation-based frameworks have been developed in conjunction with data-driven methods to efficiently estimate turbulent flows. Techniques from control theory are often used for flow estimation in the time domain (e.g., Kalman filters \citep{Che2006,Col2011,Ill2018}) and the frequency domain (e.g., Wiener filters \citep{Mar2009}). Techniques like linear stochastic estimation (LSE) \citep{Adr1988} and its spectral variant \citep{Tin2006} are rooted in conditional estimation. LSE-based techniques are often used with data-driven \citep{Pod2018,He2020} and equation-based \citep{Gup2021,Mad2019} models to augment flow reconstructions. Other methods use simplified governing models to produce forward and backward estimates that augment reconstructions from low-resolution and multiresolution temporal measurements \citep{Kri2020,Wan2021}. One central challenge in these estimation techniques is addressing the nonlinearity of the governing equations.

Resolvent analysis \citep{McS2010,McK2017} provides a powerful framework for addressing the nonlinearity of the NSE and identifying energetic linear amplification mechanisms using minimal assumptions. \rev{In a manner related to the work of \citet{Far1993}, the turbulent fluctuation dynamics are recast as a linear system forced by their nonlinearity in the resolvent framework. The resolvent operator often admits a low-rank representation, and it can be constructed using a base flow profile, which can be modeled or learned from data, e.g., via data assimilation \citep{Sym2018,Sym2020}.}

One flavor of flow reconstruction using resolvent analysis is truncated response mode estimation (TRME). TRME revolves around the singular value decomposition (SVD) of the resolvent operator, which produces orthonormal bases for the forcing and response that are related through corresponding gains, or singular values. Modal truncation is performed based on the (often) low-rank nature of the resolvent operator, which occurs when the gains associated with a small number of leading modes dominate those of the remaining (suboptimal) modes. \citet{Moa2013} demonstrated that streamwise energy amplification in each half of a turbulent channel (${\rm Re}_\tau = 2003$) is well captured by a rank-1 approximation of the resolvent operator. \citet{Moa2014} then used convex optimization techniques to compute resolvent mode weights, which encode dynamical information, using a similar low-rank approximation to capture the velocity spectra in a turbulent channel. In similar fashions, TRME was used to approximate energetic structures in flow around a cylinder \citep{Sym2020} and in lid-driven cavity flow (${\rm Re} = 1200$) with a known (2D) mean flow \citep{Gom2016}. In all of these cases, the resolvent modes are weighted using spatially isolated velocity measurements in either Fourier or physical space. In such setups, it is important that the measurements at least partially capture the most energetic regions of the flow. For example, \citet{Sym2018} showed that tailoring probe locations to capture energetic regions of both low-frequency wake modes and high-frequency shear layer modes significantly improves reconstruction accuracy. \citet{Ben2016} found a similar result by separately capturing low-frequency and high-frequency regions in a backward-facing step flow configuration. Recently, a quasisteady resolvent analysis has been used to reconstruct high-frequency fluctuations using the phase of a lower-frequency periodic motion \citep{Fra2022}. Resolvent truncations have also recently been used to design $H_2$-optimal estimators and controllers \citep{Jin2022} and select optimal sensor and actuator locations \citep{Jin2022a} in cylinder flows. 

While TRME provides an efficient and informative approximation, the assumption of a fixed-rank \footnote{In fixed-precision approximations, the rank can be tailored to efficiently satisfy a prescribed error tolerance using randomized methods \citep{Hal2011}.} truncation may not be appropriate across all frequencies. In contrast to TRME, resolvent-based estimation (RBE) revolves around the SVD of an input-output operator that modifies the resolvent operator to reflect sensor locations. Since flow sensors are usually limited and sparsely spaced, RBE has shown promise in producing reconstructions of both flow statistics \citep{Tow2020} and dynamics \citep{Mar2020} from limited data without truncating the resolvent operator \textit{a priori}. RBE \citep{Tow2020} relates response measurement statistics to observable forcing statistics via an input-output operator and (typically) models the unobservable forcing as uncorrelated with the observable forcing. The estimated forcing statistics then yield estimates of the response statistics via the resolvent operator. Building from this approach, \citet{Mar2020} derived optimal, noncausal transfer functions relating the measurements to the full response when the second-order forcing statistics and resolvent operator are known. Further, \citet{Mor2020} showed that the statistics in a turbulent channel can be reasonably modeled using a low-rank SPOD approximation \rev{of the forcing}. These studies reveal that the tradeoff between accurately capturing the forcing color and retaining a sparse measurement configuration can be somewhat alleviated by modeling the structure of the nonlinearity. \rev{For example, in turbulent channel flows, including an eddy-viscosity model \citep{Ill2018} can improve the accuracy of RBE techniques for the response \citep{Ama2020} and its statistics \citep{Mor2019,Mor2020,Tow2020} with minimal input data.}

\rev{We primarily focus on noncausal forms of linear estimators, including that derived for optimal resolvent-based estimation (ORBE) \citep{Mar2020, Ama2020}. More recently, a causal resolvent-based formulation} for estimation and control \citep{Mar2022} was developed via the Wiener-Hopf formalism. However, the numerical machinery for forming this causal estimator is typically more complex than that of the noncausal estimator due to an additional term in the estimation problem. By instead applying the noncausal formulation to the real-time reconstruction problem, we avoid these complex methods and retain a relatively simple framework. \rev{Noncausal estimation techniques (including ORBE)} are not typically amenable to real-time applications \citep{Ama2020}, but we circumvent this limitation by continuously updating the estimated temporal Fourier coefficients over a small window at every time step. Further, using this sliding window enables efficient recursive techniques for updating the coefficients \citep{Cha2022}. \rev{The implementation of this technique, which has linear algorithmic complexity, differs from the streaming Fourier sums technique of \citet{Sch2019}, which has quadratic algorithmic complexity but a smaller memory footprint.}

\rev{Nonlinear estimation techniques, especially those incorporating machine learning techniques, can also reconstruct flows with competitive accuracy. For example, \citet{Fuk2021} reported relative errors of less than 15\% for spatial superresolution of a channel flow using a coarse input with six elements across each half of the channel. Although they achieve significant data compression, the coarsened input for estimation, obtained via averaging, is not conducive to typical experimental measurements. By contrast, our investigation focuses on practical considerations for experimental implementation, including time delays, measurement configurations, scalability, and uncertainty quantification. Working in the linear estimation setting is further advantageous in that it is amenable to a wealth of estimation and control theory for linear systems. Further, in contrast to many machine learning techniques, the present linear estimation techniques are interpretable in terms of their optimality and through the basis and flow statistics imposed by the linear model. Finally, whereas machine learning techniques can take significantly longer (e.g., multiple days \citep{Fuk2021}) to train estimators on optimized hardware, we introduce estimators that can be constructed from training data within an hour on a laptop.}

%%%%%%%%%%%%%%%%%%%%%%%%%%%%%%%%%%%%%%%%%%
\subsection{Contributions}\label{sec:intro:contrib}
%%%%%%%%%%%%%%%%%%%%%%%%%%%%%%%%%%%%%%%%%%

The goal of this study is to construct a method for reconstruction of turbulent velocity fluctuations from sparse measurements, adding to the vast flow estimation literature. For this, we develop a new framework to enable efficient streaming capabilities with the potential for real-time implementation. In Sec. \ref{sec:form}, we detail the governing equations and formulate the statistical framework, \rev{based on the generalized Wiener filter,} that we use to reconstruct the fluctuations. \rev{We also detail the application of methods inspired by TRME, ORBE, and TSME to this more general framework.} In Sec. \ref{sec:meth}, we introduce efficient methods that enable streaming flow reconstructions with minimal time delay and improved scalability compared to standard techniques. When forming linear estimators from training data, we introduce blockwise inversion to efficiently compute the resolvent operator. When reconstructing the flow from testing data, we apply a recursive sliding discrete Fourier transform (SDFT) to efficiently update the inputs to the estimators in a streaming fashion. In Sec. \ref{sec:res}, we evaluate the performance of the reconstructions using each method. We specifically address (i) the fidelity of the training data and the testing data, (ii) the validity of the models underlying the estimators, (iii) the potential use cases for each reconstruction method, and (iv) the reconstruction efficiency in the context of real-time applications. \rev{In doing so, we demonstrate that practical reconstruction speeds can be achieved using modest computational resources while retaining competitive accuracies with other linear estimation techniques.} Finally, we summarize our reconstruction methods, the performance of the estimators, and promising future prospects and applications in Sec. \ref{sec:conc}.

%%%%%%%%%%%%%%%%%%%%%%%%%%%%%%%%%%%%%%%%%%
\section{Background Methodology}\label{sec:form}
%%%%%%%%%%%%%%%%%%%%%%%%%%%%%%%%%%%%%%%%%%

\subsection{Governing equations}\label{ssec:goveq}
We focus on turbulent channel flow between two parallel walls, for which we denote the streamwise, wall-normal, and spanwise directions by $x$, $y$, and $z$ and the corresponding velocities by $\mathcal{U}$, $\mathcal{V}$, and $\mathcal{W}$, respectively. This flow is governed by the nondimensional, incompressible Navier-Stokes equations (NSE)
\begin{equation}\label{NSE}
   \partial_t \boldsymbol{\mathcal{U}} + (\boldsymbol{\mathcal{U}} \cdot \nabla)\boldsymbol{\mathcal{U}}+\nabla \mathcal{P} = \frac{1}{{\rm Re}_\tau} \nabla^2 \boldsymbol{\mathcal{U}}, \quad \nabla\cdot\boldsymbol{\mathcal{U}} = 0,
\end{equation}
%nondimensionalization explanation%
where $\boldsymbol{\mathcal{U}} = [\mathcal{U}, \mathcal{V}, \mathcal{W}]^T$ is the velocity vector, $\mathcal{P}$ is pressure, and $t$ denotes time. For channel flow, ${\rm Re}_\tau = u_\tau h / \nu$ is defined in terms of the channel half-height $h$, the friction velocity $u_\tau$, and the kinematic viscosity $\nu$. Lengths, velocities, and pressures are normalized by $h$, $u_\tau$, and $\rho u_\tau^2$, respectively, where $\rho$ is the fluid density. 

We combine the velocity and pressure into a single state variable, $\boldsymbol{\mathcal{Q}} = [\mathcal{U}, \mathcal{V}, \mathcal{W}, \mathcal{P}]^T$, whose evolution is governed by the NSE. The state fluctuations are expressed by (Reynolds) decomposing the state vector as $\boldsymbol{\mathcal{Q}} = \boldsymbol{Q} + \boldsymbol{q}$, where $\boldsymbol{Q}$ and $\boldsymbol{q}$ represent the base flow and the fluctuations, respectively. Here, we specify the base flow as the mean state profile, $\boldsymbol{Q}(y) = \overline{\boldsymbol{\mathcal{Q}}}(y) = [U(y), V(y), W(y), P(y)]^T$, where $V(y) = W(y) = 0$ for channel flow \footnote{However, this assumption may be violated when averaging the velocities over finite time horizons.}. The corresponding fluctuations are Fourier transformed in time and in the spatially homogeneous and periodic directions ($x$ and $z$ here), giving
\begin{equation}\label{FT}
    \boldsymbol{\hat{q}}(y,\boldsymbol{k}) = \begin{bmatrix} \boldsymbol{\hat{u}}(y,\boldsymbol{k}) \\ \hat{p}(y,\boldsymbol{k}) \end{bmatrix} = \mathscr{F}\{\boldsymbol{q}(\boldsymbol{x},t)\} = \frac{1}{8 \pi^3}\iiint_{-\infty}^{\infty} \boldsymbol{q}(\boldsymbol{x},t){\rm e}^{{\rm i}(\omega t - k_x x - k_z z)} dt \, dx \, dz,
\end{equation}
where ${\rm i} = \sqrt{-1}$, $\boldsymbol{x} = [x,y,z]^T$ and each triplet, $\boldsymbol{k}=[k_x, k_z,\omega]^T$, contains the streamwise wave numbers, spanwise wave numbers, and (angular) temporal frequencies, respectively. The transformed differential operators are given by $\hat{\nabla} = [{\rm i} k_x, \partial_y, {\rm i} k_z]^T$ and $\hat{\nabla}^2 = \partial_{yy} - k^2$, where $k^2 = k_x^2+k_z^2$.

Applying the Reynolds decomposition and the Fourier transform defined in (\ref{FT}), (\ref{NSE}) can be used to express the state fluctuation dynamics for each triplet via a linear subsystem that is externally forced by the nonlinear advection terms. Correspondingly, each linear subsystem may be written as
\begin{equation}\label{eq:cont_sys}
    \mathcal{L}(\boldsymbol{k}) \boldsymbol{\hat{q}}(y,\boldsymbol{k}) = \begin{bmatrix}
    \mathcal{L}_B(\boldsymbol{k}) & \hat{\nabla} \\ \hat{\nabla}^T & 0 \end{bmatrix} \begin{bmatrix} \boldsymbol{\hat{u}}(y,\boldsymbol{k}) \\ \hat{p}(y,\boldsymbol{k}) \end{bmatrix} = \begin{bmatrix}
    \boldsymbol{\hat{f}}(y,\boldsymbol{k}) \\ 0 \end{bmatrix},
\end{equation}
where $\boldsymbol{k} \neq [0, 0, 0]^T$. Here, $\mathcal{L}_B$ represents the ``basic'' terms in the unsteady momentum equations, which are linear in the \textit{velocity} fluctuations, and $\boldsymbol{\hat{f}}(y,\boldsymbol{k}) =\mathscr{F}\{-(\boldsymbol{u} \cdot \nabla)\boldsymbol{u} + \overline{(\boldsymbol{u} \cdot \nabla)\boldsymbol{u}} \}$ represents the nonlinear forcing. While the subsystem for each triplet is \textit{externally} forced by velocity fluctuations satisfying $\boldsymbol{k'} \neq \boldsymbol{k}$, the system is \textit{internally} forced by its nonlinearity. In Sec. \ref{ssec:dls}, we discretize this state-space formulation of the Navier-Stokes system and express the corresponding resolvent operator in discrete form.

%%%%%%%%%%%%%%%%%%%%%%%%%%%%%%%%%%%%%%%%%%
\subsection{Discretized linear system}\label{ssec:dls}
%%%%%%%%%%%%%%%%%%%%%%%%%%%%%%%%%%%%%%%%%%

\begin{figure}[!htp]
    \centering
    \includegraphics[width=0.4\textwidth]{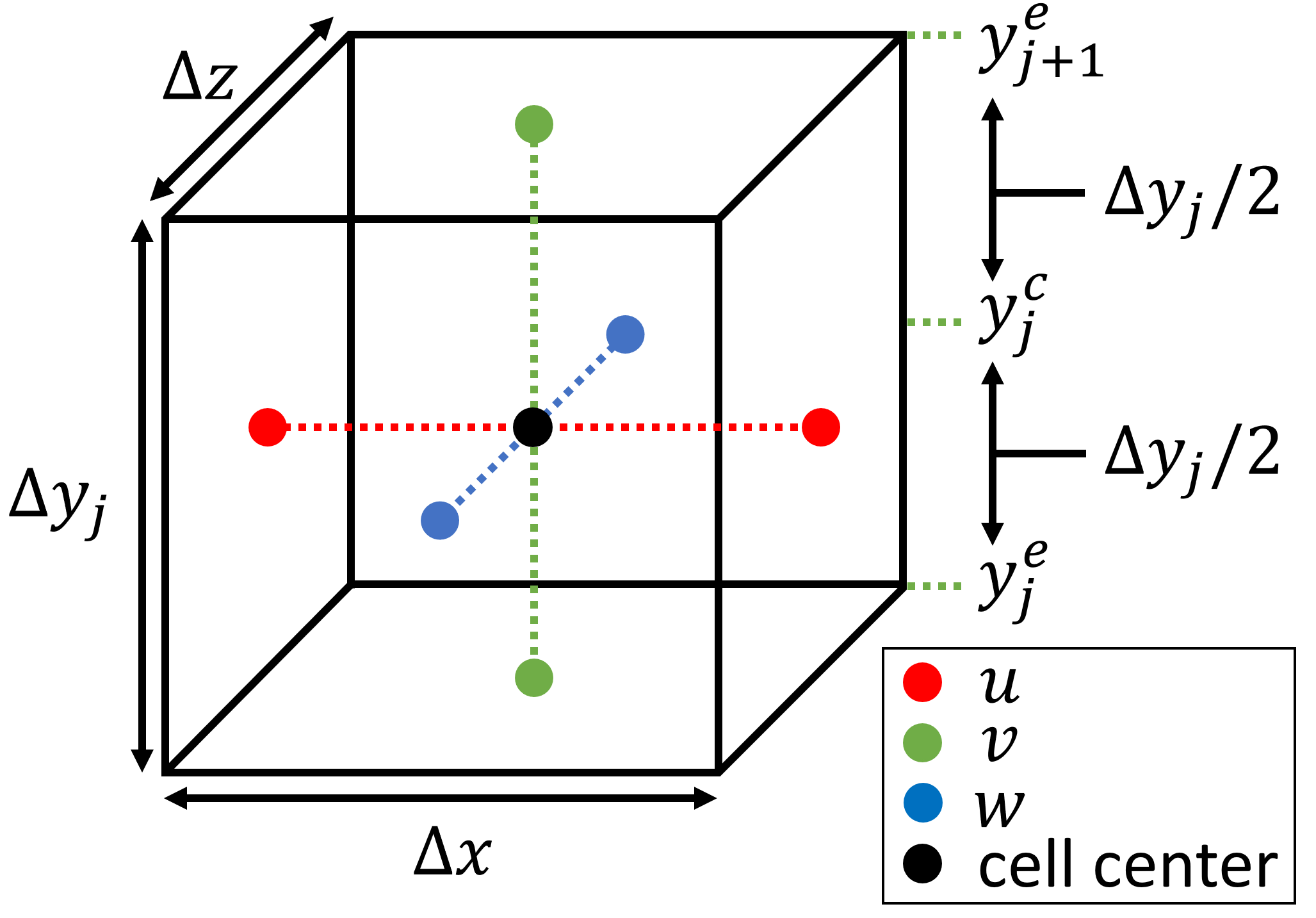}
    \caption{Staggered state variable locations on an arbitrary cell in the computational domain. This diagram is not drawn to scale since $\Delta x \neq \Delta z \neq \Delta y_j$, and $\Delta y_j$ varies in the wall-normal direction. Velocities reside on the faces of the cell and pressure resides at the cell center. The superscripts $\cdot^c$ and $\cdot^e$ represent the (wall-normal) centers and edges of the cell, respectively.}
    \label{fig:stag}
\end{figure}

When formed directly from the continuous governing equations, the linear subsystems in (\ref{eq:cont_sys}) are infinite-dimensional. However, by discretizing the equations onto a grid, they assume a finite-dimensional representation based on the spatial resolution in the wall-normal direction. As depicted in Fig. \ref{fig:stag}, the direct numerical simulation (DNS) data in the present case are staggered \footnote{Another common approach to wall-normal discretization is spectral collocation.} on a Cartesian grid.
For a wall-normal domain discretized into $N_y$ cells, the discrete representations of $\boldsymbol{\hat{u}}$ and $\boldsymbol{\hat{q}}$ are column vectors of sizes $N_u = 3 N_y + 1$ and $N_q = N_u + N_y$, respectively. The numerical quadrature (i.e., wall-normal integration) weights associated with center and edge quantities are $\Delta y_j$ and $0.5 \left(\Delta y_{j-1} + \Delta y_j \right)$, respectively.

Since state variables are Fourier transformed in $x$ and $z$, we accurately reference the cell-center values for $\hat{u}$ and $\hat{w}$ by shifting their Fourier transform origins by half a cell width in $x$ and $z$, respectively. To retain a direct analogy with the (discrete) governing equations used in simulating the flow, we maintain $\hat{v}$ and the mean shear, $\partial_y U$, at the wall-normal edges, $y^e_j$ and $y^e_{j+1}$, of each cell. Correspondingly, we represent the mean shear using a bidiagonal matrix, $\boldsymbol{\partial_y U}$, that acts to average the contributions of $\hat{v} \partial_y U$ at the cell edges to the cell centers. The diagonal matrices, $\boldsymbol{U_{c,e}}$, represent the mean profile at the cell centers and edges, respectively.

In discrete form, the linear operator, $\mathcal{L}$, is a matrix, $\boldsymbol{L} \in \mathbb{C}^{N_q \times N_q}$, of the form
\begin{equation}\label{calL}
    \boldsymbol{L}(\boldsymbol{k}) =
    \begin{bmatrix}
    \boldsymbol{L_c}(\boldsymbol{k})  & \boldsymbol{\partial_y U} & \boldsymbol{0} & {\rm i} k^*_x \boldsymbol{I_c}\\
    \boldsymbol{0} & \boldsymbol{L_e}(\boldsymbol{k}) & \boldsymbol{0} & \boldsymbol{\partial_y} \\
    \boldsymbol{0} & \boldsymbol{0} & \boldsymbol{L_c}(\boldsymbol{k}) & {\rm i} k^*_z \boldsymbol{I_c} \\
    {\rm i} k^*_x \boldsymbol{I_c} & \boldsymbol{\partial_y} & {\rm i} k^*_z \boldsymbol{I_c} & \boldsymbol{0}
    \end{bmatrix}
     = 
     \begin{bmatrix}
     \boldsymbol{L_B} & \boldsymbol{\hat{\nabla}} \\
     \boldsymbol{\hat{\nabla}}^T & \boldsymbol{0}
     \end{bmatrix},
\end{equation}
where the $\boldsymbol{0}$ matrices are conformable and $\boldsymbol{L_B}$ is the basic subblock, analogous to $\mathcal{L}_B$ in (\ref{eq:cont_sys}). The diagonal subblocks take the form
\begin{equation}
    \boldsymbol{L_{c,e}} = -{\rm i} \omega \boldsymbol{I_{c,e}} + {\rm i} k^*_x \boldsymbol{U_{c,e}} - \frac{1}{{\rm Re}_\tau}\boldsymbol{\hat{\nabla}_{c,e}^2},
\end{equation}
where $\boldsymbol{\hat{\nabla}_{c,e}^2}$ and $\boldsymbol{I_{c,e}}$ are the Laplacian and identity matrices of appropriate size. The discrete representations of $\partial_y$ and $\partial_{yy}$ are computed via central finite differences with modified boundary terms to enforce the relevant boundary conditions at the walls. In each case, boundary values can be determined by applying central finite differences on a ghost grid that symmetrically extends the center grid about the wall (edge) locations. Further,
\begin{equation}
    k^*_x = \frac{2}{\Delta x} {\rm sin} \left( \frac{k_x \Delta x}{2} \right) \quad {\rm and} \quad k^*_z = \frac{2}{\Delta z} {\rm sin} \left( \frac{k_z \Delta z}{2} \right)
\end{equation}
are the modified wave numbers corresponding to central finite differences in the streamwise and spanwise directions, respectively. The differences between $k^*_x$ and $k_x$ induce differences in the advection and Fourier phase speeds which grow as $|k_x|$ increases.

Correspondingly, the spatially discretized form of (\ref{eq:cont_sys}) is given by
\begin{equation}\label{eq:linsys}
\boldsymbol{L}(\boldsymbol{k}) \begin{bmatrix} \boldsymbol{S}(\boldsymbol{k}) & \boldsymbol{0} \\ \boldsymbol{0} & \boldsymbol{I_c} \end{bmatrix} \boldsymbol{\hat{q}}(\boldsymbol{k}) = \boldsymbol{B} \boldsymbol{S}(\boldsymbol{k}) \boldsymbol{\hat{f}}(\boldsymbol{k}),
\end{equation}
where we have assumed that the streamwise and spanwise components of $\boldsymbol{\hat{q}}$ and $\boldsymbol{\hat{f}}$ are unshifted. Here, $\boldsymbol{S}$ is a diagonal matrix that shifts the origin for staggered flow quantities (in $x$ and $z$) to the cell center. As such, it may be expressed as
\begin{equation}
    \boldsymbol{S}(\boldsymbol{k}) = \begin{bmatrix} S_x(\boldsymbol{k}) \boldsymbol{I_c} & \boldsymbol{0} & \boldsymbol{0} \\ \boldsymbol{0} & \boldsymbol{I_e} & \boldsymbol{0} \\ \boldsymbol{0} & \boldsymbol{0} & S_z(\boldsymbol{k}) \boldsymbol{I_c} \end{bmatrix}, \quad S_x = {\rm exp}\left( \frac{{\rm i} k^*_x \Delta x}{2} \right), \quad S_z = {\rm exp}\left( \frac{{\rm i} k^*_z \Delta z}{2} \right),
\end{equation}
where $\boldsymbol{S}^{-1} = \boldsymbol{S}^H$ and $(\cdot)^H$ is the conjugate transpose. The input matrix,
\begin{equation}
\boldsymbol{B} = \begin{bmatrix} \boldsymbol{I_c} & \boldsymbol{0} & \boldsymbol{0} \\ \boldsymbol{0} & \boldsymbol{I_e} & \boldsymbol{0} \\ \boldsymbol{0} & \boldsymbol{0} & \boldsymbol{I_c} \\ \boldsymbol{0} & \boldsymbol{0} & \boldsymbol{0} \\ \end{bmatrix},
\end{equation}
is used to isolate the forcing to the momentum equations. For convenience, we also use it to extract the velocity fluctuations from the state vector. For each triplet, the velocity fluctuations may then be expressed as a linear response to their nonlinear forcing via
\begin{equation}\label{eq:linsys2}
    \boldsymbol{\hat{u}}(\boldsymbol{k})= \boldsymbol{B}^T \boldsymbol{\hat{q}}(\boldsymbol{k}) =  \boldsymbol{R_u}(\boldsymbol{k}) \boldsymbol{\hat{f}}(\boldsymbol{k}),
\end{equation}
where
\begin{equation}\label{eq:oper}
    \boldsymbol{R_u} = \left( \boldsymbol{B} \boldsymbol{S} \right)^H \boldsymbol{L}^{-1} \left( \boldsymbol{B} \boldsymbol{S} \right)
\end{equation}
is the one-dimensional resolvent \footnote{Although not strictly written as $\left({\rm i}\omega \boldsymbol{I} - \boldsymbol{A} \right)^{-1}$, we colloquially refer to $\boldsymbol{R_u}$ as the resolvent operator for convenience.} operator. We directly enforce the no-penetration boundary condition in computing $\boldsymbol{L}^{-1} \boldsymbol{B} \boldsymbol{S}$.

%%%%%%%%%%%%%%%%%%%%%%%%%%%%%%%%%%%%%%%%%%
\subsection{Flow reconstruction}\label{sec:flowrec}
%%%%%%%%%%%%%%%%%%%%%%%%%%%%%%%%%%%%%%%%%%

We formulate the flow reconstruction problems for TRME (Sec. \ref{ssec:trme}), ORBE (Sec. \ref{ssec:orbe}), and TSME (Sec. \ref{ssec:tsme}) in triplet space. Correspondingly, our goal is to reconstruct velocity fluctuations in the turbulent channel using spatially sparse measurements in the wall-normal direction. We therefore consider a collection of $N_{\rm plane}$ spatially isolated planar observations,
\begin{equation}\label{eq:meas_yu}
    \boldsymbol{\hat{y}_n}(\boldsymbol{k}) = \boldsymbol{C}(\boldsymbol{k}) \boldsymbol{\hat{u}}(\boldsymbol{k}) + \boldsymbol{\hat{n}}(\boldsymbol{k}) = \boldsymbol{\hat{y}}(\boldsymbol{k}) + \boldsymbol{\hat{n}}(\boldsymbol{k}),
\end{equation}
where $\boldsymbol{C} \in \mathbb{R}^{3 N_{\rm plane} \times N_u}$ is the observation matrix, $\boldsymbol{\hat{n}}$ represents the measurement noise, and $\boldsymbol{\hat{y}} = \boldsymbol{C}\boldsymbol{\hat{u}}$ isolates the noiseless component of the observations. For all triplets of interest, $\boldsymbol{C}$ isolates \textit{collocated} velocity fluctuation measurements at cell-centered wall-parallel planes by averaging adjacent wall-normal velocities. The noise is assumed to represent a zero-mean random process that is uncorrelated with the flow. In this formulation, $\boldsymbol{\hat{n}}$ captures errors resulting from employing a discrete Fourier transform over a finite temporal window and any other sources of contamination. In an experimental setting, observations of the form in (\ref{eq:meas_yu}) may be generated by (multiplane) wall-parallel particle image velocimetry (PIV) measurements.

We formulate optimal linear transfer functions by operating under the generalized Wiener filter formalism. For each reconstruction method, we consider generic linear models for the flow and observations of the form
\begin{equation}\label{eq:un_Wiener}
    \boldsymbol{\hat{u}} = \boldsymbol{H_u} \boldsymbol{b}, \quad \boldsymbol{\hat{y}_n} = \boldsymbol{H_y} \boldsymbol{b} + \boldsymbol{\hat{n}},
\end{equation}
where $\boldsymbol{H_y} = \boldsymbol{C} \boldsymbol{H_u}$ and $\boldsymbol{b}$ is a set of unknown coefficients. In weighted form, (\ref{eq:un_Wiener}) may be written as
\begin{equation}\label{eq:we_Wiener}
    \boldsymbol{W_u} \boldsymbol{\hat{u}} = \boldsymbol{H_u^w} \boldsymbol{W_b} \boldsymbol{b}, \quad \boldsymbol{W_y} \boldsymbol{\hat{y}_n} = \boldsymbol{H_y^w} \boldsymbol{W_b} \boldsymbol{b} + \boldsymbol{W_y} \boldsymbol{\hat{n}},
\end{equation}
where $\boldsymbol{H_u^w} = \boldsymbol{W_u} \boldsymbol{H_u} \boldsymbol{W}^{-1}_{\boldsymbol{b}}$, $\boldsymbol{H_y^w} = \boldsymbol{C_w} \boldsymbol{H_u^w}$, and $\boldsymbol{C_w} = \boldsymbol{W_y} \boldsymbol{C} \boldsymbol{W}^{-1}_{\boldsymbol{u}}$. Here, we have assumed $\boldsymbol{W_y}$, $\boldsymbol{W_u}$, and $\boldsymbol{W_b}$ are Hermitian, positive-definite weighting matrices such that the norms of the unweighted vectors corresponding to their respective weighted inner products are equivalent to the Euclidean 2-norms of the weighted vectors. Unless otherwise stated, all weightings in the present investigation are real, diagonal matrices that contain the square roots of the numerical quadrature weights to ensure that the relevant norms reproduce energies integrated over the wall-normal direction.

Given a linear model, $\boldsymbol{H_u}$, and an observation matrix, $\boldsymbol{C}$, the generalized Wiener filter provides an optimal linear estimate of the coefficients, $\boldsymbol{\tilde{b}}$, assuming that the second-order statistics of the coefficients, $\boldsymbol{S_{bb}} = \mathbb{E}\left( \boldsymbol{b} \boldsymbol{b}^H \right)$, and the noise, $\boldsymbol{S_{nn}} = \mathbb{E}\left( \boldsymbol{\hat{n}}\boldsymbol{\hat{n}}^H \right)$, or models thereof are known. Optimality is typically defined in terms of minimizing the mean-squared error of the estimated coefficients. However, \citet{Mar2020} showed (in the context of RBE) that the corresponding optimal estimator is the stationary point of the entire coefficient error CSD, not just its trace. Furthermore, this stationary point also applies to the velocity fluctuation error CSD, highlighting the correspondence between optimal estimates of the coefficients and the flow. Hence, for each reconstruction method, we express the estimator via a linear transfer function, $\boldsymbol{T_u}$, that relates the noisy measurements to the flow estimates as
\begin{equation}\label{eq:lintf}
    \boldsymbol{\tilde{b}} = \boldsymbol{T_b} \boldsymbol{\hat{y}_n}, \quad \boldsymbol{\tilde{\hat{u}}} = \boldsymbol{T_u} \boldsymbol{\hat{y}_n}, 
\end{equation}
where $\boldsymbol{T_u} = \boldsymbol{H_u} \boldsymbol{T_b}$. Hereafter, unless otherwise stated, all quantities denoted as $\tilde{(\cdot)}$ represent estimates/models of true flow quantities or the building blocks thereof.

The transfer function may be explicitly written as
\begin{equation}\label{eq:tfdef}
    \boldsymbol{T_u} = \boldsymbol{S_{uy}} \boldsymbol{S}^{-1}_{\boldsymbol{yy,n}} = \boldsymbol{H_u} \boldsymbol{S_{bb}} \boldsymbol{H}^H_{\boldsymbol{y}} \left( \boldsymbol{H_y} \boldsymbol{S_{bb}} \boldsymbol{H}^H_{\boldsymbol{y}} + \boldsymbol{S_{nn}} \right)^{-1},
\end{equation}
where $\boldsymbol{S_{yy,n}} = \mathbb{E}\left( \boldsymbol{\hat{y}_n} \boldsymbol{\hat{y}}^H_{\boldsymbol{n}} \right)$ and we have used that the noise is uncorrelated with the \rev{flow such that $\boldsymbol{S_{un}} = \boldsymbol{0}$.} Absent knowledge of the noise CSD, we model it as $\boldsymbol{\tilde{S}_{nn}} = \epsilon \boldsymbol{I}$ (for small $\epsilon$) to ensure that $\boldsymbol{S}^{-1}_{\boldsymbol{yy,n}}$ remains numerically well posed while approaching the zero-noise limit. The transfer function, $\boldsymbol{T_u}$, applies to both (\ref{eq:un_Wiener}) and (\ref{eq:we_Wiener}), demonstrating that it is independent of the weightings placed on the response, coefficients, and measurements/noise. We hereafter refer to the linear models via (\ref{eq:un_Wiener}), but note that the weightings may still be incorporated implicitly in the models of $\boldsymbol{H_u}$, $\boldsymbol{S_{bb}}$, and/or $\boldsymbol{S_{nn}}$.

Importantly, (\ref{eq:un_Wiener}) is a model in the sense that $\boldsymbol{H_u} \boldsymbol{b}$ may be an ill-posed representation of the flow. Therefore, in addition to the knowledge or models of $\boldsymbol{S_{bb}}$ and $\boldsymbol{S_{nn}}$, the well-posedness of the assumed linear model, $\boldsymbol{H_u}$, is also implicit to the optimality of the linear estimator produced by the generalized Wiener filter. All methods we consider incorporate the estimated mean flow in learning $\boldsymbol{H_u}$ from training data, and TSME and ORBE further incorporate the response and (thereby) the nonlinear forcing CSDs, respectively. In what follows, we derive expressions for the transfer functions corresponding to each reconstruction method. We discuss our methodology of efficiently applying the transfer functions to streaming measurements in Sec. \ref{sec:meth}.

%%%%%%%%%%%%%%%%%%%%%%%%%%%%%%%%%%%%%%%%%%
\subsubsection{Truncated response mode estimation}\label{ssec:trme}
%%%%%%%%%%%%%%%%%%%%%%%%%%%%%%%%%%%%%%%%%%

TRME is an equation-based reconstruction framework that requires prior knowledge of the mean profile, but not of the second-order statistics. Therefore, TRME requires minimal data collection and modeling to form an estimator, but it is founded on the assumptions that a low-rank approximation of the resolvent operator can capture the relevant dynamics and that the corresponding forcing (statistics) may be modeled from first principles.

The low-rank approximation of the weighted resolvent operator is expressed by truncating its SVD as
\begin{equation}\label{eq:low_rank}
    \boldsymbol{R_u^w} = \left( \boldsymbol{W_u} \boldsymbol{\Psi} \right) \boldsymbol{\Sigma} \left(\boldsymbol{W_f} \boldsymbol{\Phi} \right)^H \approx \left( \boldsymbol{W_u} \boldsymbol{\tilde{\Psi}} \right) \boldsymbol{\tilde{\Sigma}} \left(\boldsymbol{W_f} \boldsymbol{\tilde{\Phi}} \right)^H = \boldsymbol{\tilde{R}_u^w},
\end{equation}
where
\begin{equation}
    \boldsymbol{\tilde{\Psi}} = \boldsymbol{\Psi} \boldsymbol{\tilde{M}}, \quad \boldsymbol{\tilde{\Phi}} = \boldsymbol{\Phi} \boldsymbol{\tilde{M}}, \quad \boldsymbol{\tilde{\Sigma}} = \boldsymbol{\tilde{M}}^T \boldsymbol{\Sigma} \boldsymbol{\tilde{M}}, \quad \boldsymbol{\tilde{M}} = \begin{bmatrix}
    \boldsymbol{\tilde{I}} & \boldsymbol{0}
    \end{bmatrix}^T,
\end{equation}
and $\boldsymbol{\tilde{I}}$ is the identity matrix of size $N_{\rm mode} \leq N_u$. Here, the forcing ($\boldsymbol{\phi_j} \in \mathbb{C}^{N_u}$) and response ($\boldsymbol{\psi_j} \in \mathbb{C}^{N_u}$) modes are the columns of $\boldsymbol{\Phi}$ and $\boldsymbol{\Psi}$, respectively, and their weightings ensure that orthonormality enforces an energy norm. These pairs of modes are ranked according to their gains, $\sigma_j$, which are the diagonal elements of $\boldsymbol{\Sigma}$ and ordered such that $\sigma_j \geq \sigma_{j+1}$.

The approximation in (\ref{eq:low_rank}) is founded on the premise that the gains for modes $j \leq N_{\rm mode}$ dominate those for modes $j > N_{\rm mode}$. However, response modes associated with relatively small gains can remain important to the velocity fluctuations if the forcing has a large projection onto the corresponding forcing modes. For each triplet, we collect the gain-scaled projections of the nonlinear forcing onto the $N_{\rm mode}$ selected forcing modes into a column vector of mode weights,
\begin{equation}\label{eq:mode_weights}
    \boldsymbol{c} = \boldsymbol{\tilde{\Sigma}} \left( \boldsymbol{W_f} \boldsymbol{\tilde{\Phi}} \right)^H \left( \boldsymbol{W_f} \boldsymbol{\hat{f}} \right),
\end{equation}
that characterizes the contributions of the selected response modes to the response. The linear model corresponding to the truncated flow representation is given by $\boldsymbol{\hat{u}} \approx \boldsymbol{\tilde{\Psi}} \boldsymbol{c}$.

The SVD of the resolvent operator does not provide information about $\boldsymbol{S_{cc}}$ without further assumptions. However, we may form a model of this CSD by recalling that resolvent modes and SPOD modes are identical when the resolvent mode weights are uncorrelated \citep{Tow2018}. To achieve this, we assume an uncorrelated weighted forcing, $\boldsymbol{W_f} \mathbb{E}\left( \boldsymbol{\hat{f}} \boldsymbol{\hat{f}}^H \right) \boldsymbol{W}^H_{\boldsymbol{f}} = \boldsymbol{I}$, for the present TRME-inspired estimator. Noting the orthonormality of the forcing modes, the modeled mode weight CSD is simply $\boldsymbol{\tilde{S}_{cc}} = \boldsymbol{\tilde{\Sigma}} \boldsymbol{\tilde{\Sigma}}^H = \boldsymbol{\tilde{\Sigma}}^2$. Since this statistical perspective implies that resolvent modes form an approximation of SPOD modes, we expect that a truncated SPOD mode basis is more well posed to fitting the response than a truncated (resolvent) response mode basis.

Assuming the resolvent response modes form a suitable basis for the linear model, we may form the constituents of the estimation transfer function in (\ref{eq:tfdef}) by taking
\begin{equation}
    \boldsymbol{H_u} = \boldsymbol{\tilde{\Psi}}, \quad \boldsymbol{H_y} = \boldsymbol{\tilde{\Psi}_y}, \quad {\rm and} \quad \boldsymbol{b} = \boldsymbol{c},
\end{equation}
where $\boldsymbol{\tilde{\Psi}_y} = \boldsymbol{C}\boldsymbol{\tilde{\Psi}}$. With these definitions, the transfer function is expressed as
\begin{equation}\label{eq:tftrme}
    \boldsymbol{T}^{\rm TRME}_{\boldsymbol{u}} = \boldsymbol{\tilde{\Psi}} \boldsymbol{\tilde{\Sigma}^2} \boldsymbol{\tilde{\Psi}}^H_{\boldsymbol{y}} \left( \boldsymbol{\tilde{\Psi}_y} \boldsymbol{\tilde{\Sigma}^2} \boldsymbol{\tilde{\Psi}}^H_{\boldsymbol{y}} + \boldsymbol{\tilde{S}_{nn}} \right)^{-1}.
\end{equation}
This estimation method contrasts slightly from the direct linear least-squares method of \citet{Gom2016} in that we assume that the coefficient statistics are known (and uncorrelated). However, since this method arises from the generalized Wiener filter formulation, employing a suitable model on $\boldsymbol{\tilde{S}_{cc}}$ can mitigate the instability of higher-order modal truncations associated with conventional TRME \citep{Mar2020}. Hence, this method provides an alternative to limiting modal truncations such that $N_{\rm mode} < N_{\rm plane}$, which discards sensor information.

%%%%%%%%%%%%%%%%%%%%%%%%%%%%%%%%%%%%%%%%%%
\subsubsection{Optimal resolvent-based estimation}\label{ssec:orbe}
%%%%%%%%%%%%%%%%%%%%%%%%%%%%%%%%%%%%%%%%%%

The TRME-based method in Sec. \ref{ssec:trme} uses an (unrealistic) uncorrelated forcing model to produce estimators that only require a mean profile. ORBE \citep{Mar2020} provides a practical means of improving the forcing model by estimating its CSD using limited auxiliary data. In ORBE, the constituents of the linear estimator are given by
\begin{equation}
    \boldsymbol{H_u} = \boldsymbol{R_u}, \quad \boldsymbol{H_y} = \boldsymbol{R_y}, \quad {\rm and} \quad \boldsymbol{b} = \boldsymbol{\hat{f}},
\end{equation}
where we refer to the input-output operator, $\boldsymbol{R_y} = \boldsymbol{C} \boldsymbol{R_u}$, distinctly from the resolvent operator. Hence, the ORBE transfer function is given by
\begin{equation}\label{eq:Tfn}
    \boldsymbol{T}^{\rm ORBE}_{\boldsymbol{u}} = \boldsymbol{R_u} \boldsymbol{\tilde{S}_{ff}} \boldsymbol{R}^H_{\boldsymbol{y}} \left( \boldsymbol{R_y} \boldsymbol{\tilde{S}_{ff}} \boldsymbol{R}^H_{\boldsymbol{y}} + \boldsymbol{\tilde{S}_{nn}} \right)^{-1}.
\end{equation}
Even though ORBE was originally formulated using an unweighted inner product \citep{Mar2020}, (\ref{eq:Tfn}) and (\ref{eq:we_Wiener}) reinforce that the Wiener filter origins of the transfer function imply it is invariant to nondegenerate weights imposed on the input-output formulation. We emphasize that (\ref{eq:Tfn}) is an optimal noncausal transfer function and that forming the optimal causal transfer functions derived via the Wiener-Hopf formalism \citep{Mar2022} requires more complex numerical methods. 

Instead of truncating a modal expansion of the dynamics \textit{a priori}, ORBE identifies important parts of the forcing even if they reside in linearly suboptimal resolvent modes, e.g., when their projections onto the relevant forcing bases are large. Further, while the truncations associated with TRME and TSME may result in an ill-posed basis for representing the full fluctuation dynamics, ORBE retains a full-rank (and therefore well-posed) representation of the velocity fluctuations. When all resolvent modes are retained, the TRME-based method discussed in Sec. \ref{ssec:trme} is equivalent to ORBE under the assumption of an uncorrelated weighted forcing. It is well known that this assumption is not typically valid for turbulent flows, including channel flow, since the forcing color contributes significantly to the forcing structure, e.g., via destructive interference \citep{Ros2019,Mor2020}. Correspondingly, the focus of ORBE is identifying a suitable ROM of the forcing statistics that enables accurate reconstructions.

Following previous studies \citep{Tow2020, Mar2020, Mar2022}, we estimate the forcing statistics using an auxiliary set of measurements, $\boldsymbol{\hat{y}_n'}$, with corresponding CSD, $\boldsymbol{S_{yy,n}'}$, observation matrix, $\boldsymbol{C'}$, and input-output operator, $\boldsymbol{R_{y'}} = \boldsymbol{C'} \boldsymbol{R_u}$. The forcing estimate is given by
\begin{equation}\label{eq:auxforc}
    \boldsymbol{\tilde{S}_{ff}} = \boldsymbol{R}^{\dagger, n}_{\boldsymbol{y'}} \boldsymbol{S_{yy}'} \left( \boldsymbol{R}^{\dagger, n}_{\boldsymbol{y'}} \right)^{H}, \quad \boldsymbol{R}^{\dagger, n}_{\boldsymbol{y'}} = \boldsymbol{R}^{H}_{\boldsymbol{y'}} \left( \boldsymbol{R_{y'}} \boldsymbol{R}^{H}_{\boldsymbol{y'}} + \boldsymbol{\tilde{S}_{nn}'} \right)^{-1},
\end{equation}
where $\boldsymbol{R}^{\dagger, n}_{\boldsymbol{y'}}$ maps the auxiliary observations to the smallest regularized least-squares forcing producing them and reduces to a pseudoinverse in the limit of vanishing noise. We approach this limit by using the same noise model, $\boldsymbol{\tilde{S}_{nn}'} = \epsilon \boldsymbol{I}$, as that used for the reconstruction observations. One advantage of this estimation technique is that a limited set of measurements can produce a global forcing estimate, whereas computing SPOD modes (see Sec. \ref{ssec:tsme}) typically requires knowledge of the full-field response statistics.

When full-field statistics are available, the ORBE transfer function reduces to the generalized Wiener filter transfer function, $\boldsymbol{S_{uy}}\boldsymbol{S}^{-1}_{\boldsymbol{yy,n}}$, which is independent of the governing equations. With noisy or incomplete statistics, ORBE implicitly assumes that the coefficients are the resolvent forcing. This distinction is important since the forcing statistics may not be an optimal setting for approximating $\boldsymbol{S_{uy}} \boldsymbol{S}^{-1}_{\boldsymbol{yy,n}}$ from limited data. Hence, while ORBE provides a forcing-based means of flow reconstruction, determining the optimal coefficient coordinates for flow reconstruction remains an open question.

%%%%%%%%%%%%%%%%%%%%%%%%%%%%%%%%%%%%%%%%%%
\subsubsection{Truncated SPOD mode estimation} \label{ssec:tsme}
%%%%%%%%%%%%%%%%%%%%%%%%%%%%%%%%%%%%%%%%%%

TSME provides a framework for estimating velocity fluctuations by truncating the SPOD basis to efficiently represent coherent structures in the flow. TSME requires the mean profile and second-order statistics of the velocity (response) fluctuations, but it does not consider the dynamics associated with the governing equations. For each triplet, we form the response CSD \footnote{Although $\boldsymbol{S_{uu}}$ is technically an estimate over $N_{\rm real}$ realizations, we assume these statistics to be converged and therefore treat such CSDs as exact.}, $\boldsymbol{S_{uu}},$ using an ensemble of $N_{\rm real}$ realizations of the flow. The corresponding SPOD modes, $\boldsymbol{\theta_j} \in \mathbb{C}^{N_u}$, are given by the columns of $\boldsymbol{\Theta}$ and computed via a direct eigenvalue problem,
\begin{equation}\label{eq:SPODeig}
    \boldsymbol{S_{uu}} \left(\boldsymbol{W}^{H}_{\boldsymbol{u}} \boldsymbol{W_u} \right) \boldsymbol{\Theta} = \boldsymbol{\Theta} \boldsymbol{\Lambda}.
\end{equation}
As before, the weighting matrix, $\boldsymbol{W_u}$, ensures that orthonormality enforces an energy norm. This formulation is typically employed when $N_{\rm real} \geq N_u$, whereas the method of snapshots is standard when $N_{\rm real} < N_u$. Since it is often the case that $N_{\rm real} \ll N_u$ when estimating turbulent flows, the method of snapshots is often less computationally expensive than the direct eigenvalue problem. However, since the CSD sizes are not prohibitive in the present investigation, we simply employ the direct eigenvalue formulation in (\ref{eq:SPODeig}).

The matrix $\boldsymbol{\Lambda}$ contains the eigenvalues that rank the SPOD modes in terms of the variance they capture in the response. Since a given subset of SPOD modes captures more energy than in any other orthogonal basis \citep{Sch2020}, truncating the SPOD mode basis provides a natural means of efficiently representing dominant coherent flow structures. The truncated representation of the flow is given by $\boldsymbol{\hat{u}} \approx \boldsymbol{\tilde{\Theta}} \boldsymbol{a}$, where
\begin{equation}
    \boldsymbol{\tilde{\Theta}} = \boldsymbol{\Theta} \boldsymbol{\tilde{M}}, \quad \boldsymbol{\tilde{\Lambda}} = \boldsymbol{\tilde{M}}^T \boldsymbol{\Lambda} \boldsymbol{\tilde{M}},
\end{equation}
and $\boldsymbol{\tilde{M}}$ again selects the first $N_{\rm mode}$ modes of the expansion (for $N_{\rm mode} \leq N_{\rm real}$).

The SPOD mode expansion is advantageous in that the coefficients are uncorrelated and the relevant CSD is given by $\boldsymbol{S_{aa}} = \boldsymbol{a} \boldsymbol{a}^H = \boldsymbol{\tilde{\Lambda}}$. As such, the elements of the transfer function in (\ref{eq:tfdef}) are given by
\begin{equation}
    \boldsymbol{H_u} = \boldsymbol{\tilde{\Theta}}, \quad \boldsymbol{H_y} = \boldsymbol{\tilde{\Theta}_y}, \quad {\rm and} \quad \boldsymbol{b} = \boldsymbol{a},
\end{equation}
where $\boldsymbol{\tilde{\Theta}_y} = \boldsymbol{C} \boldsymbol{\tilde{\Theta}}$. Correspondingly, the SPOD modes and energies suffice in providing both the linear model and the coefficient statistics implicit to the optimality of the linear transfer function. When all modes are retained, the TSME estimator converges to the generalized Wiener filter, $\boldsymbol{S_{uy}}\boldsymbol{S}^{-1}_{\boldsymbol{yy,n}}$. Even when no modal truncation is performed, it is important to capture sufficiently many realizations to ensure the SPOD modes provide a well-posed basis for the flow.

For TSME, the transfer function is given by
\begin{equation}\label{eq:tfspod}
    \boldsymbol{T}^{\rm TSME}_{\boldsymbol{u}} = \boldsymbol{\tilde{\Theta}} \boldsymbol{\tilde{\Lambda}} \boldsymbol{\tilde{\Theta}}^H_{\boldsymbol{y}} \left( \boldsymbol{\tilde{\Theta}_y} \boldsymbol{\tilde{\Lambda}} \boldsymbol{\tilde{\Theta}}^H_{\boldsymbol{y}} + \boldsymbol{\tilde{S}_{nn}} \right)^{-1}.
\end{equation}
If the weighted resolvent forcing were truly uncorrelated, such that $\boldsymbol{\tilde{\Sigma}}^2 = \boldsymbol{\tilde{\Lambda}}$ and $\boldsymbol{\tilde{\Psi}} = \boldsymbol{\tilde{\Theta}}$, (\ref{eq:tftrme}) and (\ref{eq:tfspod}) would be identical. As such, the TRME estimation method can be used to evaluate the validity of the assumed forcing statistics and thereby the ability of the resolvent modes to approximate SPOD modes for flow reconstruction.

%%%%%%%%%%%%%%%%%%%%%%%%%%%%%%%%%%%%%%%%%%
\section{Methods and Approach}\label{sec:meth}
%%%%%%%%%%%%%%%%%%%%%%%%%%%%%%%%%%%%%%%%%%

As a prototypical real-time application of the estimation techniques discussed in Sec. \ref{sec:form}, we reconstruct turbulent velocity fluctuations in a minimal channel by continuously incorporating streaming measurements in a manner akin to time-frequency analysis. Within this streaming context, we refer to the techniques in Secs. \ref{ssec:trme}, \ref{ssec:orbe}, and \ref{ssec:tsme} as streaming TRME (STRME), streaming ORBE (SORBE), and streaming TSME (STSME), respectively. In what follows, we detail our methods and highlight how we enable and evaluate efficient streaming flow reconstructions.

%%%%%%%%%%%%%%%%%%%%%%%%%%%%%%%%%%%%%%%%%%
\subsection{Numerical simulation details}\label{ssec:meth:nsd}
%%%%%%%%%%%%%%%%%%%%%%%%%%%%%%%%%%%%%%%%%%

We apply the flow reconstruction framework to reconstruct velocity fluctuations from a DNS \footnote{Whereas the time required to generate each DNS snapshot, which is roughly $\mathcal{O}(1) \; {\rm s}$ here, can scale rapidly with ${\rm Re}_\tau$ \citep{Cha1979, Cho2012}, this limitation is mitigated in analogous experiments.} of turbulent channel flow in a minimal flow unit \citep{Jim1991} at ${\rm Re}_{\tau} \approx 186$, as discussed previously by \citet{Bae2021}. As depicted in Fig. \ref{fig:stag}, the simulation data exist on a staggered grid with $(N_x, N_y, N_z) = (32, 129, 32)$ cells in each direction, excluding ghost cells. The grid resolution is uniform in $x$ and $z$ and follows a hyperbolic tangent distribution in $y$. The governing equations are spatially discretized using second-order central finite differences and time-marched using an explicit third-order Runge-Kutta scheme in the simulation. Table \ref{tab:simparams} summarizes relevant spatiotemporal parameters associated with the data we use from the DNS. While $\nu$ and $h$ are fixed, $u_\tau$ is estimated over a long time series and assumed to be constant. In the present work, we subsample the DNS time steps such that $\Delta t^+ > \Delta t^+_{\rm DNS}$ to efficiently process snapshots without significantly affecting the resolution of relevant temporal frequencies.
\begin{table}[!htpb]
    \centering
    \renewcommand*{\arraystretch}{1.4}
    \begin{tabular}{|c|c|c|c|c|c|c|c|}
    \hline
         $\Delta x^+$ & $\Delta z^+$ & $\Delta y^+_{\rm min}$ & $\Delta y^+_{\rm max}$ & $\Delta t^+$ & $u_\tau$ & $\nu$ &  $h$\\
         \hline
         10.3 &  5.16 & 0.172 & 7.58 & 0.532 & 0.0572 & 0.000308 & 1.00 \\
         \hline
    \end{tabular}
    \caption{Relevant parameters for data sampled from the DNS, where nondimensional parameters with superscripts $(\cdot)^+$ are normalized in viscous units (e.g., $\Delta t^+ = {\rm Re}_\tau \Delta t$). The values of $u_\tau$, $\nu$, and $h$ are dimensional, but the dimensional units are arbitrary so long as they are consistent.}
    \label{tab:simparams}
\end{table}

%%%%%%%%%%%%%%%%%%%%%%%%%%%%%%%%%%%%%%%%%%
\subsection{Measurement details}\label{ssec:meth:md}
%%%%%%%%%%%%%%%%%%%%%%%%%%%%%%%%%%%%%%%%%%

We reconstruct velocity fluctuations using triplet space measurements that may evolve over time in a manner similar to time-frequency analysis. Measurements and reconstructions are restricted in the present study to the smallest $(N_{k_x}, N_{k_z}) = (3,5)$ nonnegative streamwise and spanwise wave numbers and their complex conjugates. While this implies that we truncate more than 95\% of the Fourier modes corresponding to the spatial discretization scheme, these modes still contain a dominant portion of the response energy. The temporal Fourier modes are computed by discretizing a single eddy turnover time (i.e., $\Delta T = 1$) and consist of $N_\omega = 176$ nonnegative frequencies and their complex conjugates. These selections allow us to capture convection velocities more than $8\times$ larger than the centerline velocity, although most energetic motions propagate significantly slower. Considerable energy is contained by modes with periods longer than $\Delta T = 1$, but the present configuration represents a balance of capturing energetic Fourier modes and retaining efficient reconstructions. The zero values in each triplet are numerically set to $\mathcal{O}(10^{-10})$, and we verify that the transfer functions are invariant to this threshold.

We use simple heuristic arguments to inform sensor placement. As summarized in Table \ref{tab:meas1}, we consider a spectrum of measurement configurations ranging from relatively cheap (case A) to relatively expensive (case E). Case A represents a single sensor plane coinciding with the peak streamwise root-mean-square (rms) velocity fluctuations, $u_{\rm rms}$. It is unique in that it evaluates the efficacy of sensors localized to one half of the channel. Case B is similar but symmetrically extends the configuration about the centerline. Case C extends the configuration of Case B to also consider measurements near the peaks of $v_{\rm rms}$ and $w_{\rm rms}$. Case D adds sensor planes between these peaks and the centerline to better capture the outer structure of the fluctuations. Case E adds the centerline to evaluate whether it significantly improves the quality of the reconstructions. 

\begin{table}[!htp]
\begin{center}
\begin{tabular}{ |c|c|lll| } 
 \hline
 Case & $N_{\rm plane}$ & $y^+$ & & \\ 
 \hline
 A & 1 & 14.7  & & \\ 
 \hline
 B & 2 & 14.7 & (both sides) & \\%, 357 \\
 \hline
 C & 4 & 14.7, 56.4 & (both sides) & \\%, 316, 357 \\
 \hline
 D & 6 & 14.7, 56.4, 114 & (both sides) &\\%, 258, 316, 357 \\
 \hline
 E & 7 & 14.7, 56.4, 114 & (both sides), 186 &(centerline) \\%, 258, 316, 357 \\
 \hline
\end{tabular}
\end{center}
\caption{Measurement configurations for the testing period, including the number of wall-parallel planes ($N_{\rm plane}$) and the wall-normal locations ($y^+$) in viscous units.\label{tab:meas1}}
\end{table}

\begin{figure}[!ht]
    \centering
    \includegraphics[width=0.83\textwidth]{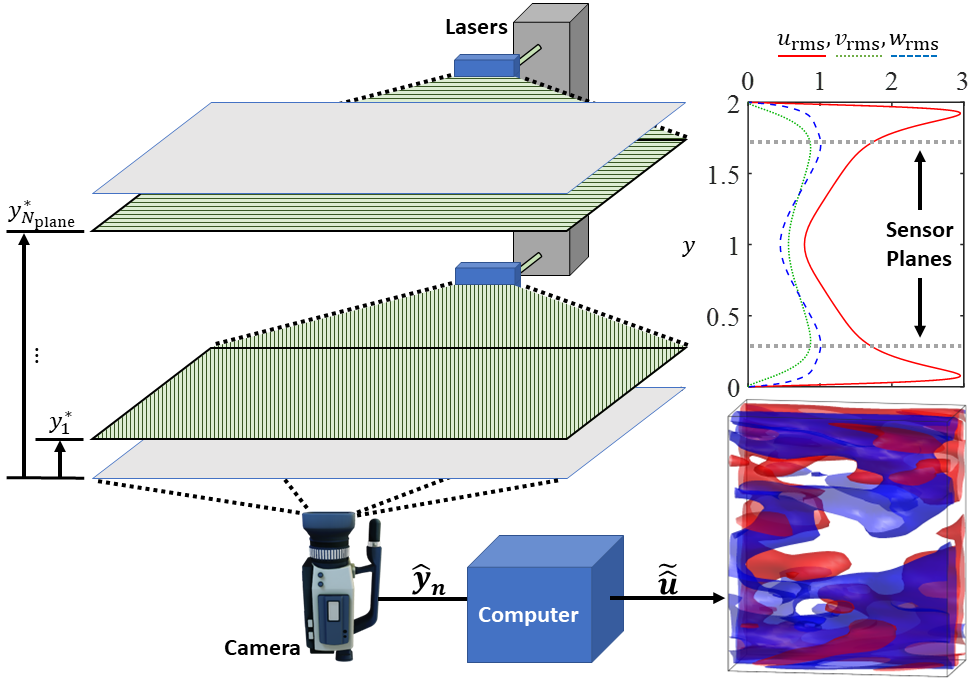}
    \caption{Analogous multiplane PIV experimental configuration that is amenable to real-time reconstructions of velocity fluctuations under the present framework. The lines along each laser sheet represent orthogonal polarizations that enable simultaneous measurements, and more than one camera may be needed to fully capture all measurement planes \citep{Lib2004,Sch2004}. The wall-normal sensor locations with superscripts $(\cdot)^*$ are selected from Table \ref{tab:meas1} and represent testing period measurements.}
    \label{fig:smallfig}
\end{figure}

Using these restricted numerical data, we simulate the process of obtaining experimentally feasible measurements. An analogous experimental configuration involves using standard multiplane PIV techniques \citep{Lib2004,Sch2004} to obtain velocity measurements in wall-parallel planes. Figure \ref{fig:smallfig} shows a diagram of this configuration in the context of real-time flow reconstructions. This configuration contrasts slightly with standard $x-y$ planar PIV setups, and inhomogeneities associated with bounding walls may need to be modeled into the noise CSD. Experimental reconstructions would require windowed spatial Fourier transforms in $x$ and $z$ since periodicity cannot be guaranteed, but this complication may also be absorbed directly into the noise vector. When designing PIV measurements, the relevant streamwise and spanwise scales of large-scale motions must be known (or modeled) to accurately capture their Fourier space representations. However, since the measurements consistently update the phase information of the estimated fluctuations, this restriction may be mitigated by absorbing unresolved large-scale motions into the streamwise-constant modes at each time step.

%%%%%%%%%%%%%%%%%%%%%%%%%%%%%%%%%%%%%%%%%%
\subsection{Training and testing periods}
%%%%%%%%%%%%%%%%%%%%%%%%%%%%%%%%%%%%%%%%%%

\begin{figure}[!ht]
    \centering
    \includegraphics[width=\textwidth]{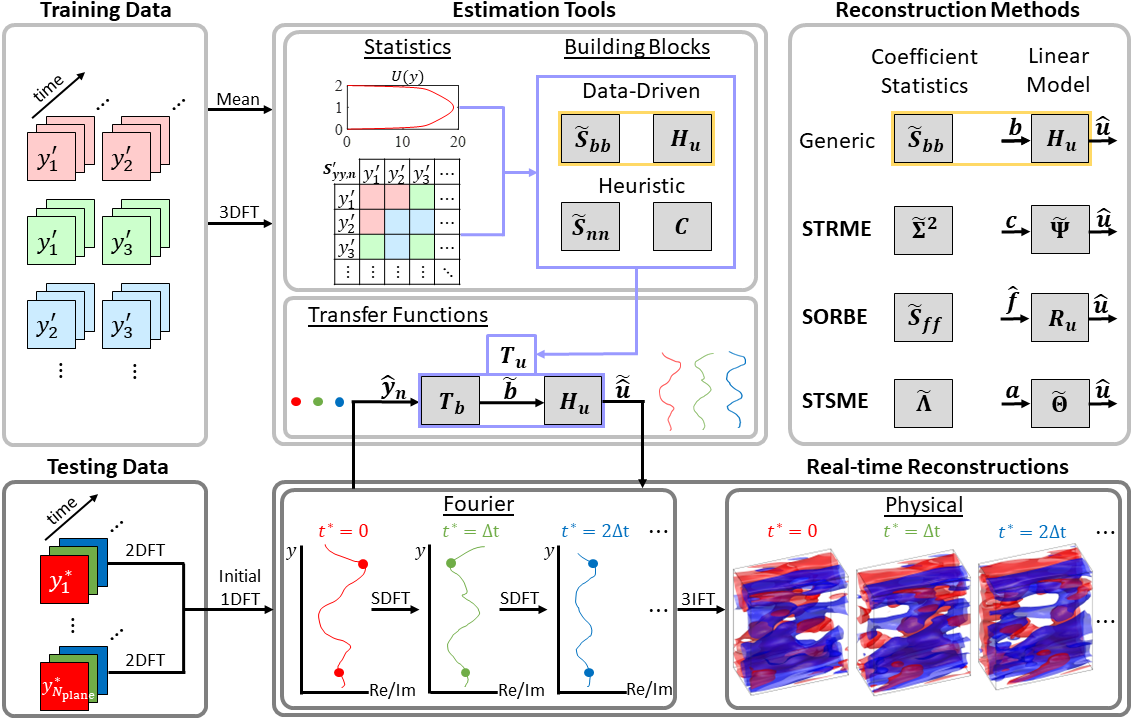}
    \caption{Diagram depicting the methods used to estimate linear transfer functions for each reconstruction method during training (top) and to reconstruct velocity fluctuations in real time from sparse measurements during testing (bottom). Discrete Fourier transforms (DFTs) and inverse discrete Fourier transforms (IFTs) are labeled by their dimensionality. SDFT refers to a (recursive) sliding temporal DFT applied every time step, as discussed in Sec. \ref{sec:sdft}.}
    \label{fig:bigfig}
\end{figure}

The reconstruction techniques we use require \textit{a priori} knowledge of various linear models and flow statistics to form linear estimators. To address this, we partition the DNS data into training and testing periods of equal size such that $\Delta T' = \Delta T^* = 80$, where $(\cdot)'$ and $(\cdot)^*$ represent training and testing period quantities, respectively. As summarized in Fig. \ref{fig:bigfig}, we form each estimator using flow statistics from the training period and evaluate the performance of streaming flow reconstructions during the subsequent testing period. While $u_\tau$ is assumed to be known prior to training, the error in $u_\tau$ estimated over the training period with respect to its assumed constant value is 0.172\%. Further, all present flow estimators and their reconstruction accuracies are relatively insensitive to $\Delta T'$ and $\Delta T^*$, respectively.

STRME, SORBE, and STSME require various degrees of training data to form the corresponding estimators. STRME requires only a mean profile to compute the resolvent operator, which may be efficiently estimated using ergodicity or modeled using scaling laws. As discussed in Sec. \ref{ssec:bi}, we introduce an efficient and accurate blockwise inversion technique to compute the resolvent operator for STRME and SORBE. SORBE requires more detailed (second-order) statistics, $\boldsymbol{S_{yy,n}'}$, for an arbitrary number of auxiliary sensor planes. By limiting the number of sensor planes considered, SORBE provides an efficient means of estimating global flow statistics from limited training data. At the high end, STSME requires full-field second-order statistics, $\boldsymbol{S_{uu}}$, to compute SPOD modes and energies. In this limit of complete statistical information, both (untruncated) STSME and SORBE approach the generalized Wiener filter transfer function, $\boldsymbol{S_{uy}}\boldsymbol{S}^{-1}_{\boldsymbol{yy,n}}$.

For SORBE and STSME, we assume 50\% overlap between adjacent $\Delta T = 1$ windows when computing the relevant CSDs. This gives $N_{\rm real} = 158$ realizations of each measurement set over the training period. We intentionally design these experiments such that $N_{\rm real} < N_u$ to evaluate the efficacy of our statistical methods under typical conditions for turbulence estimation problems. We do not use windowing in computing the CSDs to preserve the relationships associated with the forcing and response (CSDs) without considering a correction term \citep{Mor2020,Mar2020,Nog2020}.

For each measurement configuration in Table \ref{tab:meas1}, we consider estimators of varying fidelity based on auxiliary measurements (SORBE) and modal truncations (STRME and STSME). We summarize the fidelities of these estimators in Table \ref{tab:subcases}. 
\begin{table}[!htpb]
    \centering
    \begin{tabular}{|c|c|l|}
    \hline
    Method & Parameter & Values \\
    \hline
    SORBE & $y'$ & $y^*$ only \\
          &      & $y^*$ and $y^+ \leq 14.7$ \\
          &      & $y^*$ and $y^+ \leq 56.4$ \\
          &      & $y^*$ and $y^+ \leq 114$ \\
          &      & all $y$  \\
          \hline
    STRME & $N_{\rm mode}$ & 2, 8, 40, 158, 388 \\
    \hline
    STSME & $N_{\rm mode}$ & 2, 8, 40, 158 \\
    \hline
    \end{tabular}
    \caption{Subcases considered for each reconstruction method, characterized by the extent of auxiliary sensors (SORBE) and the rank of modal truncations (STRME and STSME). For each case, the subcases are listed in order of increasing fidelity. For SORBE, the $y^+$ inequalities apply to both sides of the channel.}
    \label{tab:subcases}
\end{table}

As depicted in Fig. \ref{fig:bigfig}, during training we consider various pairs of wall-normal coordinates, $(y_i', y_j')$, associated with auxiliary measurements that are distinct from the fixed measurements used for reconstruction. We may compute each element of $\boldsymbol{S_{yy,n}'}$ (for SORBE) or $\boldsymbol{S_{uu}}$ (for STSME) in parallel for all $\boldsymbol{k}$ using two simultaneous planar (PIV) sensors. In experiments, this implies that the length of the training period is dictated by the number of sensor pairs, $N_{yy}$, required to sufficiently resolve the wall-normal structure of the second-order statistics. However, under the assumption of statistical stationarity, we simultaneously collect all elements of the relevant CSDs during the training period for the present (numerical) investigations. In effect, this parallelizes the process of simulating two-plane PIV measurements for each sensor pair, which would be conducted sequentially in experiments absent extra sensors. We emphasize that $N_{yy}$ does not necessarily scale as $N_y^2$ since an underlying model with sufficient fidelity can often capture the structure of the statistics with relatively few spatial measurements \citep{Man2018,Mor2020}.

During the subsequent testing period, we use the estimators from training to reconstruct the flow with minimal time delay, as is required in real-time applications. One advantage of computing the transfer functions during training is that the flow at each triplet is efficiently reconstructed with a single application of the transfer function (via matrix multiplication). We further enable fast reconstructions by applying this operation simultaneously for all selected triplets. Since no sensitive computational techniques (e.g., inverses) are required after training, we truncate the transfer functions to single precision to improve memory usage and temporal efficiency without significantly impacting reconstruction quality. 

Since an infinite time series is not available, we estimate the temporal Fourier coefficients using a relatively small sliding window ($\Delta T = 1$). We accommodate real-time estimation problems by continuously updating the temporal Fourier coefficients using the incoming data stream. As discussed in Sec. \ref{sec:sdft}, we enable efficient streaming reconstructions by recursively updating the coefficients using the SDFT as new measurements arrive. The errors associated with the finite temporal window are implicitly accounted for in $\boldsymbol{\hat{n}}$, highlighting that the assumption that $\boldsymbol{\tilde{S}_{nn}} = \epsilon \boldsymbol{I}$ \citep{Ama2020, Mar2020} is indeed a simplified model in the present investigation. Appendix \ref{sec:app:sdft} \rev{shows that} the short-time ($\Delta T = 1$) and long-time ($\Delta T' = 80$) Fourier amplitude spectra \rev{are consistent with each other} for a selected wave number pair. For all reconstructions, we assume that $\epsilon = 10^{-8}$ to approach the zero-noise limit.

%%%%%%%%%%%%%%%%%%%%%%%%%%%%%%%%%%%%%%%%%%
\subsubsection{Blockwise inversion}\label{ssec:bi}
%%%%%%%%%%%%%%%%%%%%%%%%%%%%%%%%%%%%%%%%%%

In the present work, we invoke the resolvent operator when computing the STRME and SORBE estimators during training. More generally, computing the resolvent operator and forming its modal decomposition are central challenges in efficiently and accurately representing fluctuation dynamics. Efficient randomized \citep{Rib2020} and matrix-free \citep{Mar2021} techniques for forming these equation-based elements have been studied previously. Here, we introduce the exact blockwise representation of the linear dynamics to enable efficient and accurate computation of the channel flow resolvent operator by means of blockwise inversion. This methodology takes advantage of the simple blockwise structure associated with 1D resolvent analysis and expresses the inverse of the full linear dynamics in terms of its smaller subblock constituents.

Conventionally, the resolvent operator in (\ref{eq:oper}) is obtained by directly inverting $\boldsymbol{L}$ and requires a large matrix inverse. However, the form of $\boldsymbol{L}$ in (\ref{calL}) lends itself to blockwise inversion. We begin by blockwise inverting the basic subblock of $\boldsymbol{L}$ to give the basic resolvent operator,
\begin{equation}
    \boldsymbol{R_u^B} = \boldsymbol{L}^{-1}_{\boldsymbol{B}} =  \begin{bmatrix}
    \boldsymbol{L}^{-1}_{\boldsymbol{c}}  & -\boldsymbol{L}^{-1}_{\boldsymbol{c}}\left(\boldsymbol{\partial_y U}\right)\boldsymbol{L}^{-1}_{\boldsymbol{e}} & \boldsymbol{0} \\
    \boldsymbol{0} & \boldsymbol{L}^{-1}_{\boldsymbol{e}} & \boldsymbol{0} \\
    \boldsymbol{0} & \boldsymbol{0} & \boldsymbol{L}^{-1}_{\boldsymbol{c}}
    \end{bmatrix},
\end{equation}
which is defined here between forcing and response quantities that are aligned in $x$ and $z$ for convenience (see Sec. \ref{ssec:dls}). The simple inverse defining the basic resolvent neglects the contribution of the pressure gradient term and that only the solenoidal forcing and response are active in the input-output formulation for velocity fluctuations \citep{Ros2019,Mor2020}. Now, by blockwise inverting $\boldsymbol{L}$ and applying the definition of the resolvent in (\ref{eq:oper}), we express the resolvent operator in terms of the basic resolvent and the continuity subblock as
\begin{equation}\label{eq:conterr}
    \boldsymbol{R_u} = \boldsymbol{S}^H \left[ \boldsymbol{R_u^B} - \boldsymbol{R_u^B} \boldsymbol{\hat{\nabla}} \left( \boldsymbol{\hat{\nabla}}^T \boldsymbol{R_u^B} \boldsymbol{\hat{\nabla}} \right)^{-1} \boldsymbol{\hat{\nabla}}^T \boldsymbol{R_u^B} \right] \boldsymbol{S}.
\end{equation}
This equation may be used to rewrite the resolvent formulation in (\ref{eq:oper}) as
\begin{flalign}\label{eq:proj}
\boldsymbol{\hat{u}} &= \boldsymbol{S}^H \left[ \boldsymbol{I} - \boldsymbol{R_u^B} \boldsymbol{\hat{\nabla}} \left( \boldsymbol{\hat{\nabla}}^T \boldsymbol{R_u^B} \boldsymbol{\hat{\nabla}} \right)^{-1} \boldsymbol{\hat{\nabla}}^T \right] \boldsymbol{R_u^B} \boldsymbol{S} \boldsymbol{\hat{f}} \\
&= \boldsymbol{S}^H \boldsymbol{R_u^B} \left[ \boldsymbol{I} - \boldsymbol{\hat{\nabla}} \left( \boldsymbol{\hat{\nabla}}^T \boldsymbol{R_u^B} \boldsymbol{\hat{\nabla}} \right)^{-1} \boldsymbol{\hat{\nabla}}^T \boldsymbol{R_u^B} \right] \boldsymbol{S} \boldsymbol{\hat{f}},
\end{flalign}
where the terms in square brackets represent projection matrices since they satisfy the property that $\boldsymbol{P}^2 = \boldsymbol{P}$. These terms share a similar form with the Leray projection typically used to remove $\hat{p}$, except for the inclusion of the basic resolvent. Since this formulation also serves to enforce continuity and eliminate pressure fluctuations \footnote{The pressure can alternatively be eliminated by defining a state vector of the system in terms of the wall-normal velocity and vorticity \citep{Moa2013}, but doing so complicates data processing in analogous experiments.}, we refer to it as a ``modified projection.'' The validity of applying the projection terms both before and after applying the basic resolvent highlights the correspondence between the solenoidal forcing and response for velocity fluctuations. Similar expressions are readily derived for the continuous linear system, (\ref{eq:cont_sys}), by expressing $\boldsymbol{\hat{u}}$ in terms of $\boldsymbol{\hat{f}}$ and $\nabla \hat{p}$ and enforcing continuity to eliminate the pressure term.

There are numerous advantages to computing the resolvent operator using blockwise inversion as opposed to directly inverting $\boldsymbol{L}$. First, the matrix inversions (of $\boldsymbol{L_c}$, $\boldsymbol{L_e}$, and $\boldsymbol{\hat{\nabla}}^T \boldsymbol{R_u^B} \boldsymbol{\hat{\nabla}}$) required using blockwise inversion are roughly $4 \times 4$ smaller than that of $\boldsymbol{L}$. Further, if flow variables are collocated (such that $\boldsymbol{L_c} = \boldsymbol{L_e}$), computing the basic and full resolvent operators requires only one and two small inverses, respectively. The advantage of these smaller matrix inverses is that the smaller subblocks have smaller condition numbers than the full linear system. Therefore, blockwise inversion is typically more computationally efficient and accurate than direct inversion of the entire system. Further, the form of the correction term in (\ref{eq:conterr}) reveals a direct relationship between the basic resolvent operator, where only the terms in the momentum equations that are linear in $\boldsymbol{\hat{u}}$ are inverted, and the full resolvent operator, which includes the continuity constraint and eliminates the pressure fluctuations. Therefore, the present blockwise formulation of the resolvent operator allows for a physical characterization of the influence of various subblocks (representing different facets) of the linear dynamics. \rev{We apply this technique to the resolvent operator modified to include an eddy-viscosity model \citep{Ill2018,Ama2020} in Appendix \ref{sec:app:ev}, but we do not employ that model in the present computations.}

%%%%%%%%%%%%%%%%%%%%%%%%%%%%%%%%%%%%%%%%%%
\subsubsection{Recursive sliding discrete Fourier transform}\label{sec:sdft}
%%%%%%%%%%%%%%%%%%%%%%%%%%%%%%%%%%%%%%%%%%

During the testing period, we employ the sliding discrete Fourier transform (SDFT) \citep{Dud2010,Kol2018,Cha2022} to recursively update the temporal Fourier coefficients of the measurements at every time step, $t_i$. We thereby enable efficient updates of our triplet space measurements as new data arrive. We compute discrete Fourier coefficients using a sliding temporal window of fixed length ($\Delta T = 1$) that is discretized into $N_t$ time steps. Given a (potentially complex) scalar function, $\phi(t)$, we assume that initial discrete Fourier coefficients, $\hat{\phi}_{N_t}(\omega_m)$ for $m = 1, ... , N_t$, are computed using time steps $i = 1, ..., N_t$. The discrete frequencies are given by $\omega_m = \Delta \omega (m - 1)$ for $m \leq \lceil N_t/2 \rceil$ and $\omega_m = \Delta \omega (m - 1 - N_t)$ for $m > \lceil N_t/2 \rceil$, where $\Delta \omega = 2 \pi / N_t \Delta t$ is the frequency resolution. Using the standard SDFT, the $m{\rm th}$ discrete Fourier coefficient at a time step $i > N_t$ is computed recursively as
\begin{equation}\label{eq:sdft}
   \hat{\phi}_{i}(\omega_m) =  \underbrace{{\rm exp}\bigg( {\rm i}\omega_m \Delta t \bigg)}_{\rm Phase \; Update} \bigg( \underbrace{\vphantom{{\rm exp}\bigg( {\rm i}\omega_m \Delta t \bigg)} \hat{\phi}_{i-1}\left(\omega_m\right)}_{\rm Previous \; Value} + \underbrace{\vphantom{{\rm exp}\bigg( {\rm i}\omega_m \Delta t \bigg)} \Big( \phi \left(t_{i}\right) - \phi\left(t_{i-N_t}\right) \Big)}_{\rm Streaming \; Info}  \bigg).
\end{equation}
This process may be continuously applied as new time steps become available, thereby making it amenable to real-time computation of Fourier coefficients over a sliding temporal window.

A primary advantage of the SDFT is that a complete Fourier transform need only be computed over the initial window. Stepping forward in time, the coefficients at time step $i$ are redundant with respect to those at time step $i-1$ except for the difference between the incoming measurement (at $i$) and the outgoing measurement (at $i-N_t$). As such, instead of reusing the \texttt{fft} algorithm, for all time steps $i>N_t$ we may use the SDFT to express the Fourier coefficients recursively in terms of (i) the previous Fourier coefficients, (ii) the aforementioned difference, and (iii) a complex multiplicative correction factor. Correspondingly, the number of required operations scales as $\mathcal{O}(N_t)$ as opposed to $\mathcal{O}(N_t\log{}N_t)$ for the standard \texttt{fft}. Further, since the frequency bins may be updated independently of one another, this complexity is reduced to $\mathcal{O}(N_\omega)$ when considering a limited subset of Fourier modes (i.e., if $N_\omega < N_t$).

The standard SDFT algorithm suffers from marginal stability, moderate accuracy, and time-accumulating error (such that, eventually, the \texttt{fft} algorithm would need to be called to refresh the error). The more recent modulated SDFT (mSDFT) \citep{Dud2010} and observer-based SDFT (oSDFT) \citep{Kol2018} have more desirable stability, accuracy, and robustness to noise compared to the standard SDFT and other variants \citep{Cha2022}. However, we retain the standard SDFT due to its simple implementation and since it remains accurate and stable over sufficiently large intervals for our purposes (see Appendix \ref{sec:app:sdft}). The mSDFT, oSDFT, or some other robust method may be considered to mitigate the propagation of noise over longer time intervals.

%%%%%%%%%%%%%%%%%%%%%%%%%%%%%%%%%%%%%%%%%%
\subsection{Error metrics}\label{sec:errmet}
%%%%%%%%%%%%%%%%%%%%%%%%%%%%%%%%%%%%%%%%%%

A primary objective of our methodology is to produce efficient flow reconstructions that accurately capture the dynamics throughout the channel. Here, we define error metrics that reflect reconstruction accuracy at various levels of detail. \rev{These metrics are advantageous in that they enable comparisons of spatially integrated errors and their wall-normal distributions across different estimators. Whereas previous investigations \citep{Ill2018,Ama2020} have considered similar metrics, they focus more on errors at specific estimation planes and wave numbers.}

\begin{figure}[!htpb]
    \centering
    \includegraphics[trim=0 0 0 0, clip, width=\textwidth]{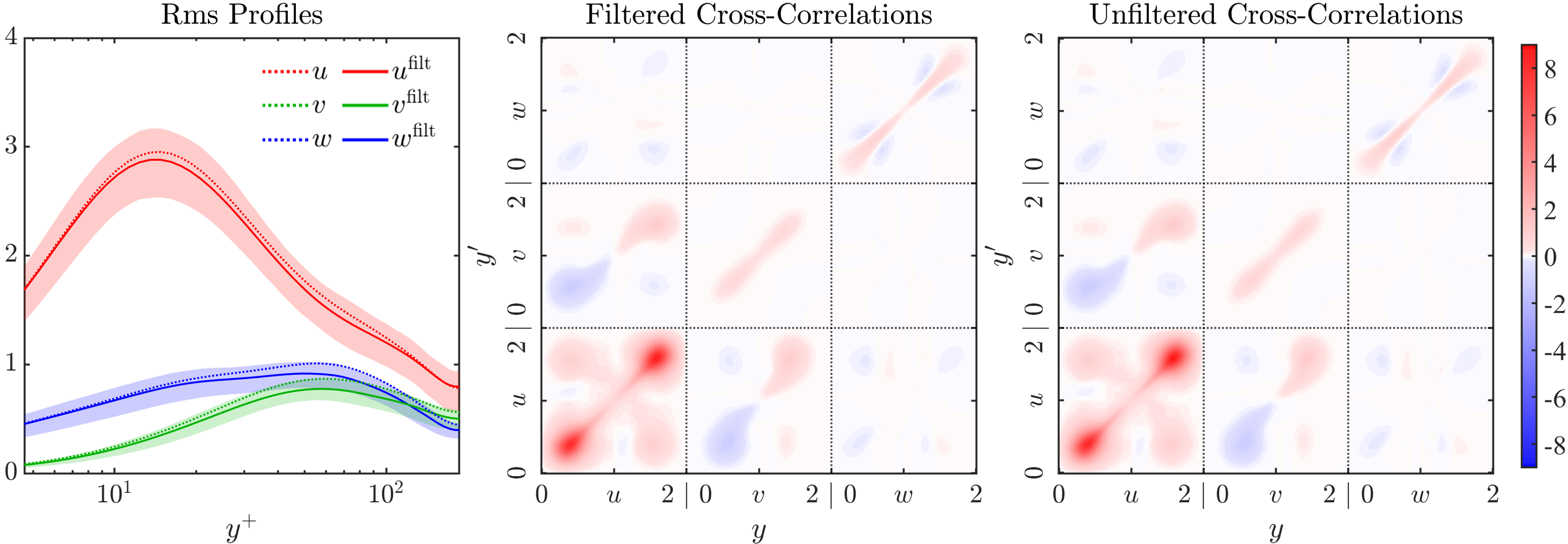}
    \caption{Comparisons between the filtered and unfiltered flow statistics for the testing period. (Left) Comparison between the filtered rms profiles during the testing period (solid curves: \rev{square root of mean variance}, shading: \rev{$20{\rm th} - 80{\rm th}$ percentiles}) and the long-time unfiltered rms profiles (dotted curves). For the testing period, the mean filtered cross-correlations (middle) are shown alongside their unfiltered counterparts (right).}
    \label{fig:1d_2d_stats}
\end{figure}

Noting that we use only a subset of roughly 5\% of the spatial Fourier modes in the channel, we begin by characterizing how well the filtered dynamics associated with these modes capture unfiltered (i.e., true) fluctuations. As shown in Fig. \ref{fig:1d_2d_stats}, the filtered dynamics excellently capture the statistical features of the unfiltered dynamics. On average, the filtered fluctuations during the testing period contain 91.1\% of the turbulence kinetic energy (TKE) of the long-time unfiltered fluctuations. Further, the (wall-normal) cross-correlations associated with filtered and unfiltered dynamics have an excellent agreement over the testing period. These results justify our compressed triplet space flow representation and motivate the definition of the following error metrics.

The error in TKE \rev{coarsely} characterizes the accuracy of the reconstructed turbulence intensity. Correspondingly, we evaluate the turbulence intensity error using the square root of the relative TKE error, given by
\begin{equation}\label{eq:errTKE}
    \epsilon_{\rm TKE}(t) = \left( \frac{\int_0^2 \left| \boldsymbol{u}_{\rm rms}(y,t) \right|^2 dy - \int_0^2 \left| \boldsymbol{\tilde{u}}_{\rm rms}(y,t) \right|^2 dy }{\int_0^2 \left| \boldsymbol{u}_{\rm rms}(y,t) \right|^2 dy} \right)^{1/2},
\end{equation}
where rms fluctuations are computed using the sliding temporal window ($\Delta T = 1$). This sliding window allows all error metrics to take into account time-varying flow statistics in streaming applications, thereby enabling quantification of temporal uncertainties. 

Beyond reconstructing flow statistics, we are primarily interested in the accuracy of the reconstructed dynamics. We consider two error profiles, $\boldsymbol{\epsilon}_{\rm filt}(y,t)$ and $\boldsymbol{\epsilon}_{\rm full}(y,t)$, that capture the wall-normal distributions of the reconstruction errors with respect to the filtered and unfiltered snapshots, respectively. These error profiles are given by
\begin{flalign}
    \boldsymbol{\epsilon}_{\rm filt}(y, t) &= \left( \frac{1}{N_x N_z} || \boldsymbol{u}_{\rm filt}(\boldsymbol{x},t) - \boldsymbol{\tilde{u}}(\boldsymbol{x},t) ||_{x,z}^2 \right)^{1/2}, \\
    \boldsymbol{\epsilon}_{\rm full}(y, t) &= \left( \frac{1}{N_x N_z} || \boldsymbol{u}(\boldsymbol{x},t) - \boldsymbol{\tilde{u}}(\boldsymbol{x},t) ||_{x,z}^2 \right)^{1/2},
\end{flalign}
where $\boldsymbol{u}_{\rm filt}$ represents the filtered snapshots and $|| \cdot ||_{x,z}$ represents the $L^2$ norm computed over the streamwise and spanwise directions.

We also evaluate time-varying full-field errors by integrating the error variance profiles over the wall-normal direction. The corresponding integrated error metrics, which are given by
\begin{equation}\label{eq:errFF}
    \epsilon_{\rm filt}(t) = \left( \frac{\int_0^2 \left| \boldsymbol{\epsilon}_{\rm filt}(y,t) \right|^2 dy}{\int_0^2 \left| \boldsymbol{u}_{\rm rms}(y,t) \right|^2 dy} \right)^{1/2}, \quad \epsilon_{\rm full}(t) = \left( \frac{\int_0^2 \left| \boldsymbol{\epsilon}_{\rm full}(y,t) \right|^2 dy}{\int_0^2 \left| \boldsymbol{u}_{\rm rms}(y,t) \right|^2 dy} \right)^{1/2},
\end{equation}
evaluate (under the square root) the error energies with respect to the TKE. Importantly, we use the same (unfiltered) TKE in both $\epsilon_{\rm filt}(t)$ and $\epsilon_{\rm full}(t)$ to simplify the visual comparisons between both metrics, but using the filtered TKE for $\epsilon_{\rm filt}(t)$ also provides a useful metric.

Since we primarily reconstruct the flow using a small number of sparsely sampled measurements, it may not be reasonable to expect highly accurate reconstructions far from the measurement planes. This premise is especially true for flow structures that are localized in the wall-normal direction. To address this issue, we introduce truncated error metrics, $\epsilon^*_{\rm TKE}(t)$, $\epsilon^*_{\rm filt}(t)$, and $\epsilon^*_{\rm full}(t)$, that evaluate the errors in regions localized about the measurement planes. These metrics are given by
\begin{flalign}
    \epsilon^*_{\rm TKE}(t) &= \left( \frac{\int_{y_{\rm local}} \left| \boldsymbol{u}_{\rm rms}(y,t) \right|^2 dy - \int_{y_{\rm local}} \left| \boldsymbol{\tilde{u}}_{\rm rms}(y,t) \right|^2 dy }{\int_{y_{\rm local}} \left| \boldsymbol{u}_{\rm rms}(y,t) \right|^2 dy} \right)^{1/2}, \label{eq:TKEstar} \\
    \epsilon_{\rm filt}^*(t) &= \left( \frac{\int_{y_{\rm local}} \left| \boldsymbol{\epsilon}_{\rm filt}(y,t) \right|^2 dy}{\int_{y_{\rm local}} \left| \boldsymbol{u}_{\rm rms}(y,t) \right|^2 dy} \right)^{1/2}, \quad \epsilon_{\rm full}^*(t) = \left( \frac{\int_{y_{\rm local}} \left| \boldsymbol{\epsilon}_{\rm full}(y,t) \right|^2 dy}{\int_{y_{\rm local}} \left| \boldsymbol{u}_{\rm rms}(y,t) \right|^2 dy} \right)^{1/2}, \label{eq:ffstar}
\end{flalign}
where $y_{\rm local}$ includes all cells within $N_{\rm local}$ cells from each measurement plane. These definitions allow us to balance reconstruction accuracy with the fraction of the channel captured. Moreover, since $\Delta y_j$ is smaller close to the walls and larger close to the centerline, these metrics provide a simple means of adapting the extent over which the errors are integrated based on the expected scale of the corresponding flow structures.

\rev{One unique advantage of the $\epsilon_{\rm TKE}$ metrics in (\ref{eq:errTKE}) and (\ref{eq:TKEstar}) is that they have the potential to be evaluated in experimental settings. By contrast, the $\epsilon_{\rm filt}$ and $\epsilon_{\rm full}$ metrics in (\ref{eq:errFF}) and (\ref{eq:ffstar}) require access to instantaneous 3D flow fields, which are typically not available in experiments. However, one disadvantage of the $\epsilon_{\rm TKE}$ metrics in isolation is that their values do not reflect the wall-normal distribution of TKE. Nevertheless, the TKE-based error metrics remain useful when supplemented with the dynamics-based error metrics since, together, they encode information regarding the spatially integrated correlations between the reconstructions and the DNS.}

%%%%%%%%%%%%%%%%%%%%%%%%%%%%%%%%%%%%%%%%%%
\section{Results}\label{sec:res}
%%%%%%%%%%%%%%%%%%%%%%%%%%%%%%%%%%%%%%%%%%

In this section, we evaluate the accuracy and efficiency of flow reconstructions using STRME, SORBE, and STSME.

%%%%%%%%%%%%%%%%%%%%%%%%%%%%%%%%%%%%%%%%%%
\subsection{Equation-based estimation: STRME}\label{sec:res:strme}
%%%%%%%%%%%%%%%%%%%%%%%%%%%%%%%%%%%%%%%%%%

\begin{figure}[!htpb]
    \centering
    \includegraphics[trim=0 25 0 50, clip, width=0.869\textwidth]{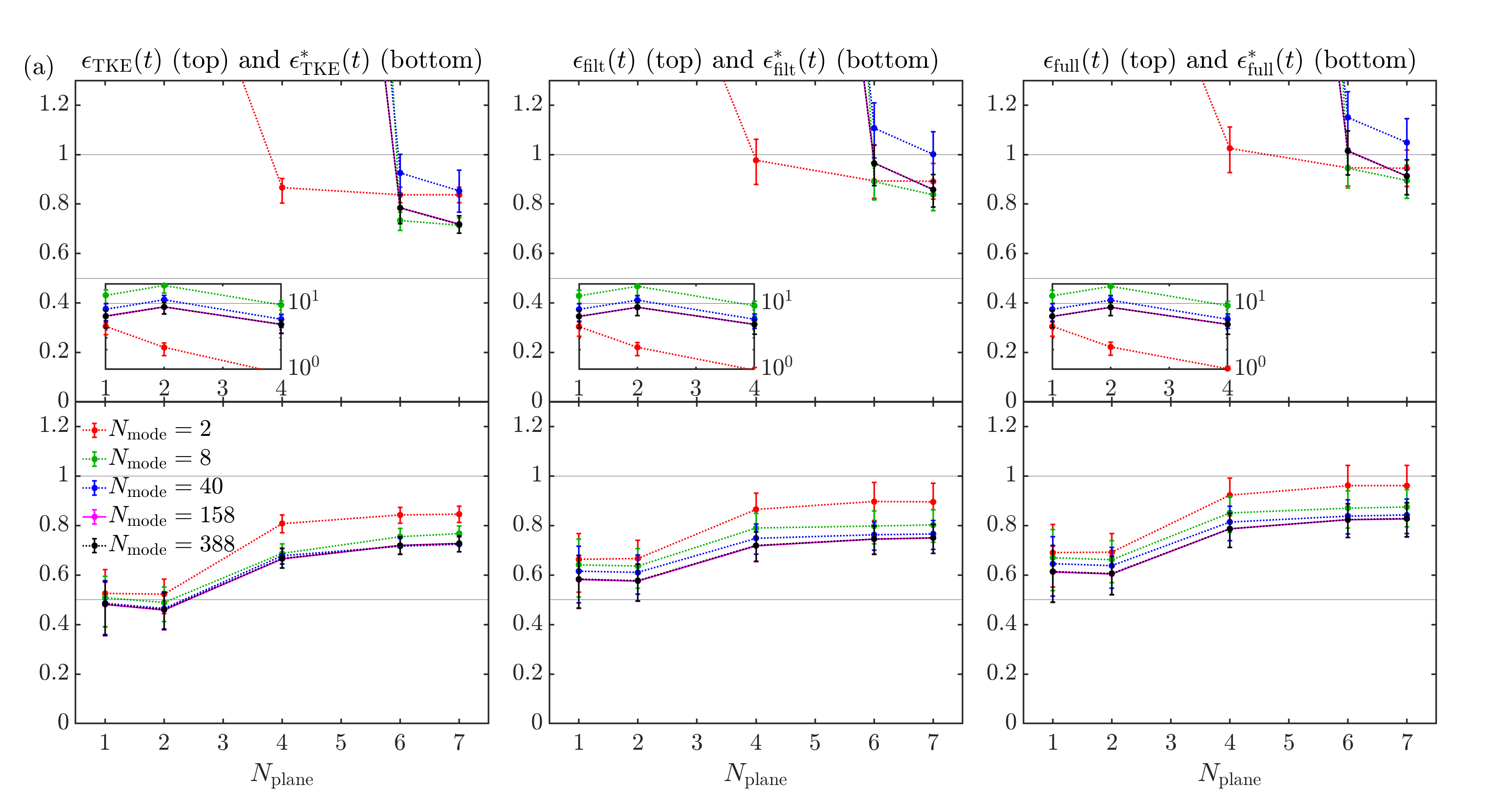}
    \includegraphics[trim=0 25 0 50, clip, width=0.869\textwidth]{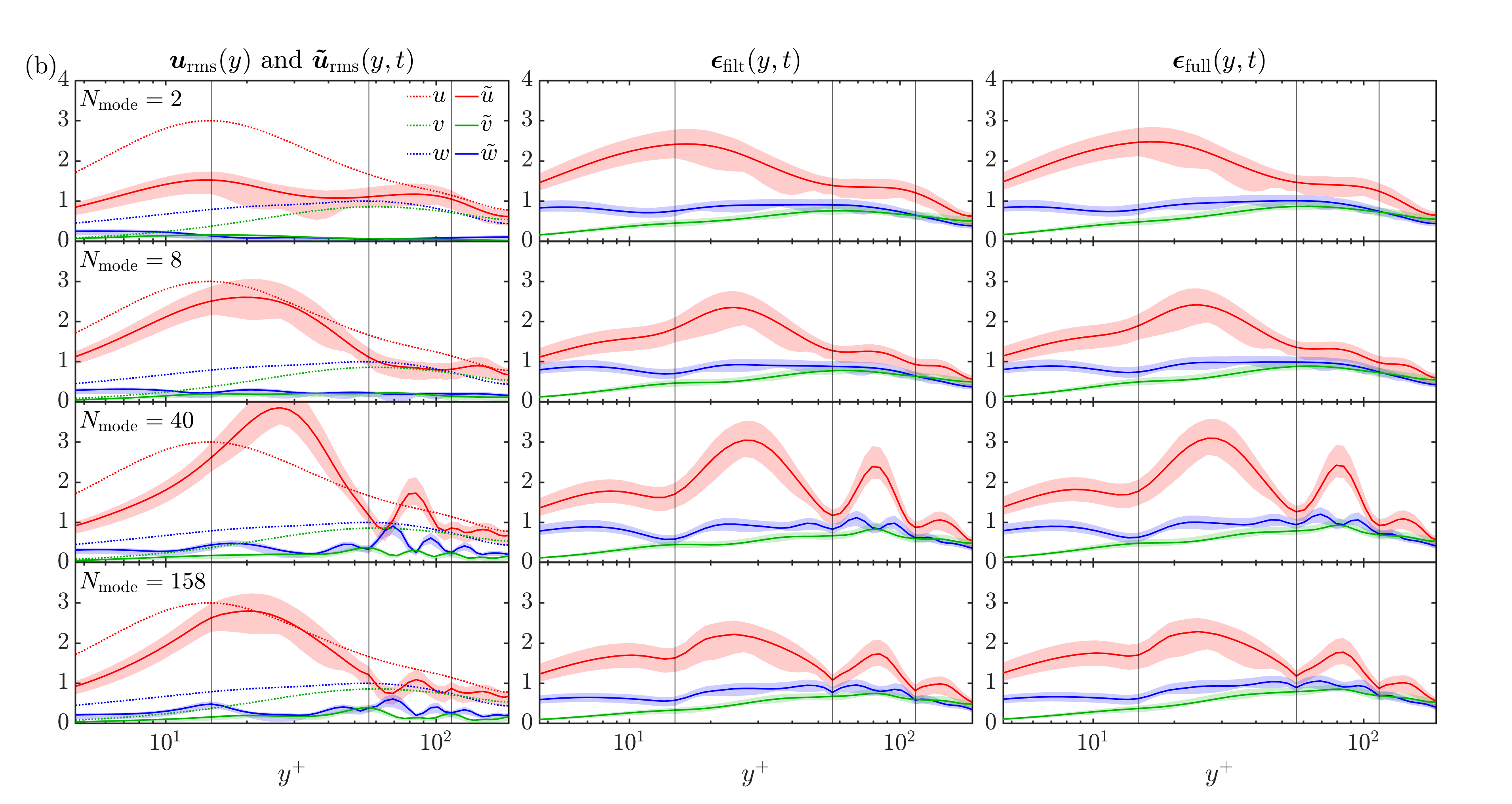}
    \includegraphics[trim=0  0 0 50, clip, width=0.869\textwidth]{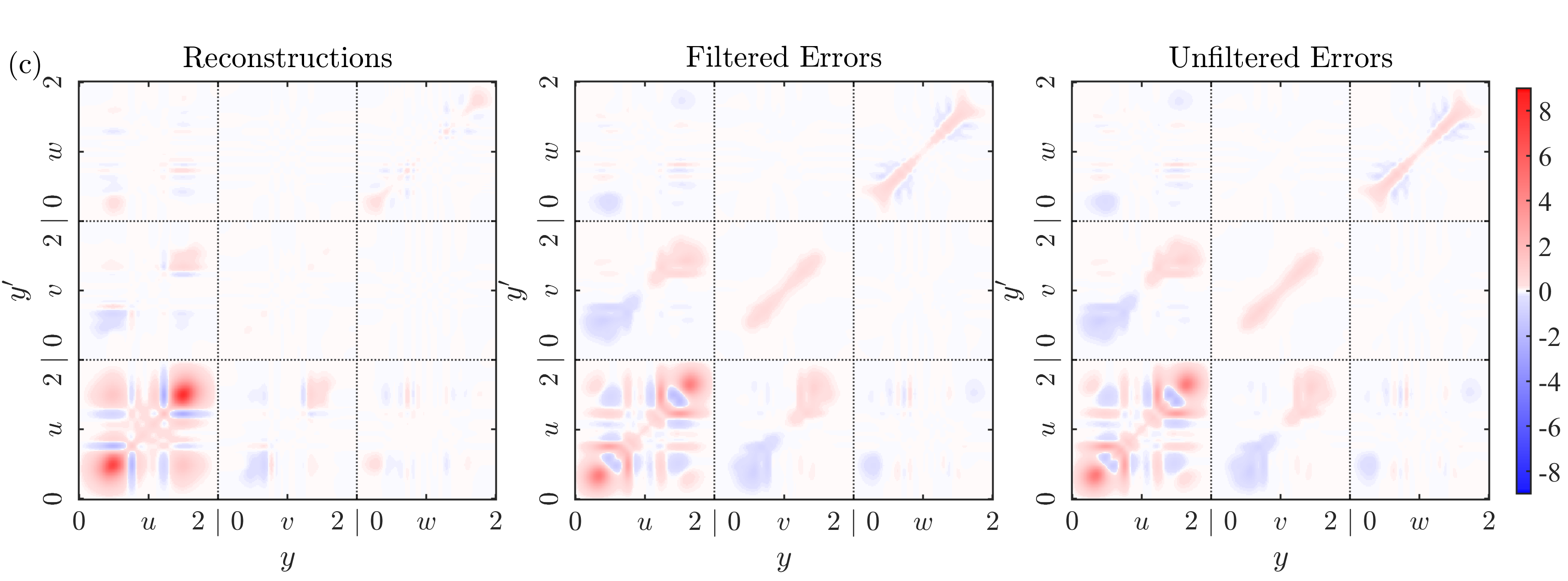}
    \caption{\rev{Summary of reconstruction accuracies for the STRME estimators. (a) Full-field (top) and truncated (bottom) error statistics, where $N_{\rm local} = 2$ is used for the truncated metrics. (b) RMS velocity profiles and error profiles for reconstructions using different modal truncations and $N_{\rm plane} = 7$. The vertical lines represent the sensor locations.} (c) Reconstructed cross-correlations (left) and their mean errors with respect to the filtered (middle) and unfiltered (right) cross-correlations for $N_{\rm plane} = 7$ and $N_{\rm mode} = 158$. \rev{The averages in (a) and (b) are computed prior to applying the square roots in the error metrics and the uncertainties extend to the $20{\rm th}$ and $80{\rm th}$ percentiles.}}
    %\vspace{-22pt}% to remove warning of overfull page (REMOVE LATER?)
    \label{fig:res:strme:1}
\end{figure}

\begin{figure}[!htpb]
    \centering
    \includegraphics[trim=25 100 25 100, clip, width=0.92\textwidth]{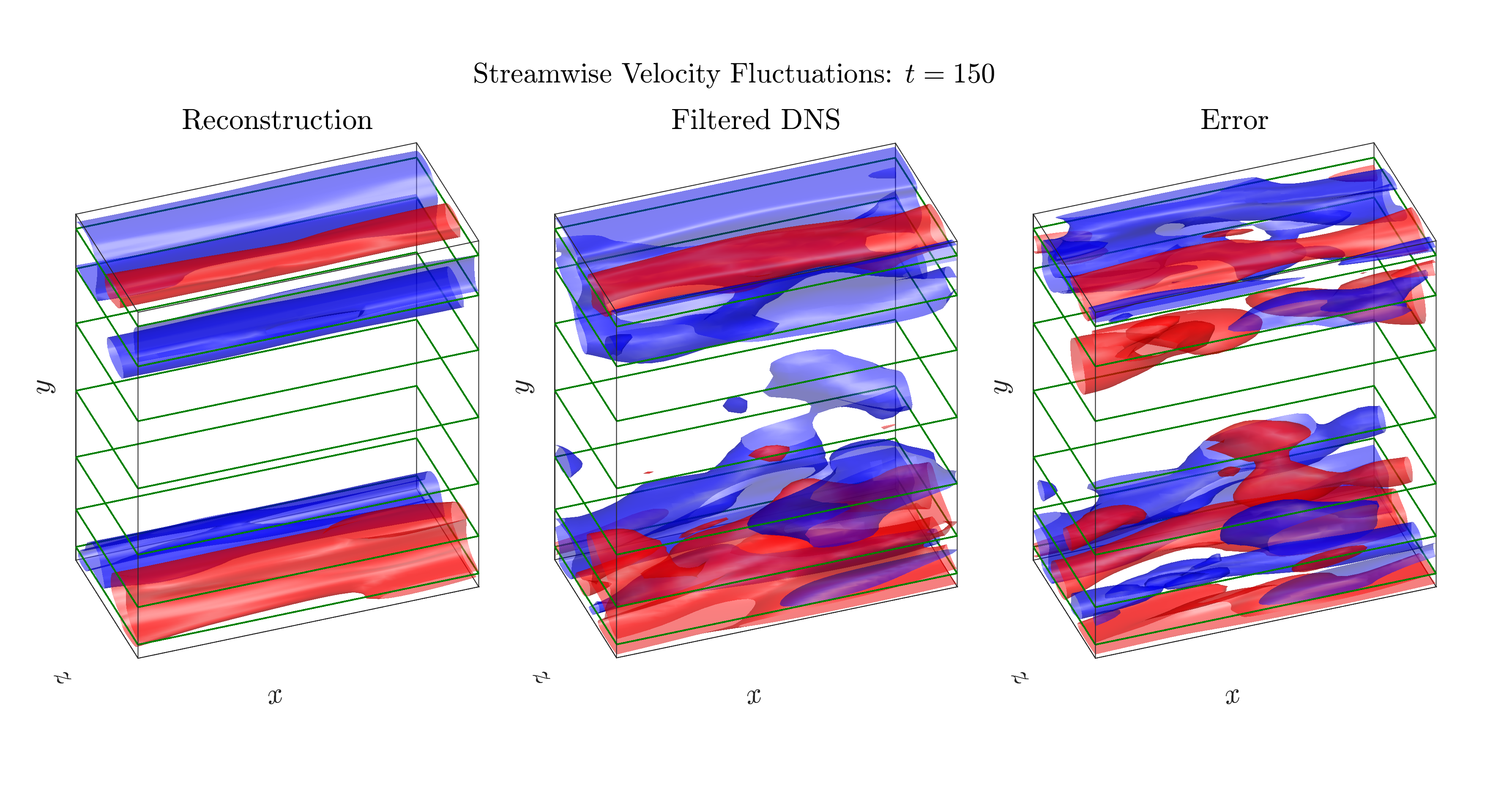}
    \includegraphics[trim=25 100 25 100, clip, width=0.92\textwidth]{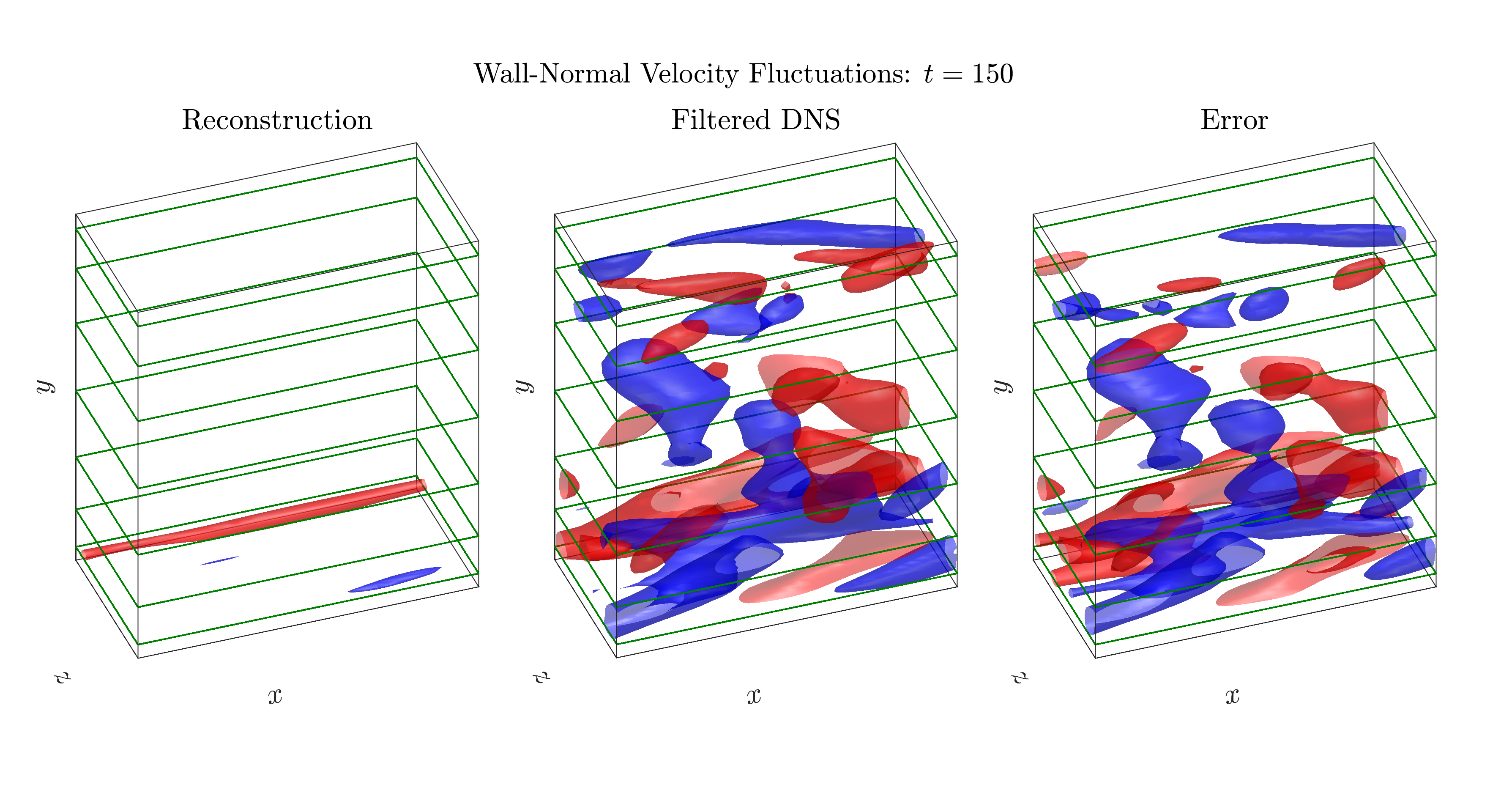}
    \includegraphics[trim=25 100 25 100, clip, width=0.92\textwidth]{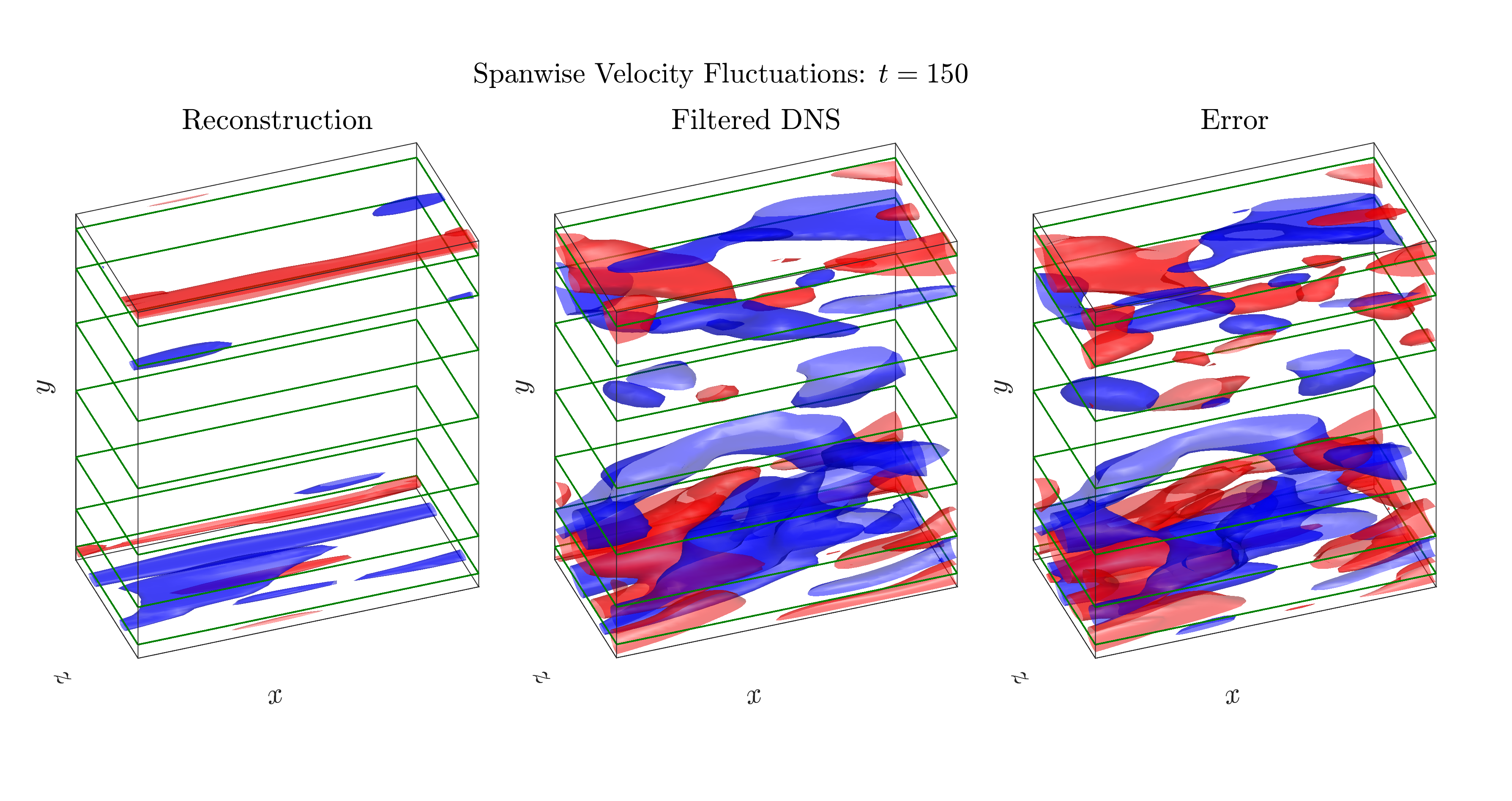}
    \caption{Comparisons of the reconstructions, filtered DNS snapshots, and their differences for the streamwise (top), wall-normal (middle), and spanwise (bottom) velocity fluctuations at $t = 150$. The comparisons here are shown for STRME with $N_{\rm plane} = 7$ and $N_{\rm mode} = 158$, for which $\epsilon_{\rm filt}(t) = 0.708$ and $\epsilon_{\rm full}(t) = 0.737$. Isosurfaces ($+$: red, $-$: blue) are based on the TKE of each velocity component and have magnitudes $\left\langle u^2_{\rm rms} \right\rangle^{1/2} = 1.58$ (top), $\left\langle v^2_{\rm rms} \right\rangle^{1/2} = 0.68$ (middle), and $\left\langle w^2_{\rm rms} \right\rangle^{1/2} = 0.77$ (bottom), where $\left\langle \cdot \right\rangle$ represents the mean over the wall-normal direction.}
    \label{fig:res:strme:2}
\end{figure}

Using no training data except a mean profile, STRME forms a simple, informative framework for flow reconstruction. However, its accuracy relies on an unrealistic model of an uncorrelated forcing. Figures \ref{fig:res:strme:1} and \ref{fig:res:strme:2} summarize the accuracy of the STRME estimators for each modal truncation. Due to the unrealistic uncorrelated forcing model, the global errors are relatively large for all modal truncations.

For $N_{\rm plane} \geq 6$, the $N_{\rm mode} = 8$ ROM produces the most accurate full-field reconstructions in a relatively economical fashion. However, for $N_{\rm plane} \leq 6$ the $N_{\rm mode} = 2$ ROM outperforms all higher-order truncations considered. These results suggest that the uncorrelated forcing assumption is most valid for the gain-dominant resolvent modes, for which the gains are likely to contribute significantly to the mode weight statistics. Since the validity of the forcing assumption underlies correspondence between resolvent modes and SPOD modes, our results are consistent with previous observations \citep{Tow2018} that this correspondence is most appropriate for dominant resolvent modes. The low-rank truncations for STRME and STSME are qualitatively similar in the present investigation, but the STSME estimators are more accurate as they are derived from higher-fidelity training data (see Sec. \ref{sec:res:stsme}).

The higher-order truncations are nearly identical visually (e.g., for $N_{\rm mode}$ = 158 and 388) since the corresponding gains are too small to impact the estimator. Especially for truncations with $N_{\rm mode} > N_{\rm plane}$, we expect that improved models of the mode weight statistics are likely to improve estimator performance. Further, in the limit of including all $N_u = 388$ modes, STRME represents a form of SORBE under the uncorrelated forcing assumption. One natural, data-driven extension of STRME involves estimating the mode weight statistics using limited training data (as is done for the forcing in SORBE). However, as shown in Appendix \ref{sec:app:fvsxi}, the forcing-based SORBE estimator empirically outperforms the mode weight-based estimator inspired by STRME in the present investigation. In the context of the open question of optimal coefficients for flow reconstruction, this justifies our use of the forcing over the mode weights in the SORBE formulation.

Despite relatively poor full-field accuracy, the STRME estimators are more accurate in the vicinity of the measurement planes. This effect is especially pronounced for $N_{\rm plane} \leq 2$, although a relatively small portion of the channel is captured in these cases. The spatial profiles for $N_{\rm plane} = 7$ confirm that reconstructions tend to be more accurate near the measurement planes in that the errors associated with higher-order modal truncations are concentrated far from the measurement planes. These results indicate that targeted reconstructions of localized flow structures are more feasible than accurate full-channel reconstructions and may require considerably fewer measurements to provide meaningful (partial) flow information.

The STRME reconstructions of $u_{\rm rms}$ are remarkably good considering the training data includes only a (streamwise) mean profile, but the reconstructions of $v_{\rm rms}$ and $w_{\rm rms}$ are still missing considerable energy for all STRME estimators. Consistent with previous findings \citep{Moa2013}, these results suggest that a broadband forcing assumption can be used to garner significant information about $u_{\rm rms}$ given only a mean profile (via the resolvent formulation). The reconstructed mean cross-correlations reinforce this notion, as the most energetic correlations captured involve the streamwise fluctuations. Further, the snapshots in Fig. \ref{fig:res:strme:2} confirm that energetic flow structures are better captured for the streamwise fluctuations than for the wall-normal or spanwise fluctuations.

In summary, STRME can leverage the governing equations to produce informative localized reconstructions, especially of streamwise statistics, when no training data are available.

%%%%%%%%%%%%%%%%%%%%%%%%%%%%%%%%%%%%%%%%%%
\subsection{Hybrid estimation: SORBE}\label{sec:res:sorbe}
%%%%%%%%%%%%%%%%%%%%%%%%%%%%%%%%%%%%%%%%%%

\begin{figure}[!htpb]
    \centering
    \includegraphics[trim=0 25 0 50, clip, width=0.869\textwidth]{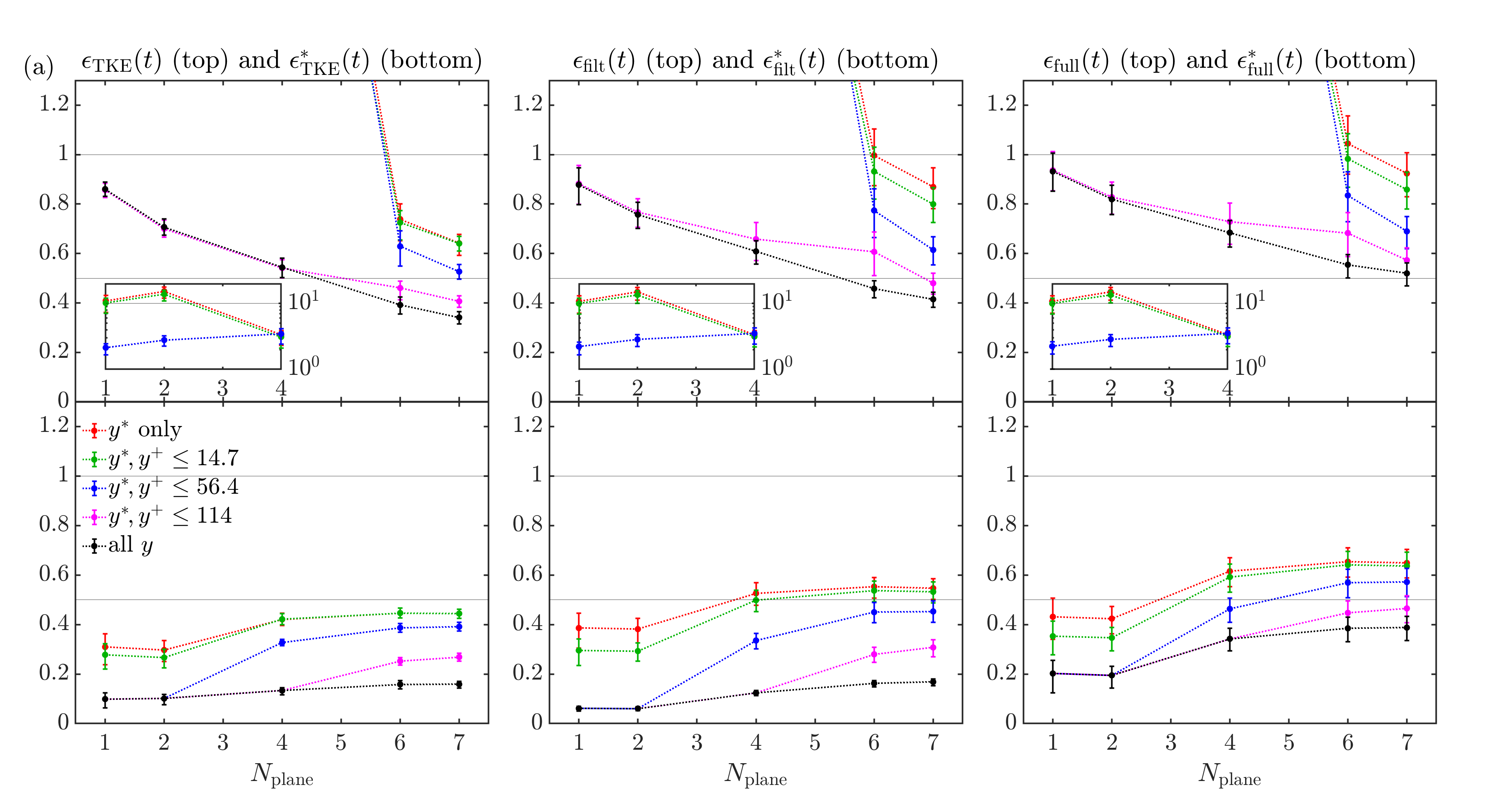}
    \includegraphics[trim=0 25 0 50, clip, width=0.869\textwidth]{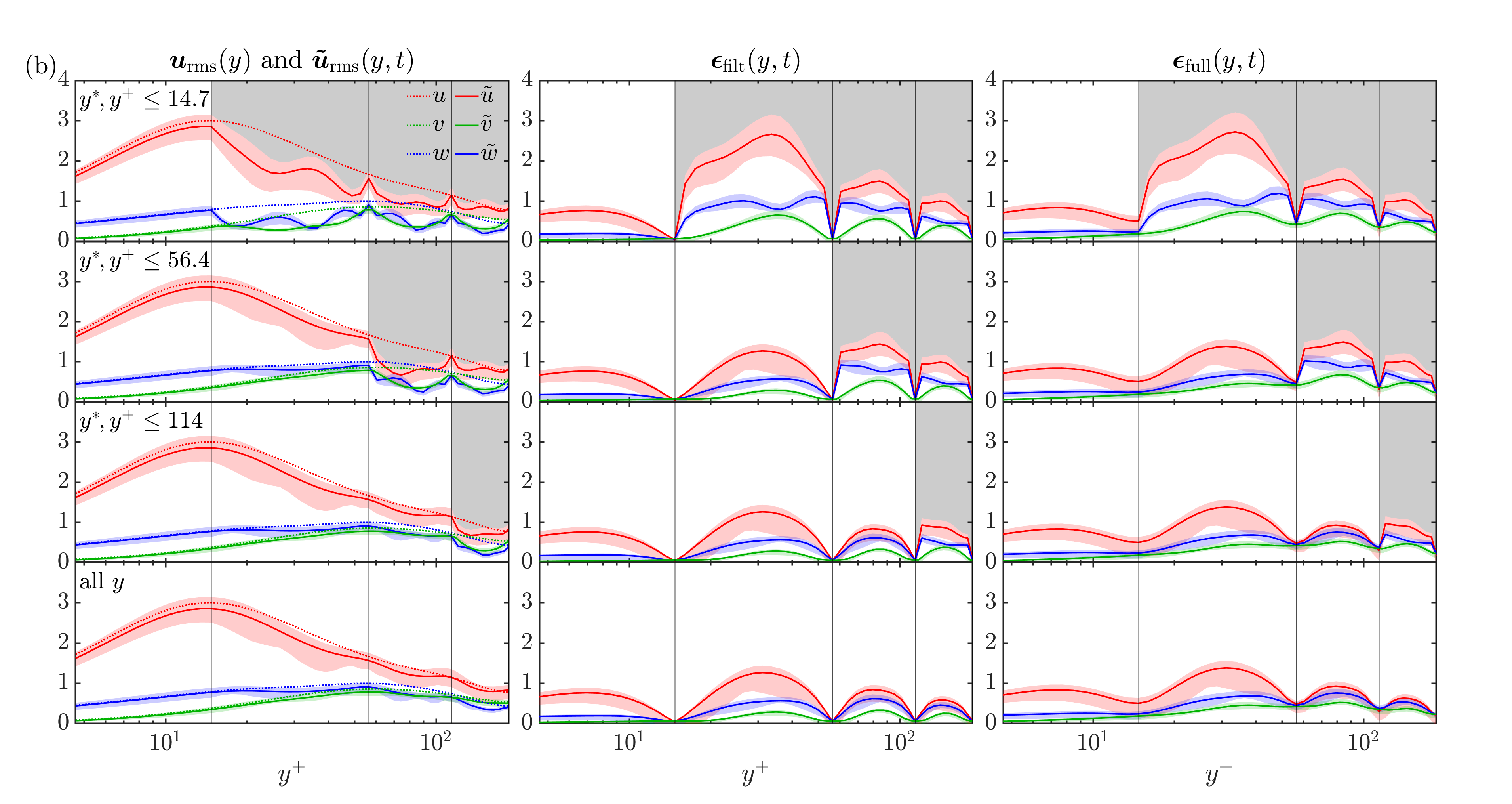}
    \includegraphics[trim=0  0 0 50, clip, width=0.869\textwidth]{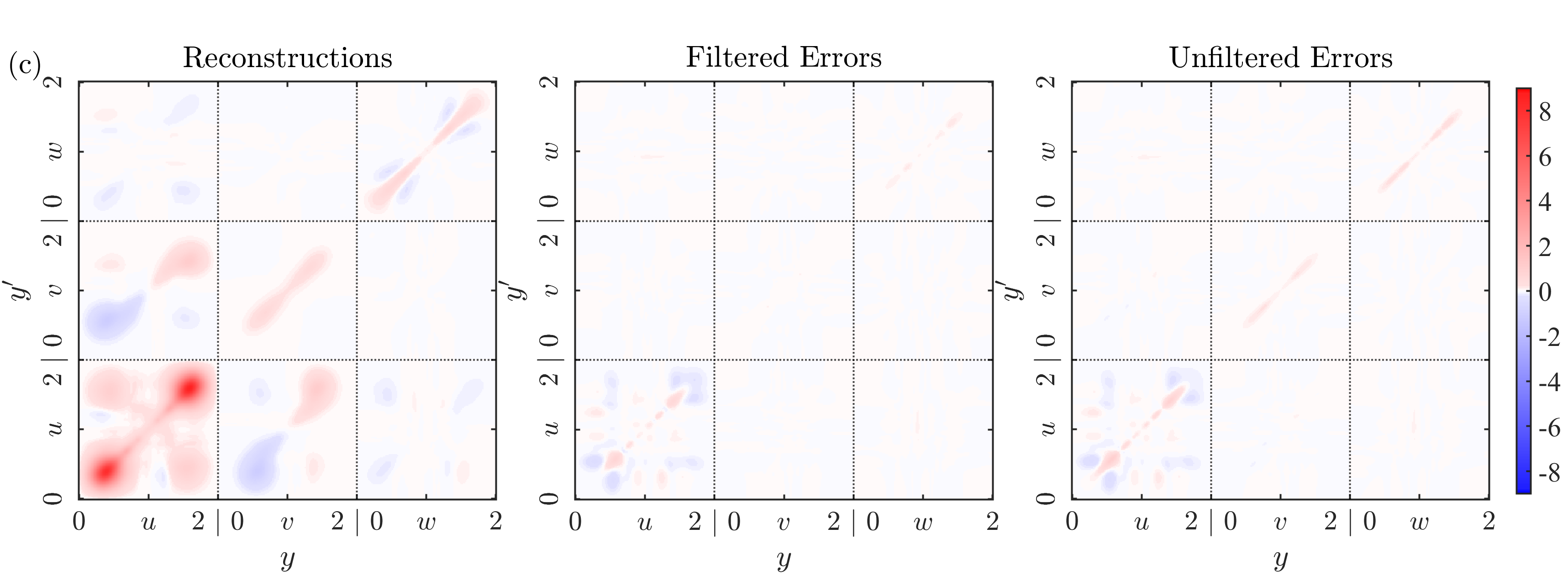}
    \caption{Summary of reconstruction accuracies for the SORBE estimators in the style of Fig. \ref{fig:res:strme:1}. Here, the spatial profiles (for $N_{\rm plane} = 7$) are shown for various extents of auxiliary training data, where regions with gray shading are not captured during training. The cross-correlations are shown for the $N_{\rm plane} = 7$ case with full-field training data and represent the generalized Wiener filter.}
    \label{fig:res:sorbe:1}
\end{figure}

\begin{figure}[!htpb]
    \centering
    \includegraphics[trim=25 100 25 100, clip, width=0.92\textwidth]{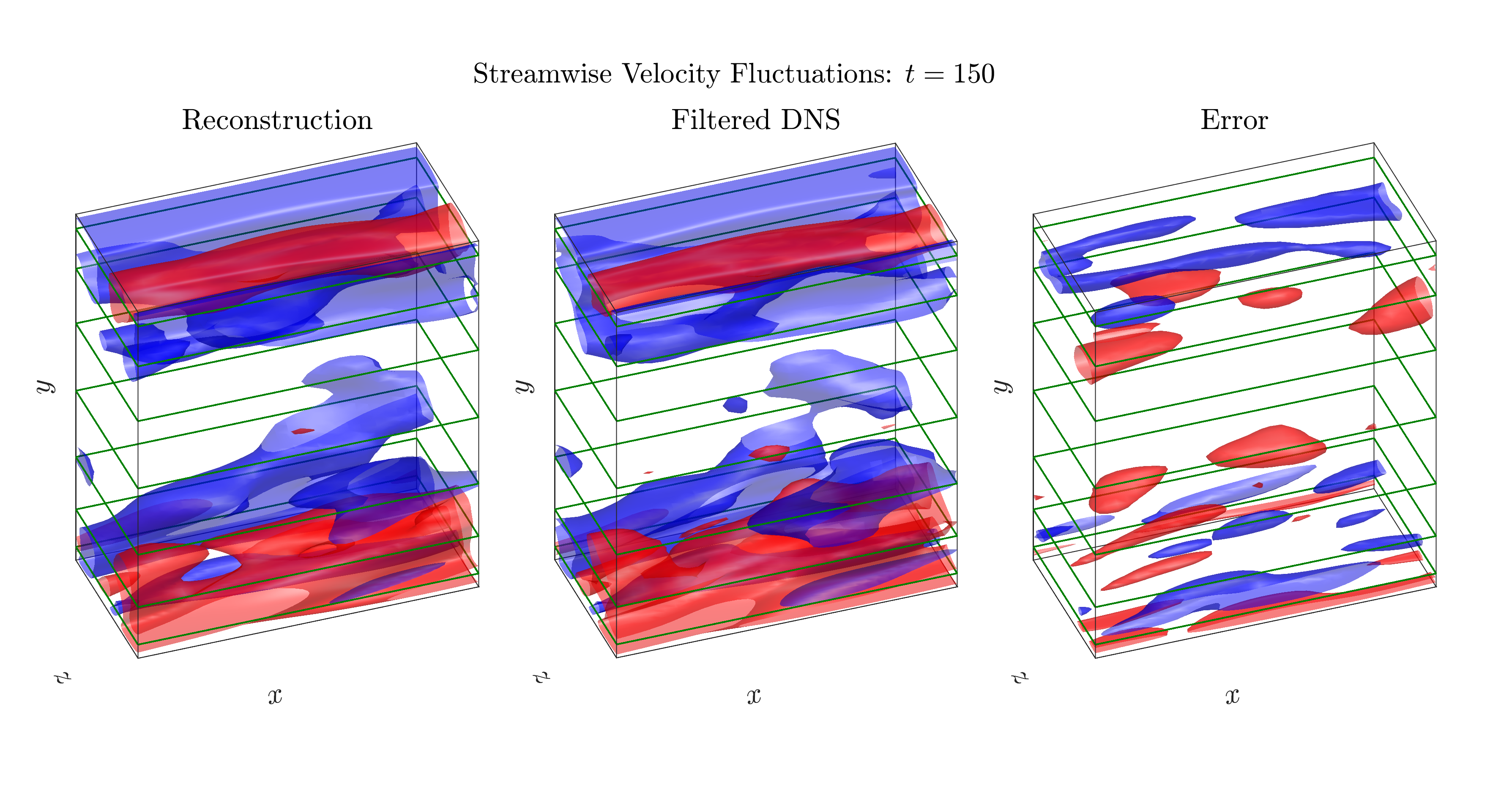}
    \includegraphics[trim=25 100 25 100, clip, width=0.92\textwidth]{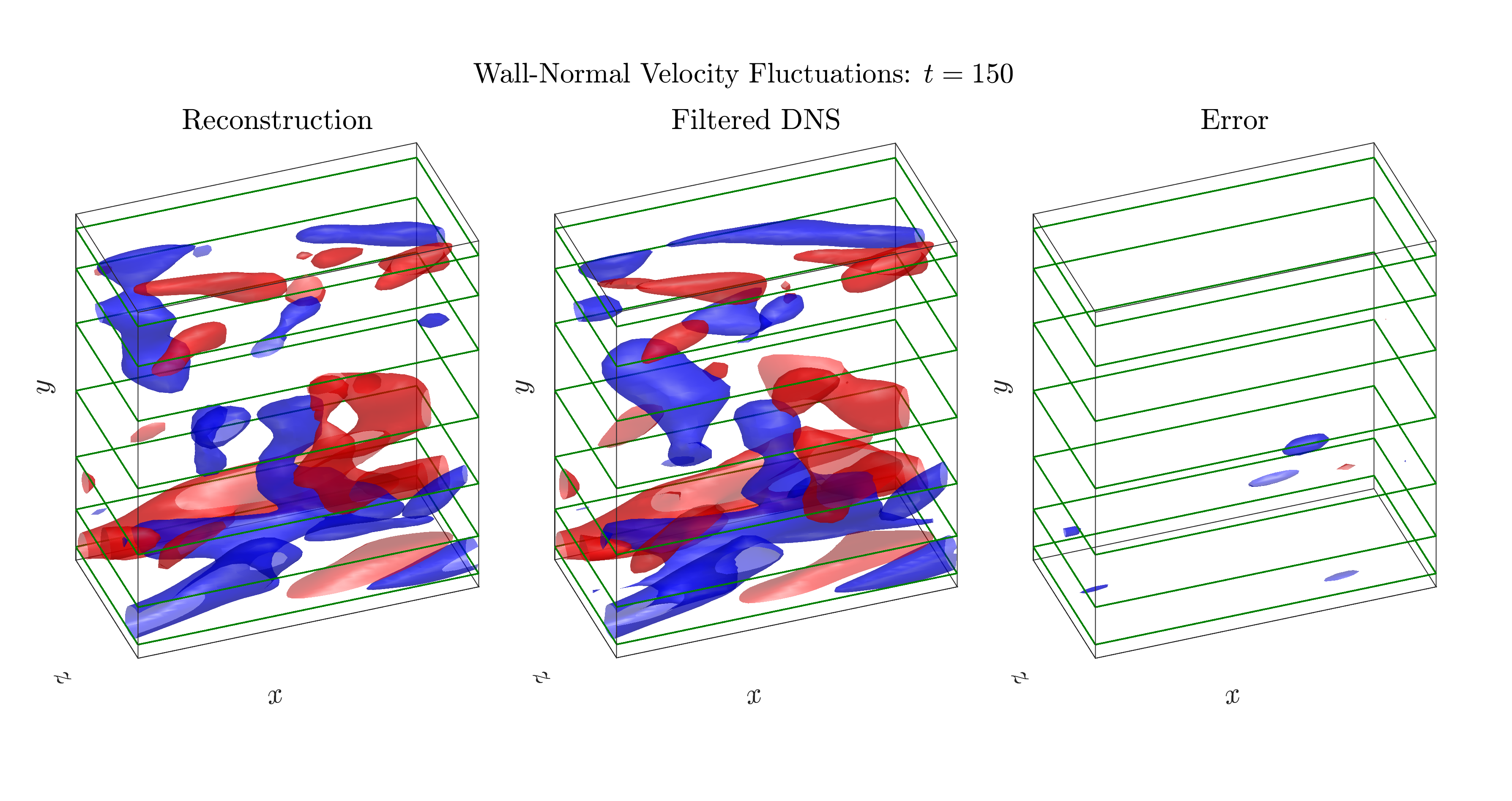}
    \includegraphics[trim=25 100 25 100, clip, width=0.92\textwidth]{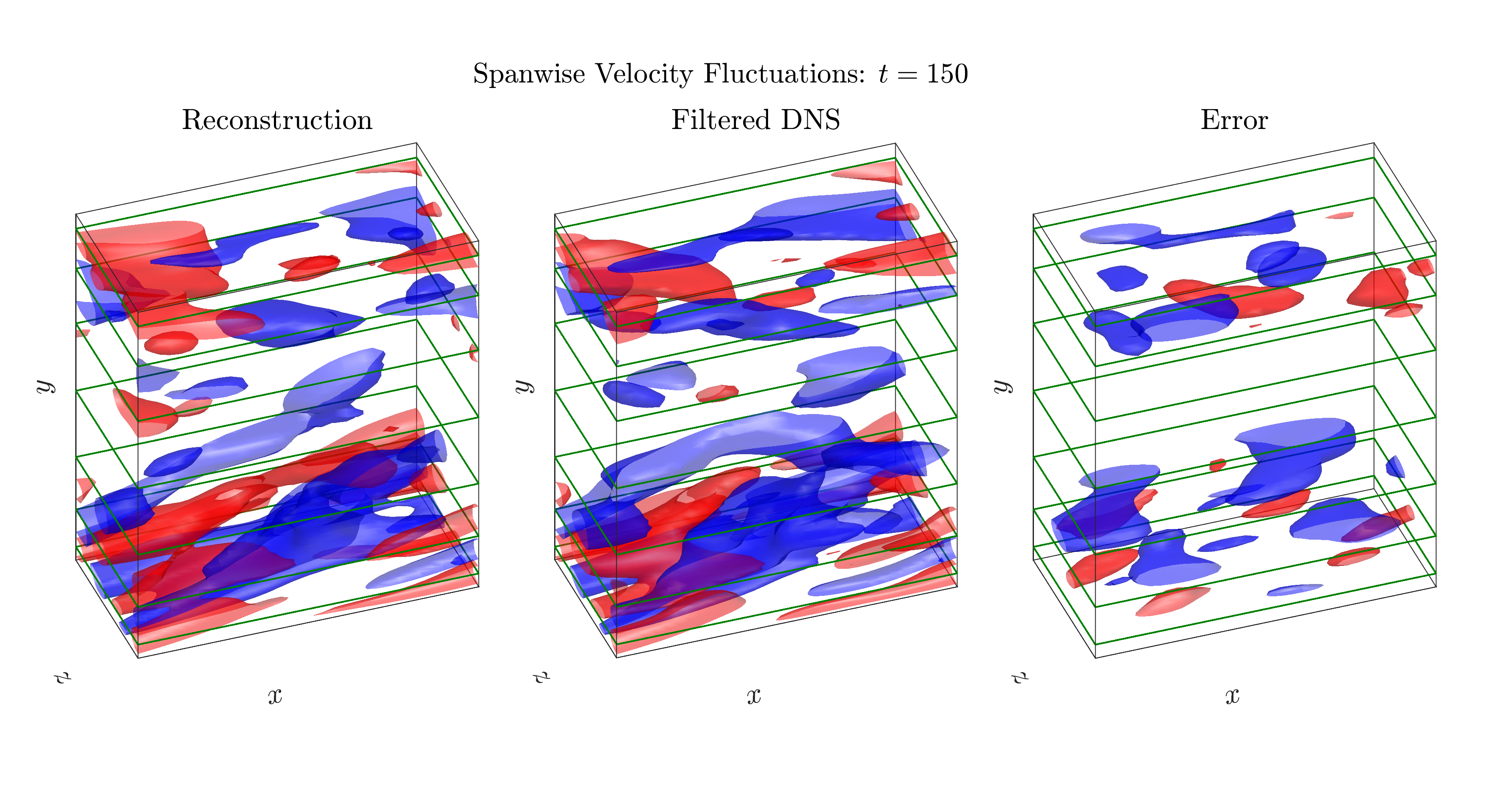}
    \caption{Comparisons of the reconstructions, filtered DNS snapshots, and their differences in the style of Fig. \ref{fig:res:strme:2}. The comparisons here are shown for SORBE with $N_{\rm plane} = 7$ and full-field training data (i.e., the generalized Wiener filter), for which $\epsilon_{\rm filt}(t) = 0.399$ and $\epsilon_{\rm full}(t) = 0.448$. Videos depicting the temporal evolutions of these comparisons are provided in the Supplemental Material \citep{SuppMat}.}
    \label{fig:res:sorbe:2}
\end{figure}

SORBE provides a practical approach to improve estimator performance by using spatially limited data to produce more realistic forcing statistics than the uncorrelated forcing CSD assumed in STRME. This hybrid framework provides a natural means of incorporating training measurements in conjunction with the governing equations. It allows a balance between cost and accuracy along the spectrum of training data fidelity. Figures \ref{fig:res:sorbe:1} and \ref{fig:res:sorbe:2} summarize the accuracy of the SORBE estimators computed using various extents of training data.  

For all measurement configurations, increasing the spatial extent of training data used for the forcing estimate broadly improves reconstruction accuracy. When using relatively few measurement planes ($N_{\rm plane} \leq 4$), the forcing fidelity has a large influence on reconstruction accuracy. By contrast, using more measurement planes ($N_{\rm plane} \geq 6$) produces relatively smooth and more incremental (but still significant) reductions in error. These behaviors highlight two forms of the cost-accuracy tradeoff associated with augmenting the SORBE estimator using measurements. When the testing measurements are sparse, high-fidelity training data can drastically improve reconstruction accuracy. When higher-fidelity testing measurements are available, lower-fidelity training data are required to achieve similar accuracy, and high-fidelity training data further improve reconstruction accuracy.

The errors localized about each measurement plane are even lower than the full-field errors for all SORBE estimators considered. Since reconstructions are not limited by an ill-posed flow representation, the filtered errors \rev{are small in the vicinity of the measurement planes}. Correspondingly, the unfiltered errors at these planes are also \rev{relatively small} due to the \rev{minimal} energy associated with the excluded Fourier modes. Further, as depicted in the profiles for $N_{\rm plane} = 7$, errors tend to grow with the distance from the measurement planes. Between the measurement planes, the errors for the streamwise and spanwise fluctuations grow more rapidly than for the wall-normal fluctuations \rev{since they} are (i) larger in magnitude and (ii) more localized in the wall-normal direction.  

The errors in the regions between measurements are primarily attributed to \rev{the extent of the training data and the fidelity of the streaming measurements during testing}. When only partial training data are used, errors grow in the gray regions (see Fig. \ref{fig:res:sorbe:1}) where measurements were not used to inform forcing statistics. This error source reflects that the least-squares forcing statistics informed by partial training data do not fully capture the structure of the true forcing statistics. However, even when the estimator incorporates full-field statistics from training, the errors still grow with the distance from the measurements. \rev{In the noncausal estimation setting, this error source formally reflects that the flow is poorly correlated with the measurements far from the sensor planes.} Hence, while it is unreasonable to expect perfect global reconstructions from sparse measurements, this behavior can be used to inform the number of and spacings between sensors used for flow reconstruction. \rev{Moreover, as discussed more extensively in Sec. \ref{sec:res:sumsec}, this error source is augmented by the application of the formally noncausal estimator in a streaming setting, where the flow statistics vary in time.}

In the limit of full training data, SORBE produces results that converge to those produced by a generalized Wiener filter. When the generalized Wiener filter is used with $N_{\rm plane} = 7$ (as in Case E), the mean values of $\epsilon_{\rm filt}(t)$ and $\epsilon_{\rm full}(t)$ are 41.5\% and 52.0\%, respectively. \rev{Since these errors are averaged prior to applying the square roots in their definitions}, the corresponding error energies are 17.2\% and 27.0\% of the TKE, respectively. Further, the reconstructed mean cross-correlations are significantly more energetic than their corresponding errors for this estimator, indicating the second-order flow statistics during the testing period are well captured.

Figure \ref{fig:res:sorbe:2} further reinforces that this estimator performs well throughout the channel, as the dominant global flow structures are well reconstructed for all components of the velocity fluctuations. The energetic flow structures are best captured for the wall-normal fluctuations ($v$) since they are less localized in $y$ than the streamwise ($u$) and spanwise ($w$) fluctuations. Further, the reconstructions of structures of $u$ and $w$ tend to accumulate errors between sensor planes since the corresponding structures tend to be more localized in $y$. These observations are consistent with the error profiles for each component and the reductions in error when considering localized wall-normal subdomains.

In summary, the SORBE estimator provides a practical and effective means of improving reconstruction accuracy by balancing the cost-accuracy tradeoff associated with increasing training (and testing) data fidelity. In the full training data limit, a relatively small number of sensors can reproduce the dominant statistical and dynamical flow features with high accuracy.

%%%%%%%%%%%%%%%%%%%%%%%%%%%%%%%%%%%%%%%%%%
\subsection{Data-driven estimation: STSME}\label{sec:res:stsme}
%%%%%%%%%%%%%%%%%%%%%%%%%%%%%%%%%%%%%%%%%%

\begin{figure}[!htpb]
    \centering
    \includegraphics[trim=0 25 0 50, clip, width=0.869\textwidth]{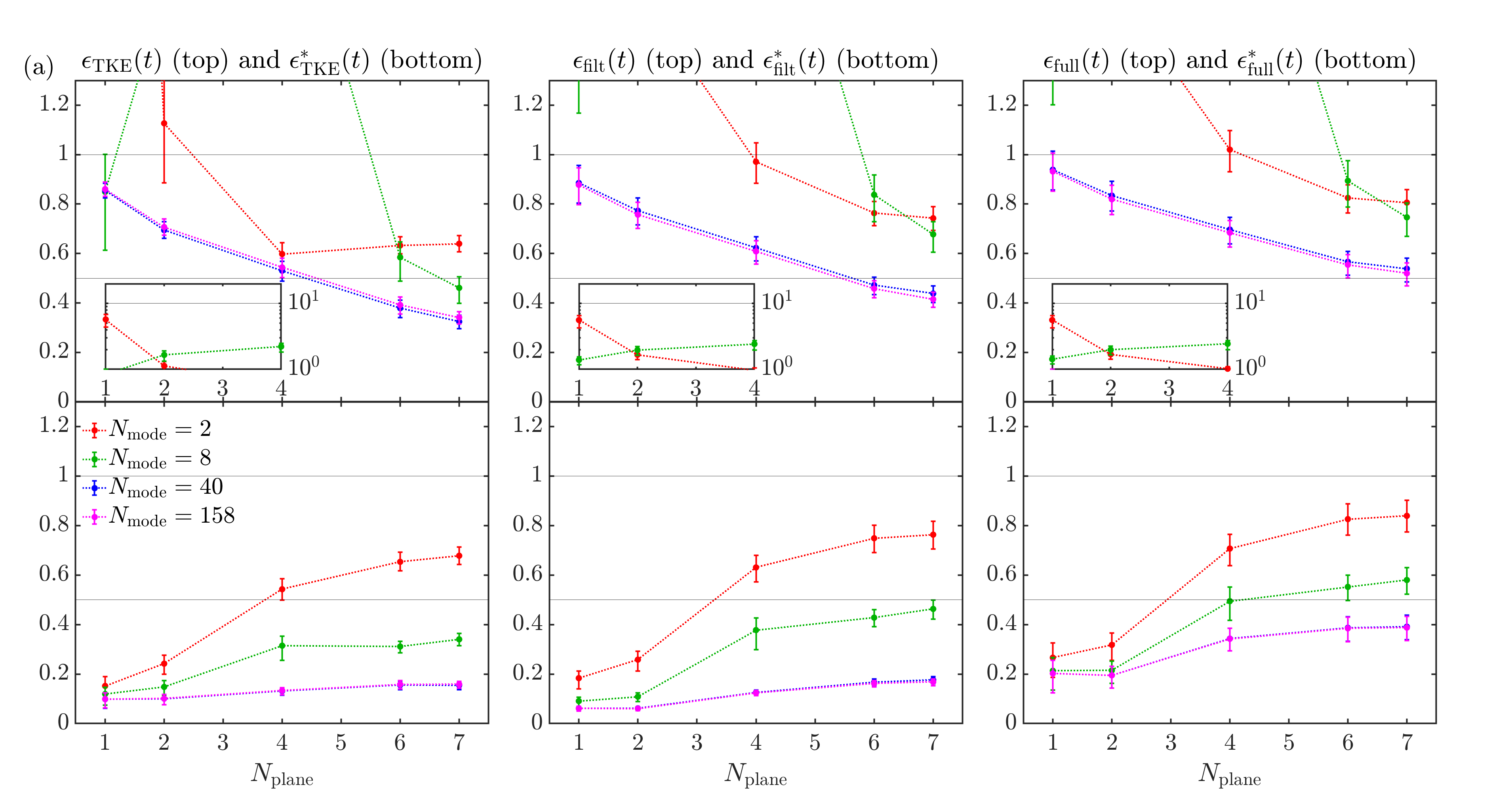}
    \includegraphics[trim=0 25 0 50, clip, width=0.869\textwidth]{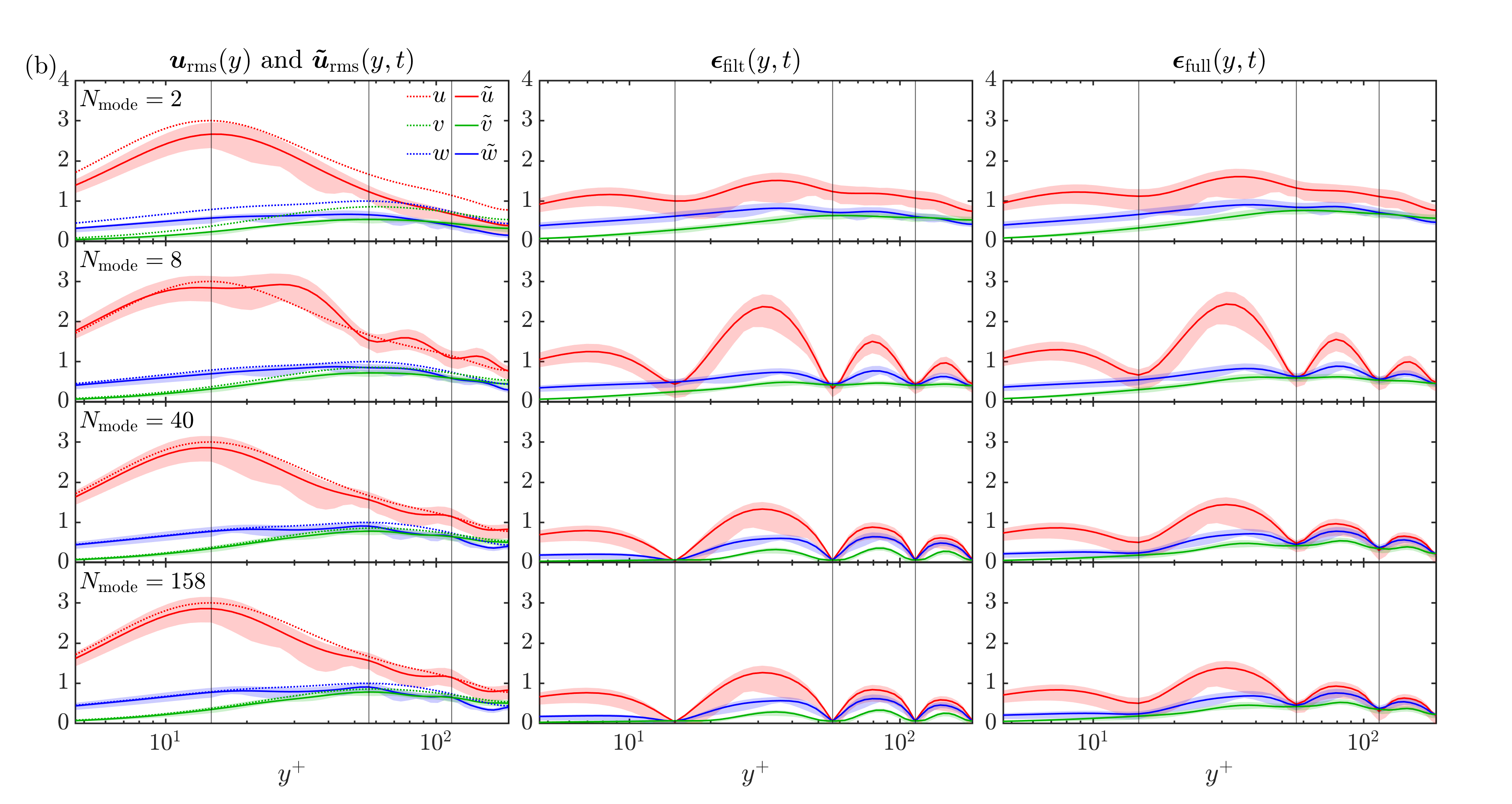}
    \includegraphics[trim=0  0 0 50, clip, width=0.869\textwidth]{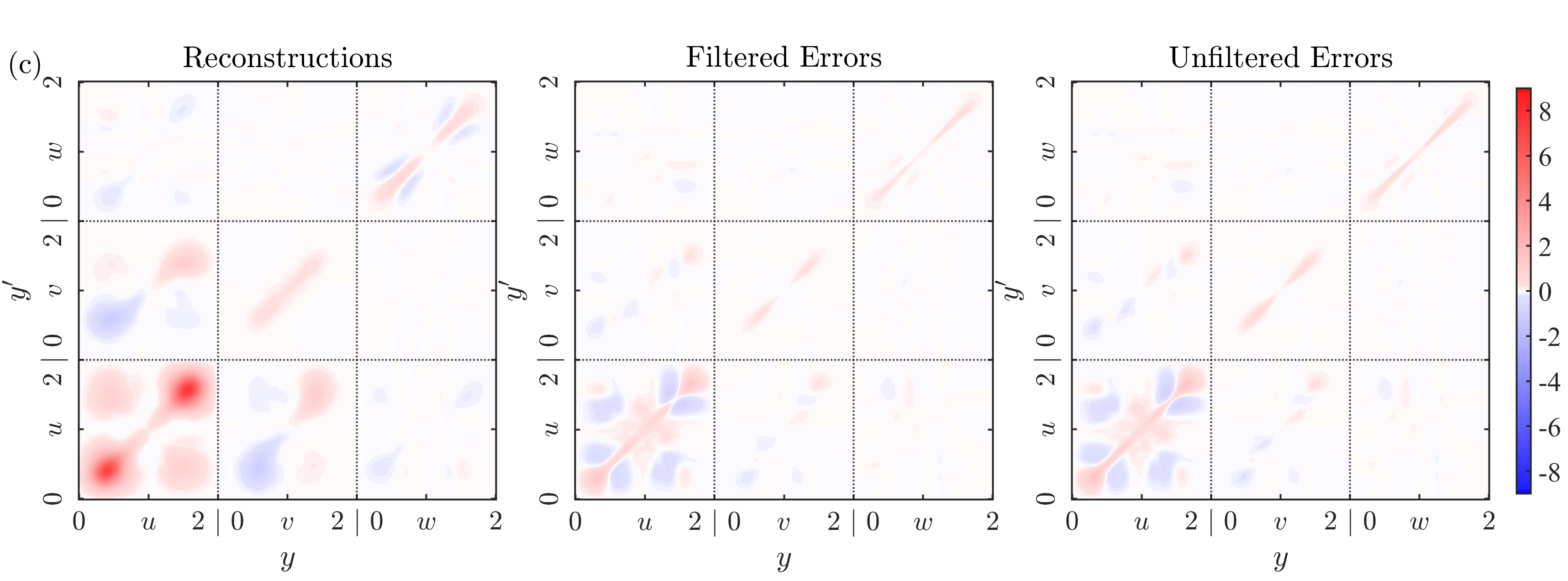}
    \caption{Summary of reconstruction accuracies for the STSME estimators in the style of Figs. \ref{fig:res:strme:1} and \ref{fig:res:sorbe:1}. Here, the spatial profiles (for $N_{\rm plane} = 7$) are shown for various modal truncations and the cross-correlations (for $N_{\rm plane} = 7$) are shown for the case with $N_{\rm mode} = 2$.}
    \label{fig:res:stsme:1}
\end{figure}

\begin{figure}[!htpb]
    \centering
    \includegraphics[trim=25 100 25 100, clip, width=0.92\textwidth]{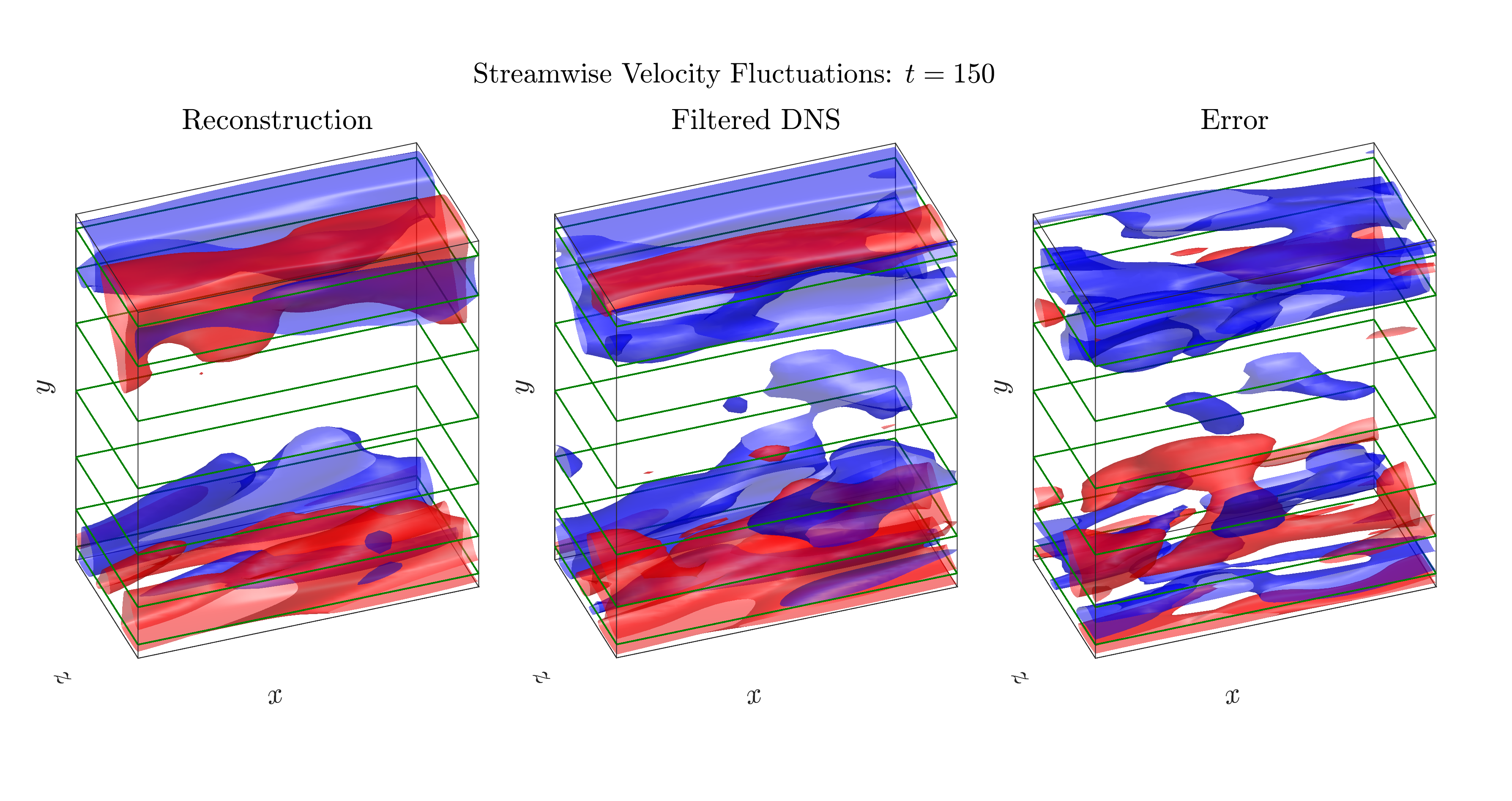}
    \includegraphics[trim=25 100 25 100, clip, width=0.92\textwidth]{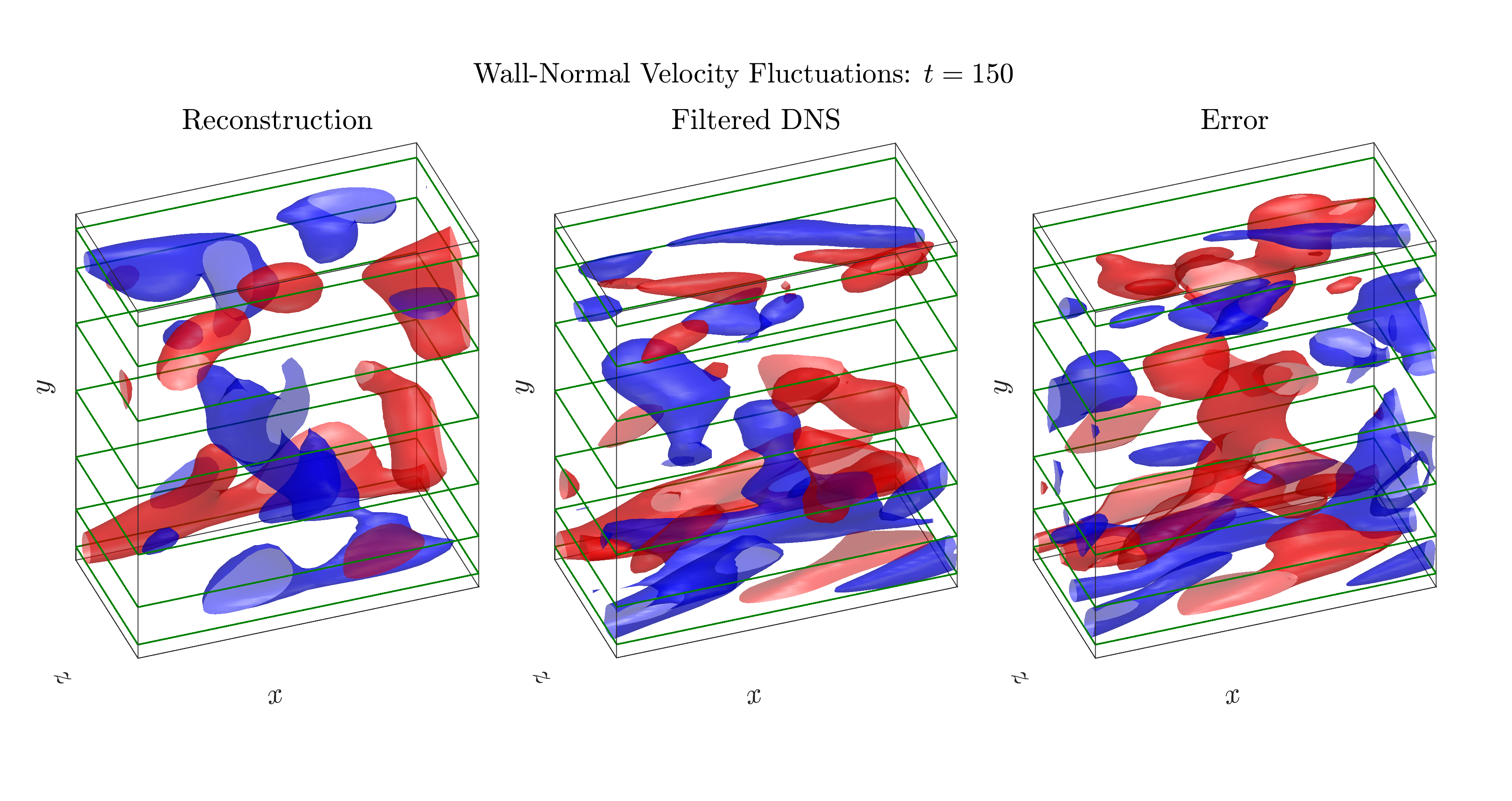}
    \includegraphics[trim=25 100 25 100, clip, width=0.92\textwidth]{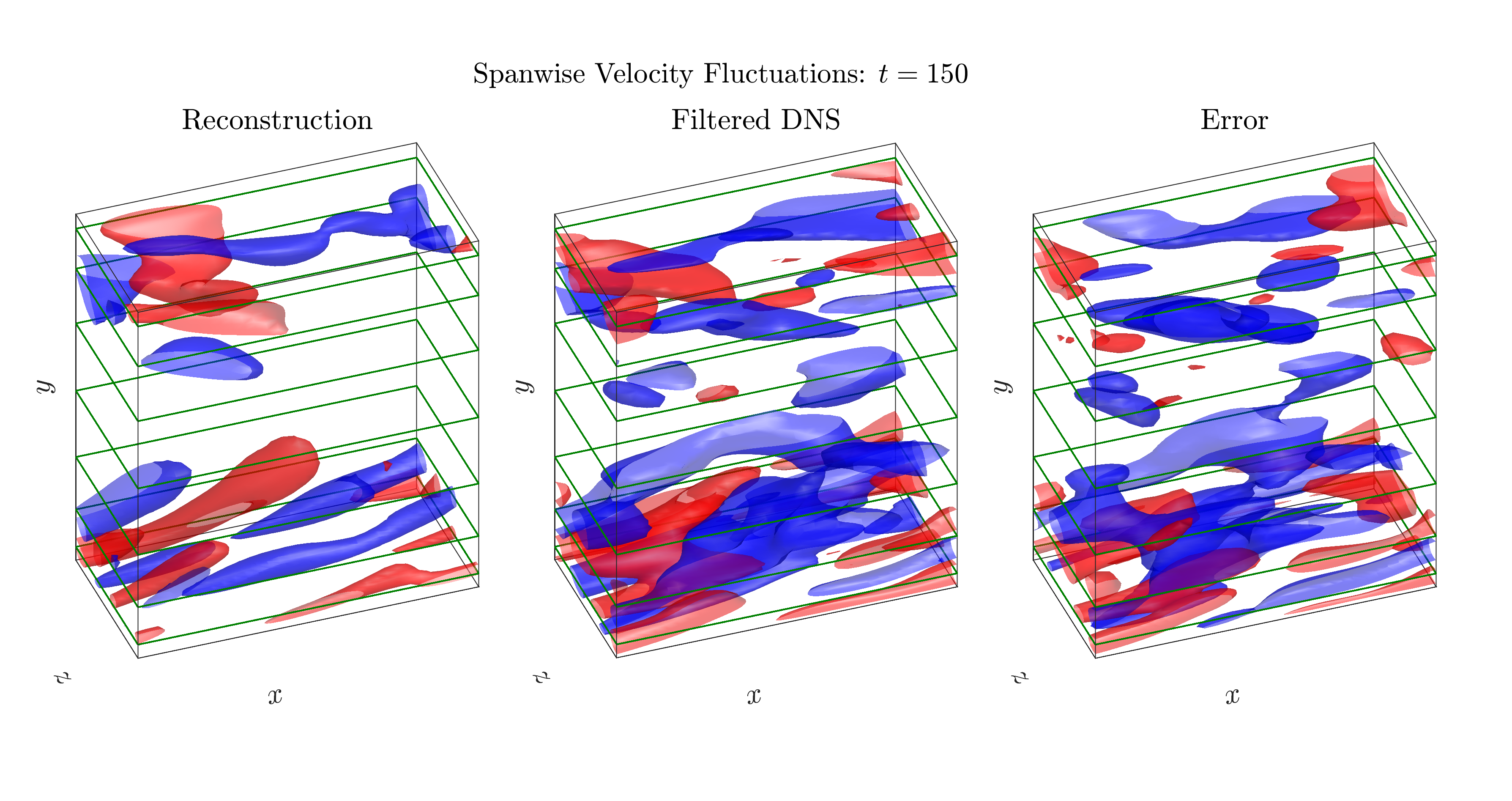}
    \caption{Comparisons of the reconstructions, filtered DNS snapshots, and their differences in the style of Figs. \ref{fig:res:strme:2} and \ref{fig:res:sorbe:2}. The comparisons here are shown for STSME with $N_{\rm plane} = 7$ and $N_{\rm mode} = 2$, for which $\epsilon_{\rm filt}(t) = 0.681$ and $\epsilon_{\rm full}(t) = 0.711$.}
    \label{fig:res:stsme:2}
\end{figure}

When full-field statistics from training data are available, STSME provides a data-driven method for flow reconstruction that efficiently captures energetic structures on a modal basis. Figures \ref{fig:res:stsme:1} and \ref{fig:res:stsme:2} summarize the accuracy of the STSME estimators for each modal truncation.

%%% First two figures go here (errors and snapshots)

\begin{figure}[!htpb]
    \centering
    \includegraphics[width=0.7\textwidth,trim=70 30 0 50,clip]{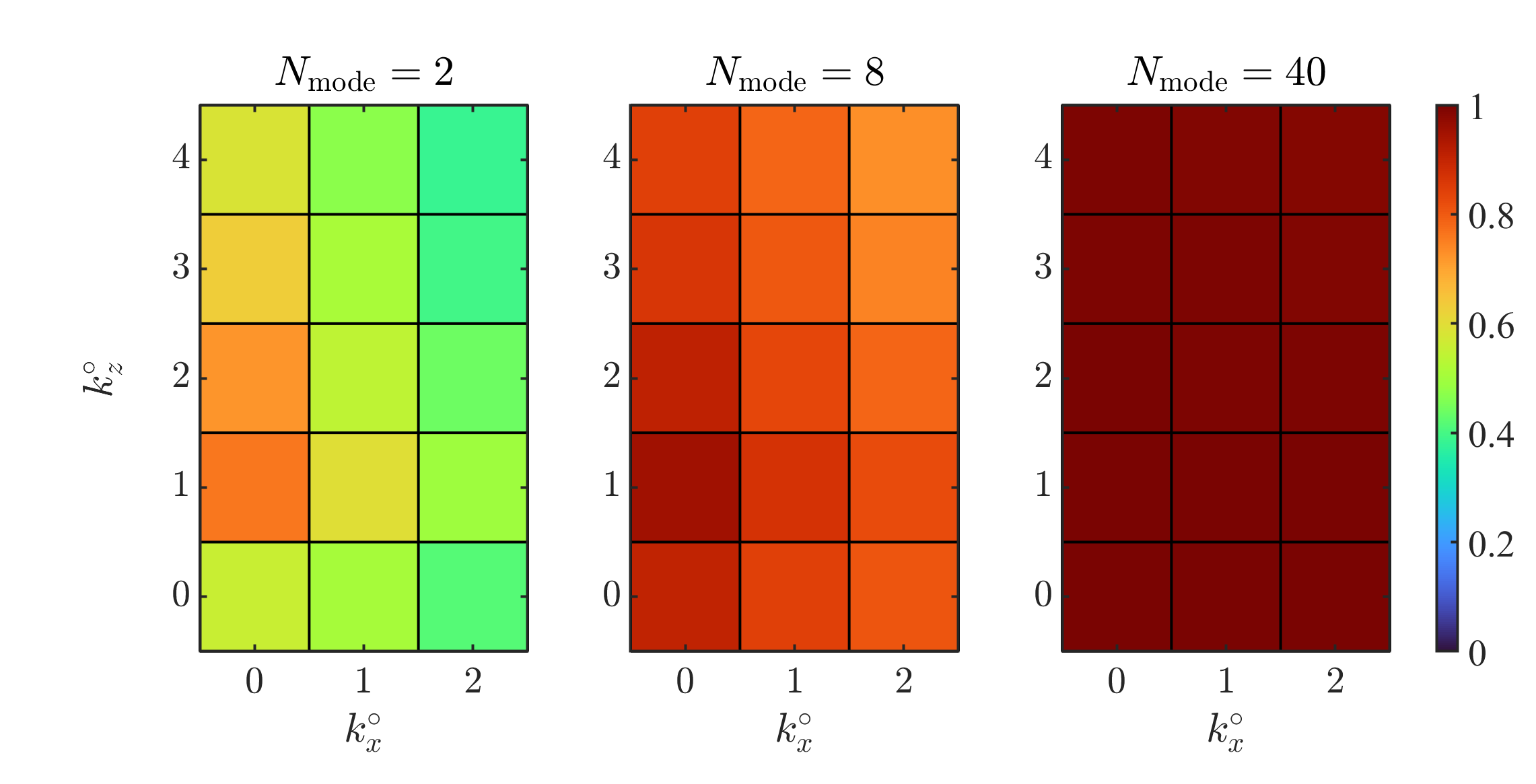}
    \caption{\rev{The fraction of the kinetic energy captured by the first $N_{\rm mode}$ SPOD modes for the $N_{\rm mode} =$ 2, 8,  and 40 truncations. The fraction is shown as a function of $k_x^\circ = k_x \Delta x N_x/ 2 \pi \approx k_x/3.54$ and $k_z^\circ = k_z \Delta z N_z / 2 \pi \approx k_z/7.08$, and the energies are summed over all temporal frequencies. The results are symmetric about $k_x^\circ = 0$ and $k_z^\circ = 0$.}}
    \label{fig:res:SPOD_compare}
\end{figure}

Similar to the STRME cases, the $N_{\rm mode} = 2$ estimator can outperform the $N_{\rm mode} = 8$ estimator when relatively few measurement planes are used. However, in contrast to the STRME cases, the higher-order truncations ($N_{\rm mode} \geq 40$) produce similar, relatively accurate reconstructions for all measurement configurations. \rev{This behavior and the approximate stationarity of the estimation errors during the testing period suggest that the SPOD modes, and the flow statistics underlying their computation, are approximately converged. Hence, the coefficient statistics given by the SPOD eigenvalues accurately weight the SPOD modes, unlike in the STRME method. The performance of the $N_{\rm mode} = 40$ estimator also suggests that the first 40 SPOD modes form an approximately well-posed basis for the flow. This observation is supported by Fig. \ref{fig:res:SPOD_compare}, which shows that the $N_{\rm mode} = 40$ truncation captures nearly all of the kinetic energy of the velocity fluctuations. By contrast, the $N_{\rm mode} = 2$ and $N_{\rm mode} = 8$ truncations capture significant but relatively incomplete portions of the kinetic energy, especially at smaller spatial scales.} 

%%% Third figure goes here (SPOD modes)

Since the SPOD modes remain energetically relevant up to approximately mode 40, the lower-rank modal truncations provide ill-posed bases for representing the flow. Correspondingly, whereas the localized errors are excellent for $N_{\rm mode} \geq 40$, they are considerably larger for the lower-order truncations. Similarly, as shown for $N_{\rm plane} = 7$, the filtered reconstruction errors at the measurement planes are only near-zero for $N_{\rm mode} \geq 40$ and they are visibly nonzero for the lower-order cases. \rev{As observed for SORBE,} even with a well-posed basis, the reconstruction errors still grow far from the measurements due to the locality of the correlations \rev{and the influence of time-varying flow statistics (see Sec. \ref{sec:res:sumsec}).}

When all SPOD modes are retained, the STSME estimator converges to the generalized Wiener filter (as did the SORBE estimator with full training data). Correspondingly, the transfer functions computed using full-field training data for SORBE and an untruncated basis for STSME produce identical error profiles and integrated errors. As $N_{\rm plane}$ increases, these errors decay smoothly since \rev{the flow becomes increasingly well correlated with the measurements.}

Besides the generalized Wiener filter, even the lower-order truncations produce impressive reconstructions when $N_{\rm plane} = 7$. Even though using $N_{\rm mode} = 2$ provides a highly compressed representation of the flow (retaining only 1.2\% of the modes), it produces very good reconstructions of various flow features during the testing period. For example, it produces good reconstructions of the rms fluctuations, especially for the streamwise direction. Further, the dominant features of the mean cross-correlations are very well captured during the testing period. The snapshots in Fig. \ref{fig:res:stsme:2} further support this observation. Whereas the errors are not negligible with respect to the flow structures, the reconstructions qualitatively capture most relevant flow structures for each velocity component. Hence, if leading SPOD modes can be computed using alternative means to alleviate the training data requirement, STSME has the potential to provide efficient reconstructions of energetically relevant flow structures. However, as presently formulated, the spatial extent of the training statistics is likely a prohibitive limitation to applying the STSME estimators in practice. 

%%%%%%%%%%%%%%%%%%%%%%%%%%%%%%%%%%%%%%%%%%
\subsection{Reconstruction accuracy discussion}\label{sec:res:sumsec}
%%%%%%%%%%%%%%%%%%%%%%%%%%%%%%%%%%%%%%%%%%

While the STRME, SORBE, and STSME estimators considered share many themes regarding experimental design, they each have distinct use cases based on the fidelity of the available training data. When no training data are available, STRME provides an equation-based means of flow reconstruction that critically relies on the validity of the assumed mode weight statistics. The uncorrelated forcing assumption may be informative when only localized reconstructions or streamwise rms reconstructions are required. However, beyond these limited settings, second-order statistics help to inform more accurate reconstructions, especially for the wall-normal and spanwise fluctuations. When spatially limited training data are available, SORBE provides an efficient means of improving reconstruction accuracy by incorporating the data with the governing equations. Designing the estimator involves cost-accuracy tradeoffs for the fidelity of training data and the measurements. When full-field training data are available, STSME provides a data-driven framework that efficiently captures energetic flow structures on a modal basis. Practical constraints often dictate the available data and, correspondingly, the estimator used for flow reconstruction.

\rev{The linear model and training data used to form the selected estimator are critical to its behavior in constrained (e.g., experimental) settings.} One common theme for the estimators we consider is that errors grow large when relatively few ($N_{\rm plane} \leq 4)$ measurements are used in conjunction with an ill-posed modal truncation (STRME and STSME) or incomplete training data (SORBE). \rev{Whereas the errors decay monotonically with $N_{\rm plane}$ for well-posed STSME estimators ($N_{\rm mode} \gtrsim 40$), the STRME errors behave nonmonotonically even with all $N_{\rm mode} = 388$ resolvent modes due to ill-posed nature of the uncorrelated forcing assumption. By contrast, the SORBE errors decay monotonically in the generalized Wiener filter limit since the statistics from training data more accurately represent the flow.} \rev{These results reflect} that \rev{robust}, high-accuracy, full-field reconstructions require, within practical limitations, a tailored combination of (i) sufficiently high-fidelity estimators and (ii) sufficiently many measurements (e.g., $N_{\rm plane} \geq 6$) spread throughout the channel.

%%%%%%%%%%%%%%%%%%%%%%%%%%%%%%%%%%%%%%%%%%
\subsubsection{Reconstruction performance with a single plane}\label{sec:res:singleplane}
%%%%%%%%%%%%%%%%%%%%%%%%%%%%%%%%%%%%%%%%%%

\rev{While multiplane PIV is capable of simultaneously measuring multiple planes in an experimental setting, single-plane PIV represents a more conventional experimental setup. Here, we discuss the errors for our $N_{\rm plane} = 1$ generalized Wiener filter, which is} particularly informative of estimator performance when measurements are isolated to one half of the channel (at $y^+ \approx 14.7$).

\begin{figure}[!tpb]% was [!htpb], but that looks weird unfortunately
    \centering
    \includegraphics[trim=25 25 25 100, clip, width=0.869\textwidth]{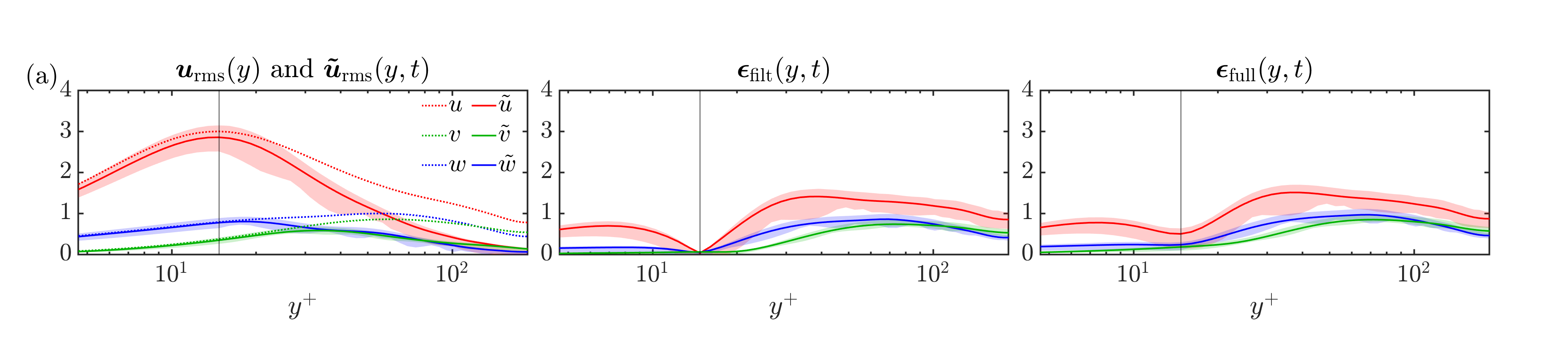}
    \includegraphics[trim=0 0 0 50, clip, width=0.869\textwidth]{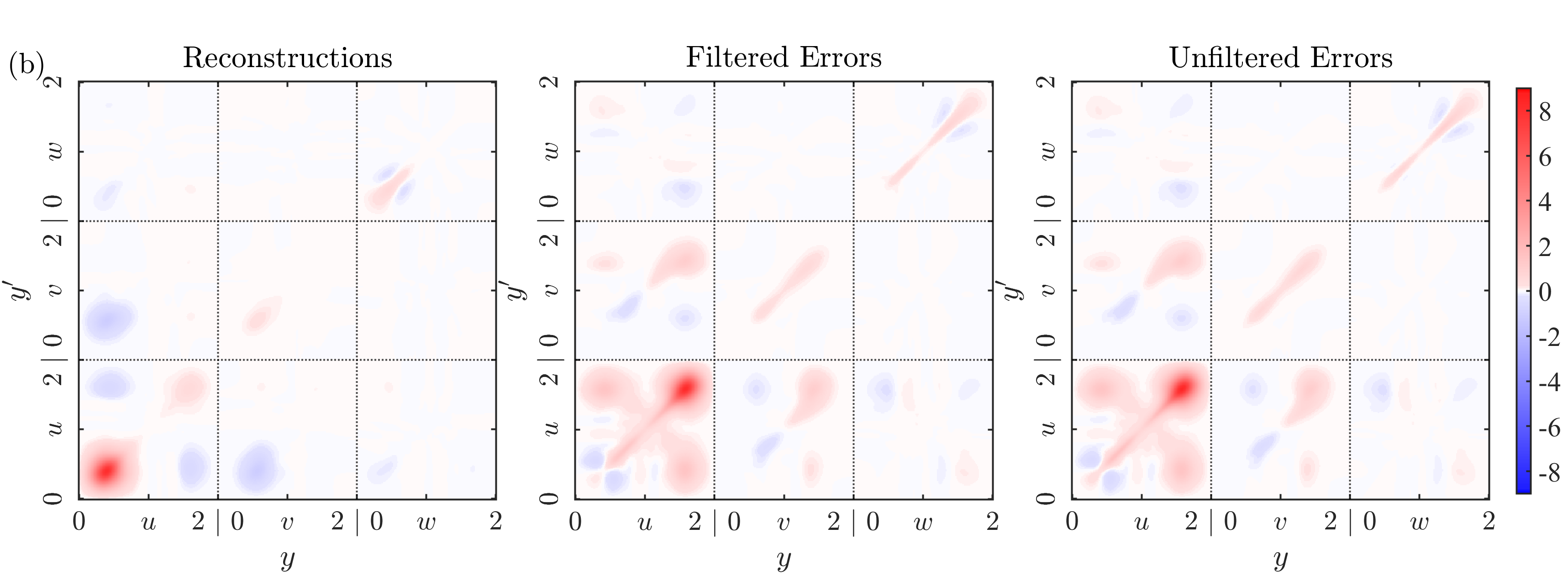}
    \caption{Summary of the reconstruction accuracies (top: spatial profiles, bottom: cross-correlations) for the generalized Wiener filter using $N_{\rm plane} = 1$. The plots are in the style of Figs. \ref{fig:res:strme:1}, \ref{fig:res:sorbe:1}, and \ref{fig:res:stsme:1}.}
    \label{fig:app:1plane_profiles}
\end{figure}

\begin{figure}[!htpb]
    \centering
    \includegraphics[trim=25 100 25 100, clip, width=0.92\textwidth]{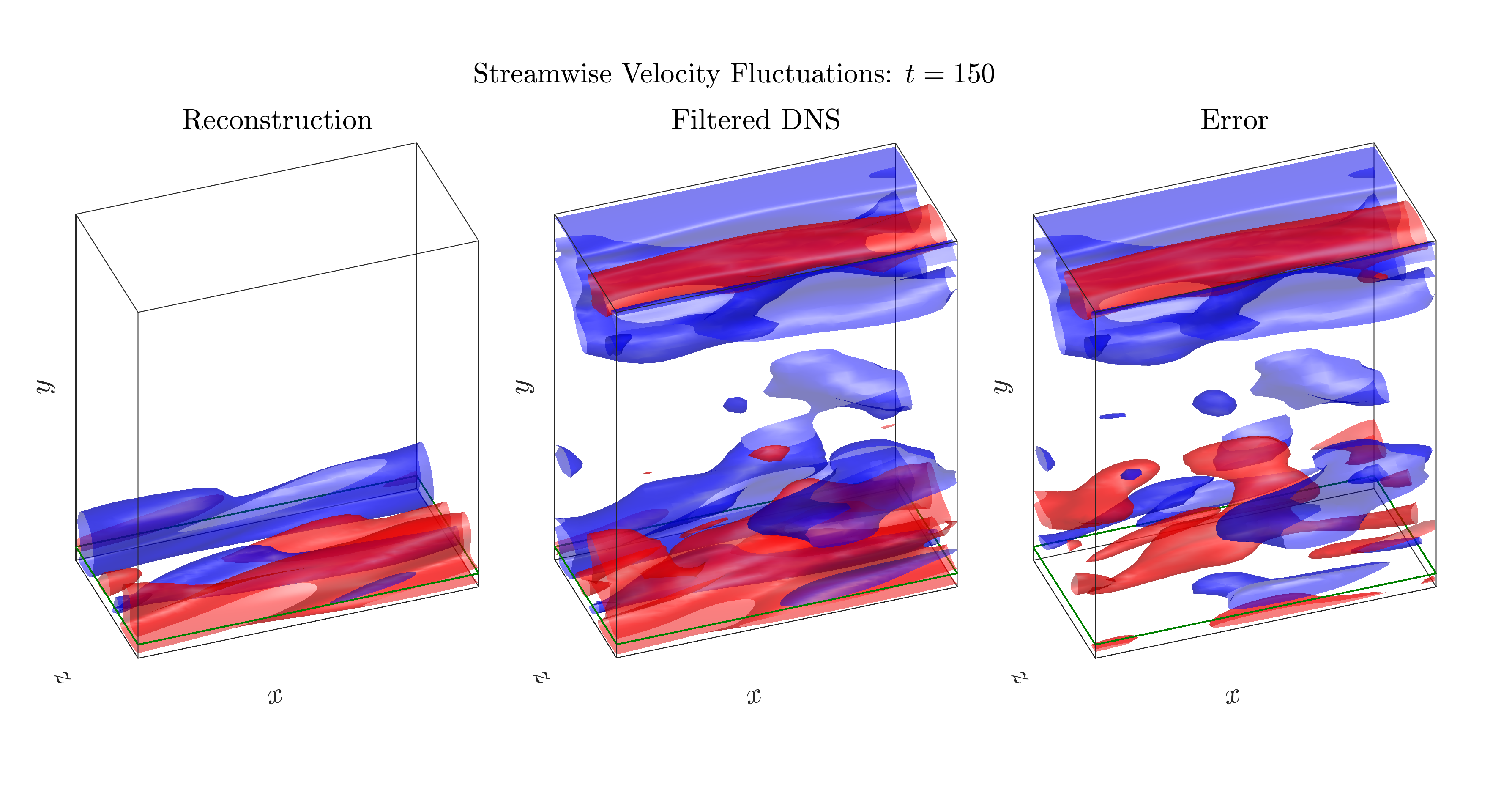}
    \includegraphics[trim=25 100 25 100, clip, width=0.92\textwidth]{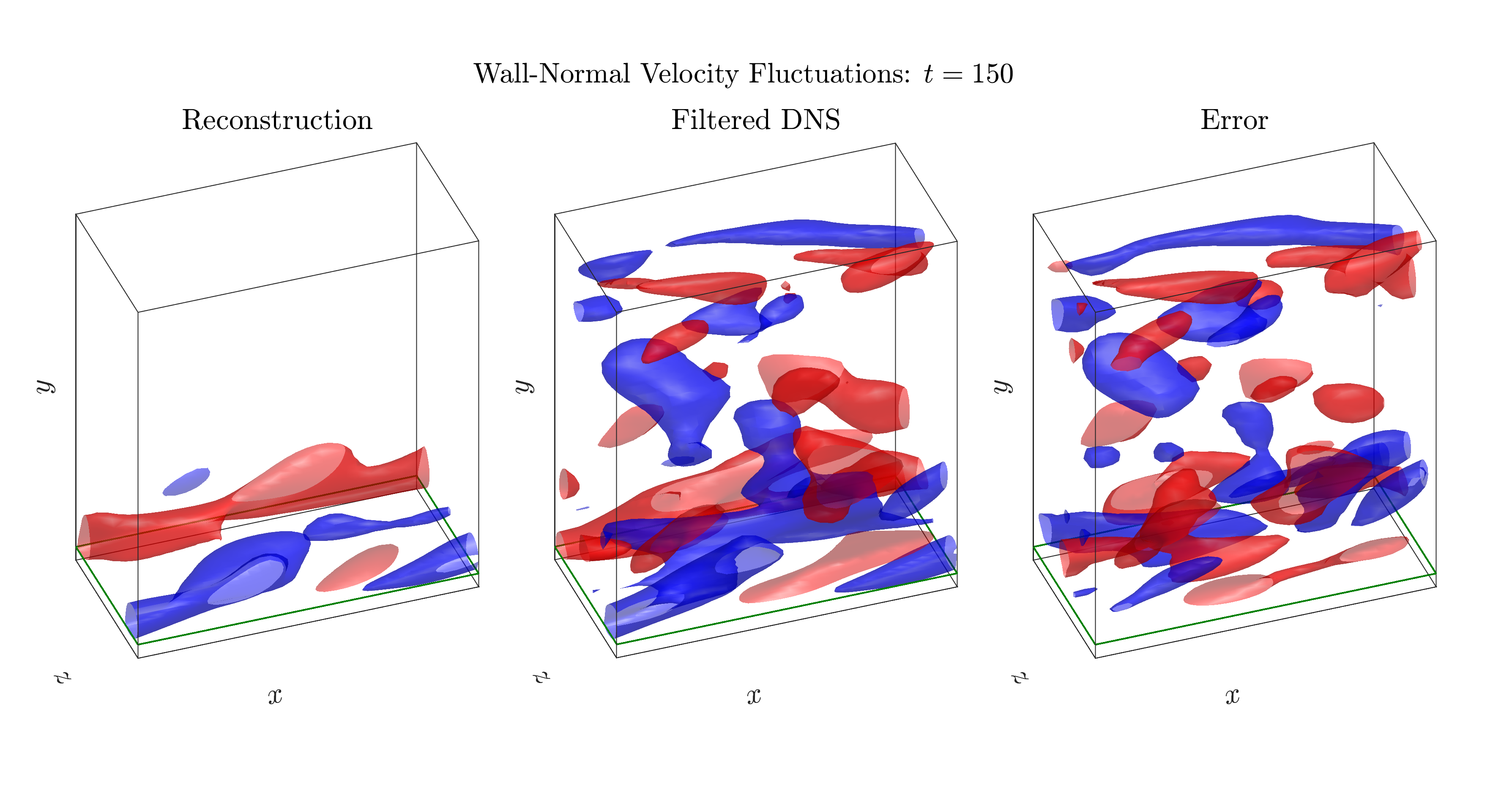}
    \includegraphics[trim=25 100 25 100, clip, width=0.92\textwidth]{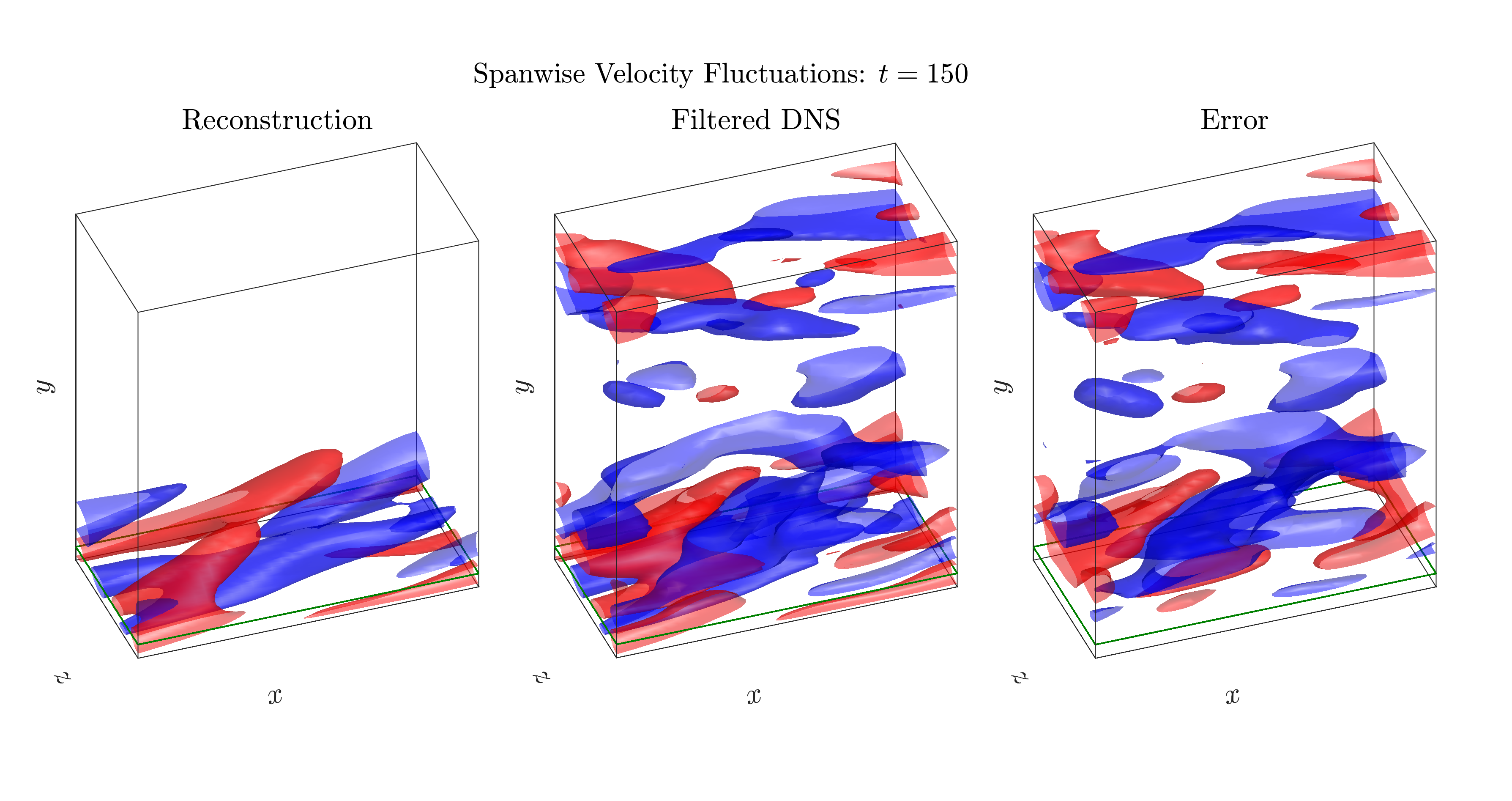}
    \caption{Comparisons of the reconstructions, filtered DNS snapshots, and their differences in the style of Figs. \ref{fig:res:strme:2}, 
    \ref{fig:res:sorbe:2}, and \ref{fig:res:stsme:2}. The comparisons here are shown for the generalized Wiener filter with $N_{\rm plane} = 1$, for which $\epsilon_{\rm filt}(t) = 0.840$ and $\epsilon_{\rm filt}(t) = 0.865$. Videos depicting the temporal evolutions of these comparisons are provided in the Supplemental Material \citep{SuppMat}.}
    \label{fig:app:1plane_snaps}
\end{figure}

% FROM APPENDIX E
Figure \ref{fig:app:1plane_profiles} shows the spatial structure of the errors associated with the $N_{\rm plane} = 1$ generalized Wiener filter. This estimator produces highly accurate reconstructions in the vicinity of the measurement plane and, more generally, in the near-wall region. However, limited information far from the wall can be inferred from the single measurement plane. As expected by the channel symmetry, the most energetic correlations with the flow in the opposite half of the channel occur at the same wall-normal location. However, these correlations are considerably less energetic than those in the vicinity of the measurement plane. These results demonstrate that the near-wall fluctuations in each half of the channel are relatively decoupled from one another. These observations are qualitatively confirmed by the reconstruction snapshots in Fig. \ref{fig:app:1plane_snaps}. While energetic flow structures are reconstructed for each velocity component, these structures are isolated to the bottom half of the channel. Since we place the single measurement plane at the peak of $u_{\rm rms}$, it represents a logical choice for reconstructing the near-wall flow associated with the corresponding streamwise-dominated flow structures.

\begin{figure}[!htpb]
    \centering
    \includegraphics[width=\textwidth,trim=125 0 0 0,clip]{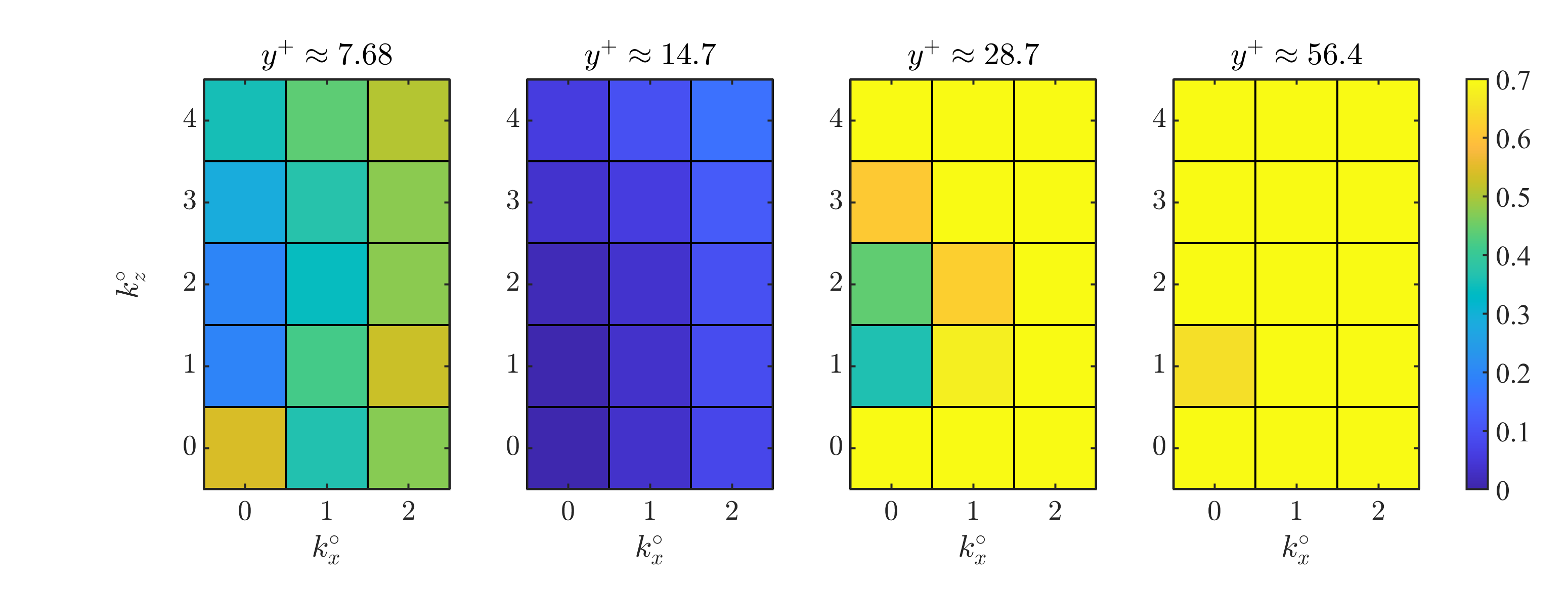}
    \caption{\rev{Relative error levels, $\gamma$, at each estimation plane, plotted using a color axis that is consistent with previous investigations \citep{Ill2018,Ama2020}. The normalized wave numbers, $k_x^\circ$ and $k_z^\circ$, are the same as those in Fig. \ref{fig:res:SPOD_compare}. The squared errors are averaged over positive and negative $k_x^\circ$ and computed prior to averaging wall-normal velocities to the cell centers.}}
    \label{fig:IllAmaCompare}
\end{figure}

% FROM REVIEWER 2 RESPONSE
\rev{Beyond providing a practical benchmark for conventional experiments, the $N_{\rm plane} = 1$ case also enables comparisons with previous studies \citep{Ill2018,Ama2020} that employ linear estimators in similar turbulent channel flows. Consistent with these studies, we consider the error metric
\begin{equation}
    \gamma(y,k_x,k_z) = \left( \frac{\int_{\Delta T^*} \left| \mathscr{F}_{x,z} \{ \boldsymbol{u}_{\rm filt}(\boldsymbol{x},t) - \boldsymbol{\tilde{u}}(\boldsymbol{x},t) \} \right|^2 dt}{\int_{\Delta T^*} \left| \mathscr{F}_{x,z} \{ \boldsymbol{u}_{\rm filt}(\boldsymbol{x},t) \} \right|^2 dt} \right)^{1/2},
\end{equation}
where the integrals are taken over the testing period and $\mathscr{F}_{x,z} \{ \cdot \}$ represents the spatial Fourier transform. This metric represents the distribution of relative errors over the spatial Fourier modes considered, and it is typically evaluated at estimation planes that are distinct from the measurement planes. \citet{Ill2018} provides a detailed discussion of estimator performance through the lens of this metric in the context of linear models.

Figure \ref{fig:IllAmaCompare} shows the relative error distributions at various planes in the same half-channel as the measurement plane. These errors are qualitatively similar to those reported by \citet{Ama2020} using wall measurements at ${\rm Re}_\tau \approx$ 500 and 1000. Specifically, the present relative error levels at $y^+ \approx 28.7$ are similar to those reported previously \citep{Ama2020} at a comparable distance (in outer units) above the measurement plane. Those errors levels, which represent ORBE in a noncausal setting, were also shown \citep{Ama2020} to be favorable to those incurred by the relevant estimator of \citet{Ill2018}. The results of these related investigations also suggest that, while not investigated here, employing an eddy viscosity has the potential to further improve the estimation errors. However, care should be taken when comparing the present errors to those of the previous studies \citep{Ill2018,Ama2020} since they were conducted at ${\rm Re}_\tau > 186$. 

}

%%%%%%%%%%%%%%%%%%%%%%%%%%%%%%%%%%%%%%%%%%
\subsubsection{Spatial structure of reconstruction errors}\label{sec:res:spatstruct}
%%%%%%%%%%%%%%%%%%%%%%%%%%%%%%%%%%%%%%%%%%

\begin{figure}[!htpb]
    \centering
    \includegraphics[width=\textwidth]{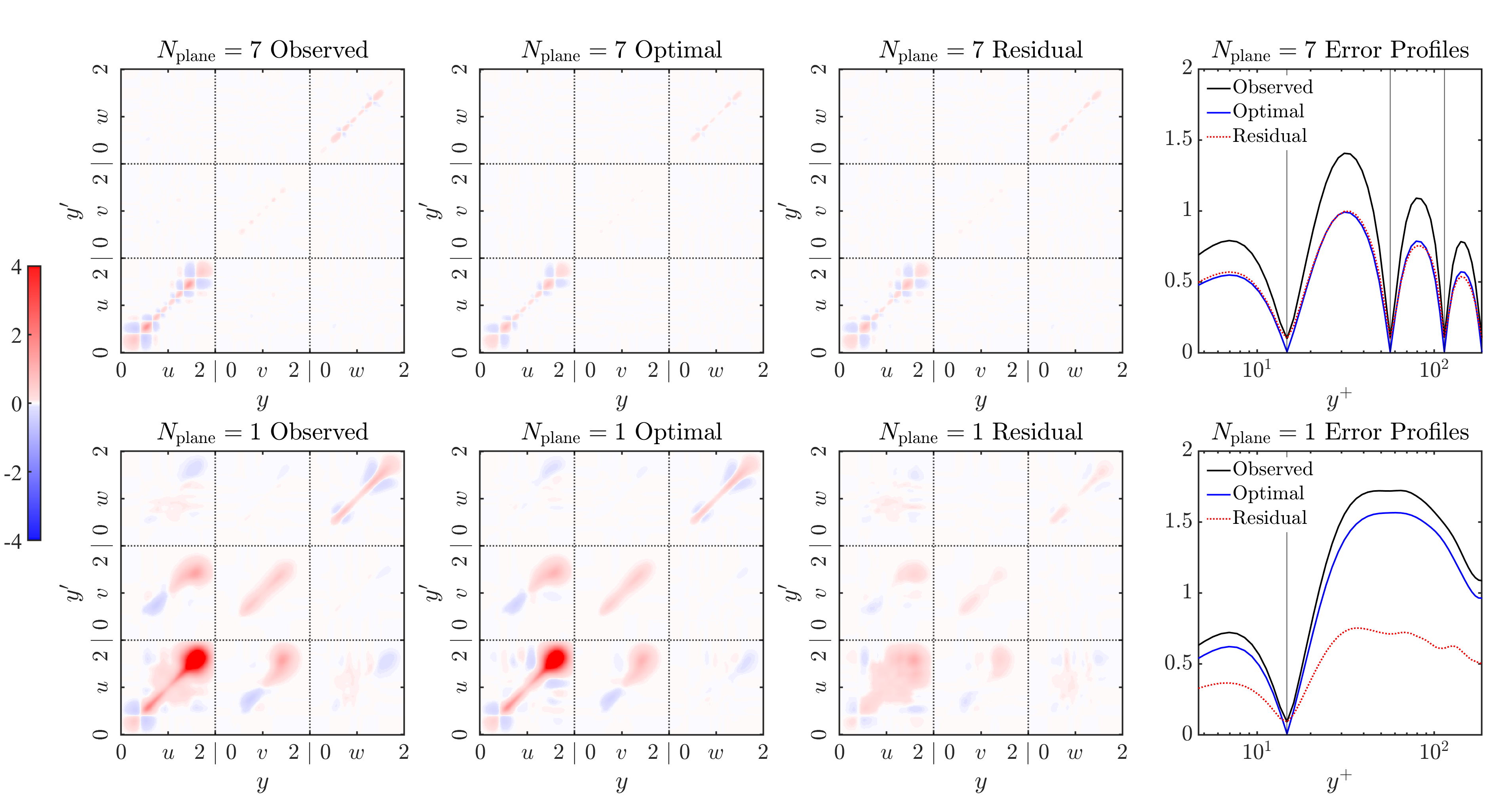}
    \caption{\rev{Time-averaged cross-correlations of the estimation errors with respect to $\boldsymbol{u}_{\rm filt}$ during the testing period (column 1), the contributions associated with the optimal error CSDs in (\ref{eq:Wiener_err}) (column 2), and those associated with the residual error CSDs in (\ref{eq:Wiener_res}) (column 3). Column 4 depicts the rms error profiles, the squares of which represent the diagonals of the cross-correlations in columns 1-3, summed over all three components. These error profiles are consistent with those of $\boldsymbol{\epsilon}_{\rm filt}$ shown for generalized Wiener filter with $N_{\rm plane} = 7$ (see Fig. \ref{fig:res:sorbe:1}) and $N_{\rm plane} = 1$ (see Fig. \ref{fig:app:1plane_profiles}).}}
    \label{fig:rev1_error_source}
\end{figure}

\rev{The spatial structure of the reconstruction errors, particularly in the wall-normal direction, provides insight into how streaming measurements, sensor placement, and flow statistics impact estimator performance. This structure can be formally characterized using the CSD of the estimation error, which is given by
\begin{equation}\label{eq:Wiener_err}
     \mathbb{E} \left[ \left( \boldsymbol{\hat{u}} - \boldsymbol{\tilde{\hat{u}}} \right)\left( \boldsymbol{\hat{u}} - \boldsymbol{\tilde{\hat{u}}} \right)^H \right] = \boldsymbol{S_{uu}} - \boldsymbol{S_{uy}} \boldsymbol{T_u}^H - \boldsymbol{T_u} \boldsymbol{S_{uy}}^H + \boldsymbol{T_u} \boldsymbol{S_{yy,n}} \boldsymbol{T_u}^H.
\end{equation}
As discussed previously in the context of ORBE \citep{Tow2020,Mar2020}, the error CSD expression is based on the observable subspace of linear model coefficients $\boldsymbol{b}$, as determined by the selected measurement planes. However, the expression in (\ref{eq:Wiener_err}) assumes that $\boldsymbol{S_{uu}}$ is fixed, consistent with the noncausal formulation of the generalized Wiener filter that formally considers a temporal Fourier transform of infinite length.

Since our streaming reconstructions employ a small sliding temporal window ($\Delta T = 1$), there is a residual in the error CSD due to the temporal variations in the streaming flow statistics. During the testing period, the streaming response CSD is instantaneously given by $\boldsymbol{S_{uu}}^*(y,\boldsymbol{k},t) = \boldsymbol{\hat{u}}\boldsymbol{\hat{u}}^H$, which is computed without averaging. This streaming CSD has a residual with respect to the true response CSD of $\boldsymbol{\varepsilon_{uu}}^*(y,\boldsymbol{k},t) =  \boldsymbol{S_{uu}}^*(y,\boldsymbol{k},t) - \boldsymbol{S_{uu}}(y,\boldsymbol{k})$, and we employ analogous definitions for $\boldsymbol{\varepsilon_{uy}}^*$ and $\boldsymbol{\varepsilon_{yy,n}}^*$. The corresponding residual in the error CSD with respect to the expression in (\ref{eq:Wiener_err}) is given by
\begin{equation}\label{eq:Wiener_res}
    \left( \boldsymbol{\hat{u}} - \boldsymbol{\tilde{\hat{u}}} \right)\left( \boldsymbol{\hat{u}} - \boldsymbol{\tilde{\hat{u}}} \right)^H -  \mathbb{E} \left[ \left( \boldsymbol{\hat{u}} - \boldsymbol{\tilde{\hat{u}}} \right)\left( \boldsymbol{\hat{u}} - \boldsymbol{\tilde{\hat{u}}} \right)^H \right] = \boldsymbol{\varepsilon_{uu}}^* - \boldsymbol{\varepsilon_{uy}}^* \boldsymbol{T_u}^H - \boldsymbol{T_u} {\boldsymbol{\varepsilon_{uy}}^{* H}}  + \boldsymbol{T_u} \boldsymbol{\varepsilon_{yy,n}}^* \boldsymbol{T_u}^H.
\end{equation}
Here, the design of the sliding temporal window primarily impacts the contribution of $\boldsymbol{\varepsilon_{uu}}^*$. The last three terms in (\ref{eq:Wiener_res}) characterize how that contribution propagates through linear estimation framework, and they depend critically on the selected measurements. Altogether, the residual in (\ref{eq:Wiener_res}) characterizes how applying a formally noncausal estimator in a setting with time-varying flow statistics contributes to the estimation errors.

Figure \ref{fig:rev1_error_source} depicts how the cross-correlations of the observed errors during testing are partitioned between the optimal and residual cross-correlations associated, via Parseval's theorem, with the CSDs in (\ref{eq:Wiener_err}) and (\ref{eq:Wiener_res}), respectively. As expected, the error variance at the sensor planes directly results from the residual errors since the optimal errors decay to zero at these locations. For $N_{\rm plane } = 7$, the variances associated with the optimal and residual errors are roughly equipartitioned away from the sensor planes. In this case, optimizing the design of the sliding Fourier transform window has the potential to further reduce errors by producing flow statistics with reduced temporal variations. By contrast, for $N_{\rm plane} = 1$, the optimal errors dominate the residual errors far from the measurement plane. Since the true generalized Wiener filter is invariant of the linear model, this case reflects a regime in which spatially constrained measurements hinder estimator performance more than temporal uncertainty in flow statistics. 
}

The spatial structures of the reconstruction accuracies also share various features across each linear model we consider. Far from measurements, the streamwise errors tend to grow larger than the spanwise errors, which tend to be larger than the wall-normal errors. This reflects (i) the typical magnitudes of these fluctuations and (ii) that low-error regions are more localized about measurements for $u$ and $w$ than for $v$. Physically, it reflects that the streamwise and spanwise flow structures are more localized in $y$, whereas the wall-normal flow structures are more persistent in $y$. The maximum errors (far from measurements) do not grow larger than the rms fluctuations when the estimators and measurements are of sufficiently high fidelity. Especially for SORBE and STSME, increasing the fidelity of the estimator reduces error levels throughout the channel. In the limit of full training data, the highest fidelity SORBE and STSME estimators are identical and converge to the generalized Wiener filter \rev{with the caveat of nonoptimality imposed by time-varying flow statistics}.

% FROM ORIGINAL MANUSCRIPT
While the full-field errors, $\epsilon_{\rm TKE}(t)$, $\epsilon_{\rm filt}(t)$, and $\epsilon_{\rm full}(t)$, are informative of the global accuracy of each reconstruction method, the spatially truncated errors, $\epsilon^*_{\rm TKE}(t)$, $\epsilon^*_{\rm filt}(t)$, and $\epsilon^*_{\rm full}(t)$, provide a more localized picture of reconstruction accuracy. As expected, these localized errors are considerably smaller than the full-field errors for all cases considered. However, the localized errors also tend to increase when increasing the number of sensor planes. These behaviors reflect that the larger full-field errors are concentrated far from the measurement planes and that truncated errors are computed over smaller fractions of the channel when fewer sensor planes are used. Correspondingly, whereas adding additional sensors can marginally increase errors near the sensor planes, this effect is outweighed globally by the reduction in errors far from the sensor planes.

In the present study, we heuristically limit the spatial integration of localized errors to within $N_{\rm local} = 2$ cells from each measurement plane. In Appendix \ref{sec:app:tempsig}, we further probe how the spatial extents of the localized error domains affect reconstruction accuracy by comparing the errors, $\epsilon^*_{\rm filt}(t)$ and $\epsilon^*_{\rm full}(t)$, for $N_{\rm local}$ = 1, 2, and 3. The results highlight that the spatial structure of the errors can be used to limit reconstructions and thereby improve their accuracy in a manner relatively invariant to the extent of truncation.

%%%%%%%%%%%%%%%%%%%%%%%%%%%%%%%%%%%%%%%%%%
\subsection{Towards real-time reconstructions}\label{sec:realtime}
%%%%%%%%%%%%%%%%%%%%%%%%%%%%%%%%%%%%%%%%%%

\begin{figure}[!htb]
    \centering
    \includegraphics[trim=0 25 100 50, clip, width=\textwidth]{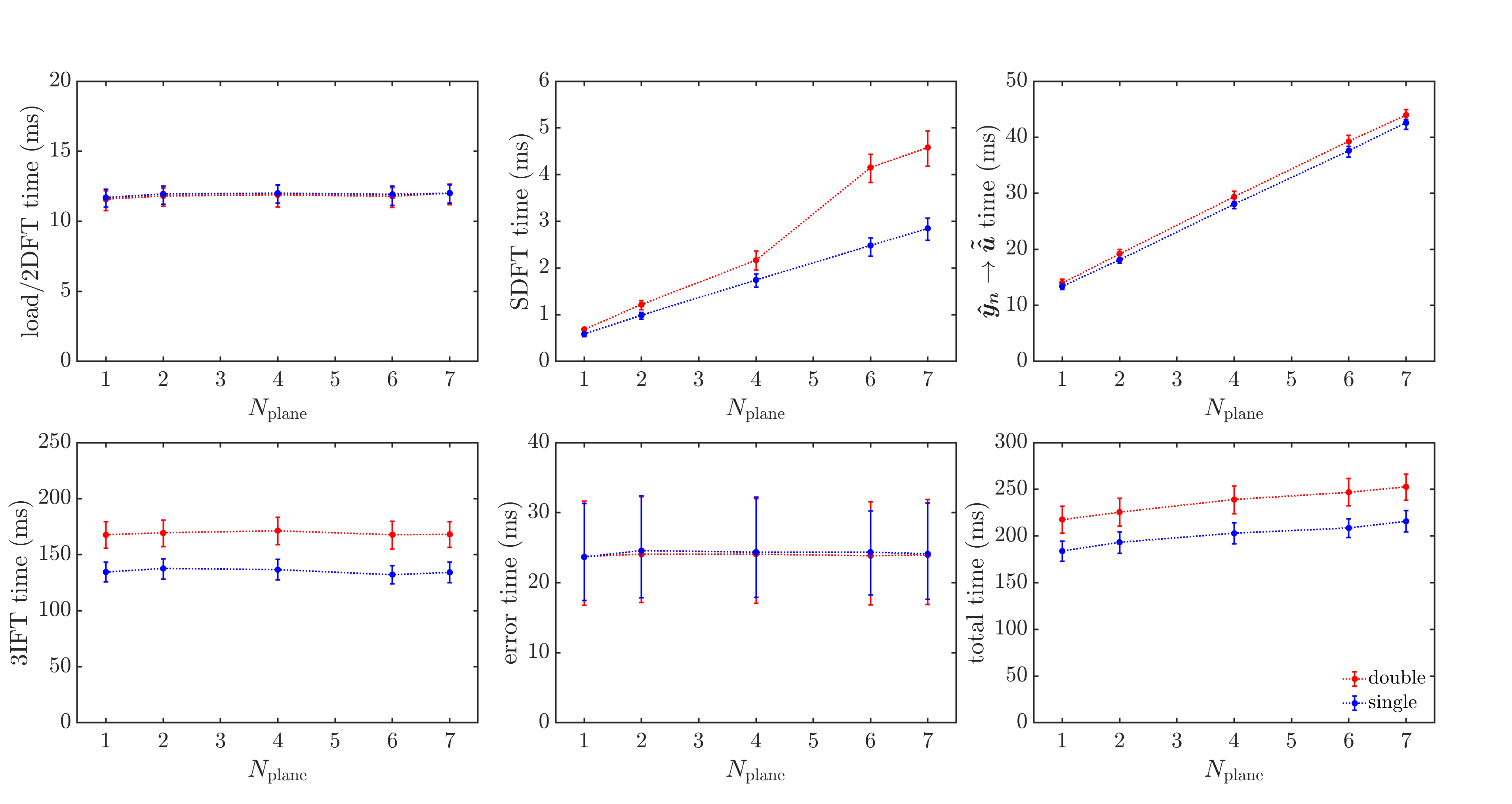}
    \caption{Wall times required (per time step, $\Delta t = 2.86 \times 10^{-3}$) to complete each stage of the real-time flow reconstruction framework employed during the testing period. \rev{The uncertainties about the averages extend to the $20{\rm th}$ and $80{\rm th}$ percentiles.} For $\boldsymbol{\hat{y}_n} \rightarrow \boldsymbol{\tilde{\hat{u}}}$, $\boldsymbol{T_u}$ is single precision for both ``double'' and ``single'' precision computations to save space. We exclude the time required to produce graphics for each snapshot.}
    \label{fig:sep_times}
\end{figure}

In addition to reconstruction accuracy, we also evaluate the feasibility of implementing the present framework in real time (e.g., for experimental flows). We primarily focus on using the time delay required for computationally reconstructing the flow to determine relevant experimental parameters that would enable real-time reconstructions. Figure \ref{fig:sep_times} provides a summary of the wall times required for each step of the reconstructions when running the (relatively unoptimized) code implementing the present frameworks on a laptop.
The minimal difference between the single and double precision times when applying $\boldsymbol{T_u}$ arises since $\boldsymbol{T_u}$ is always single precision and only the precision of $\boldsymbol{\hat{y}_n}$ varies. One way that the present code is unoptimized is that converting $\boldsymbol{\tilde{\hat{u}}}$ to $\boldsymbol{\tilde{u}}$ via the inverse Fourier transform (3IFT) is performed after estimating the fluctuations in Fourier space. Since these estimates are computed in parallel for each triplet, a more efficient method would simply transform and add the contributions of each triplet to the physical-space estimates as they are computed (in Fourier space). This modification will be helpful in that the 3IFT presently composes the largest fraction of the time required to reconstruct each snapshot. Further speedups may be garnered by exploiting the Hermitian symmetry of the Fourier transform when applied to real signals.

\begin{figure}
    \centering
    \includegraphics[trim=0 0 0 0, clip, width=0.8\textwidth]{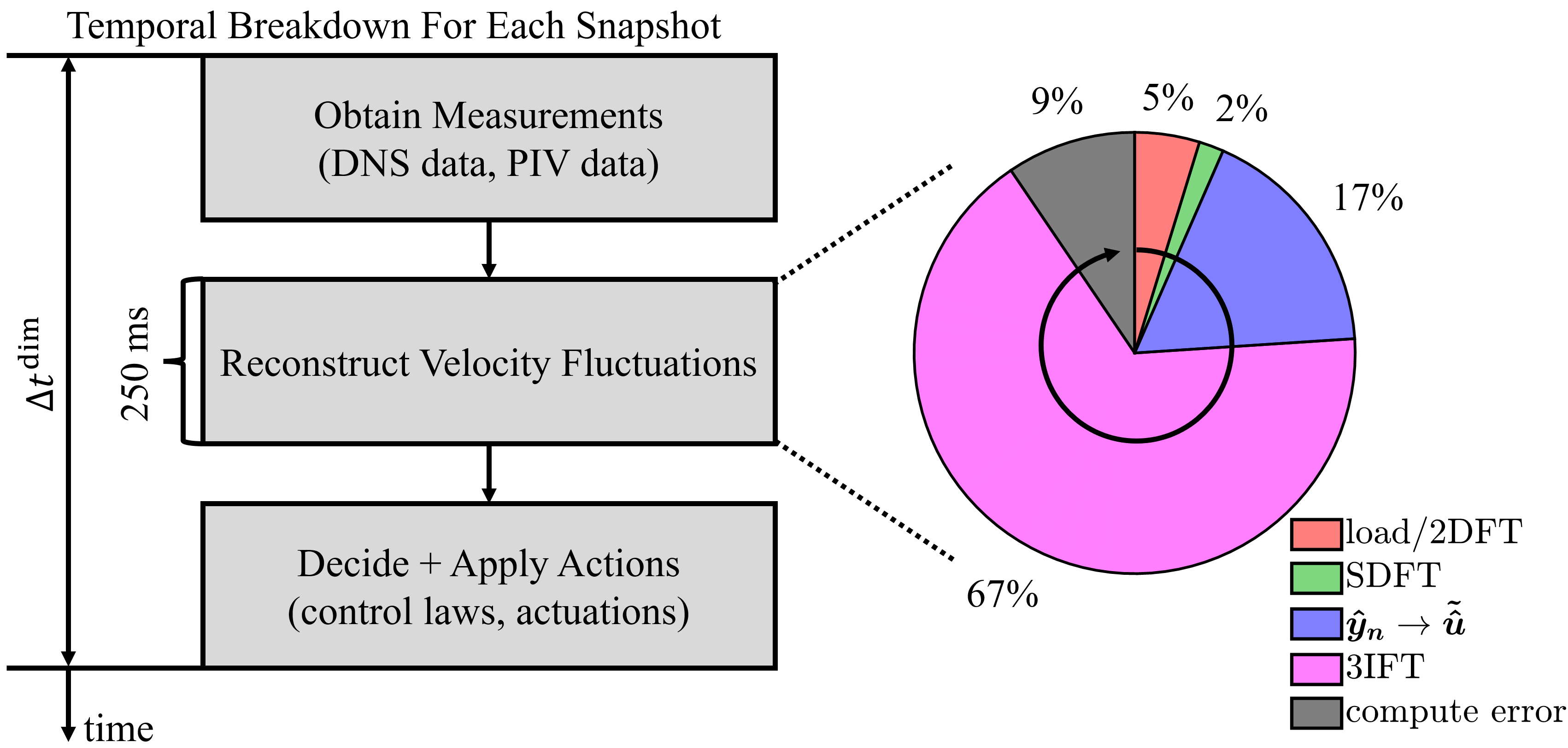}
    \caption{Breakdown of the wall time required to achieve real-time reconstructions in an experimental setting. The pie chart shown is based on the mean wall times for double precision reconstructions using $N_{\rm plane} = 7$, as in Fig. \ref{fig:sep_times}.}
    \label{fig:piechart}
\end{figure}

As expected, the scaling of the total reconstruction time (per snapshot) with $N_{\rm plane}$ results from applying the SDFT and applying $\boldsymbol{T_u}$. At the high end, the present framework reconstructs snapshots in approximately 250 ms, and we use this as our benchmark for experimental design. Figure \ref{fig:piechart} summarizes how this benchmark time can be used in designing an experimental protocol for real-time reconstructions. Beyond reconstruction wall time, real-time schemes must also consider the time required to (i) obtain velocity measurements (e.g., from raw images in PIV) and (ii) use the reconstructions for related tasks (e.g., flow control). \rev{Regarding (i), hardware implementations of PIV computations have computed flow fields from raw data at 15 Hz \citep{HYu2006} and sparse processing techniques have been used to accelerate computations to up to 2000 Hz by reducing their size \citep{Kan2022}. Nevertheless, extending real-time PIV techniques to multiplane PIV measurements may require further modifications, e.g., to accommodate multiple cameras.}

We \rev{consider} a hypothetical \rev{and somewhat na\"{\i}ve} experiment by fixing ${\rm Re}_\tau = 186$ and assuming flow in a water channel at room temperature ($20 ^\circ$ C), such that $\nu = 1.00$ mm$^2$/s. We further define the channel aspect ratio as $\xi = L_z/2h = 10$, where $L_z$ is the spanwise extent of the channel, to ensure a 1D canonical mean flow. Correspondingly, the volumetric flow rate in the channel, $Q_b$, may be expressed in terms of the nondimensional bulk velocity, $U_b$, as
\begin{equation}
    Q_b = 4 \xi U_b u_\tau h^2 = 4 \xi U_b {\rm Re}_\tau \nu h, \quad U_b = \frac{1}{2} \int_0^2 U(y) dy.
\end{equation}
Using the true mean profile, $U(y)$, estimated over an interval much longer than the training data, $U_b = 16.0$ for the present investigation. Similarly, the (dimensional) sampling time between snapshots required to match that of the present investigations is given by 
\begin{equation}
\Delta t^{\rm dim} = \frac{\Delta t}{{\rm Re}_\tau} \frac{h^2}{\nu}.
\end{equation}

\begin{figure}[!htpb]
    \centering
    \includegraphics[trim=0 0 0 0, clip, width=0.356\textwidth]{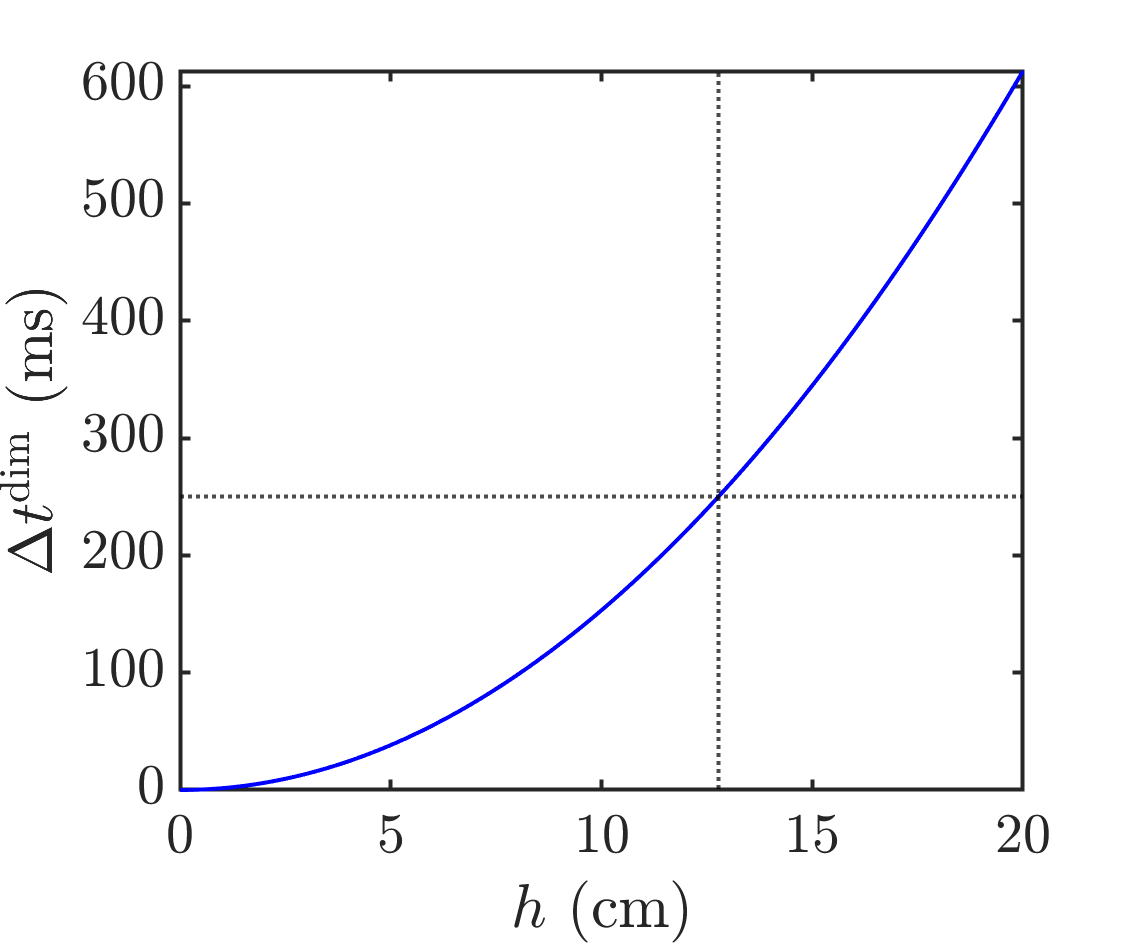}
    \includegraphics[trim=0 0 0 0, clip, width=0.356\textwidth]{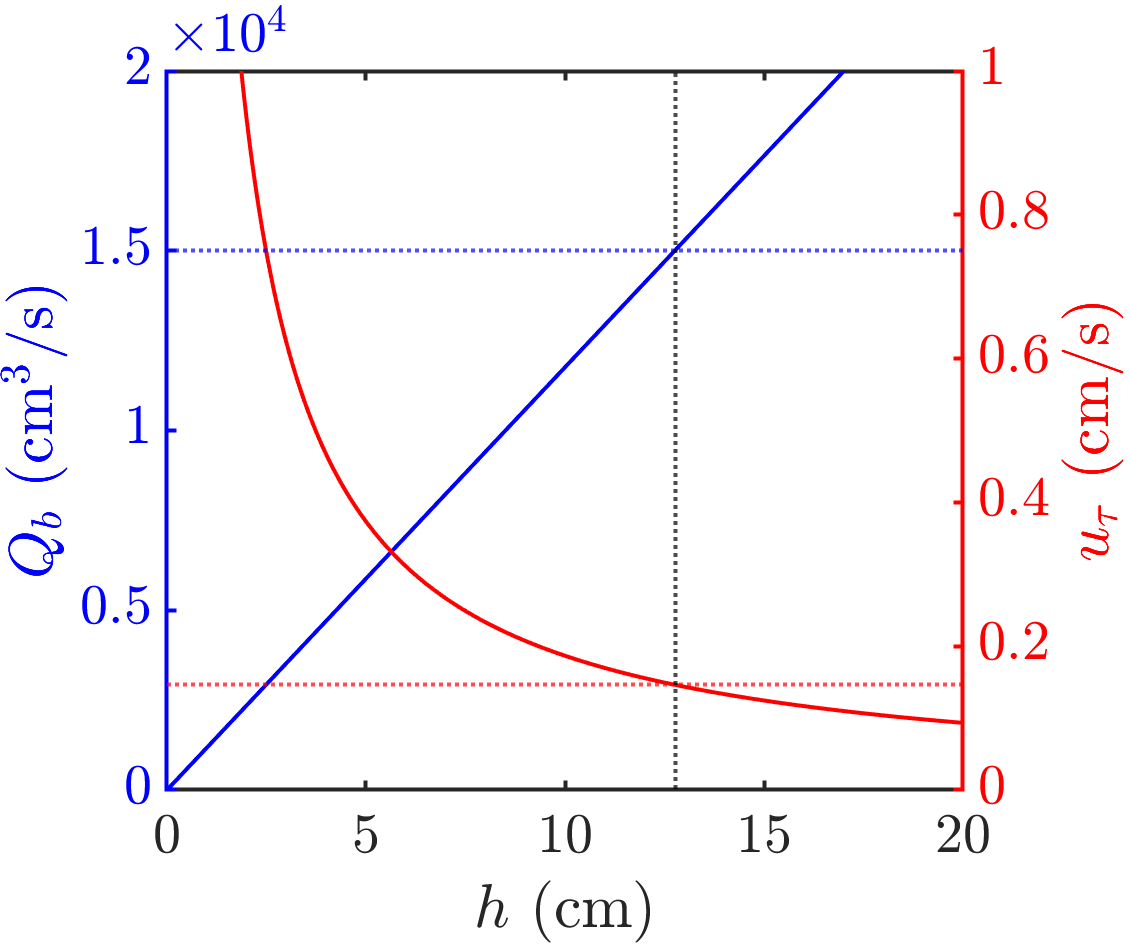}
    \caption{Required dimensional sampling time (left), required flow rate for $\xi = 10$ (right: blue), and required friction velocity (right: red) as functions of the channel half-height. Vertical dotted lines represent the half-height corresponding to the approximate computation time to reconstruct each snapshot (250 ms).}
    \label{fig:timefig}
\end{figure}

We examine the quadratic, linear, and inverse dependencies of $\Delta t^{\rm dim}$, $Q_b$, and $u_\tau$, respectively, on $h$ in Fig. \ref{fig:timefig}.
In order to have sufficient time to reconstruct each snapshot in real time (i.e., before the next measurement is taken), the minimum channel half-height is $h = 12.8$ cm and the corresponding flow rates are $\mathcal{O}(10^4)$ cm$^3$/s. Both of these parameters are feasible in standard water channel configurations with sufficient pumping power, \rev{but the physical setup must be tailored to be compatible with the imaging and computation specifications.} Since we fix ${\rm Re}_\tau$, the friction velocity must decrease as the size of the channel increases, and, hence, the wall roughness must be carefully designed to accommodate for the selected scale of the channel. Finally, we note that sampling every $\mathcal{O}(100)$ ms is \rev{a realistic benchmark for modern} PIV setups. Taken together, these rough estimates indicate that the present methodology has the potential for real-time implementation in an experimental turbulent channel flow \rev{and provide motivation for evaluating the present techniques in that setting.}

%%%%%%%%%%%%%%%%%%%%%%%%%%%%%%%%%%%%%%%%%%
\section{Conclusions}\label{sec:conc}
%%%%%%%%%%%%%%%%%%%%%%%%%%%%%%%%%%%%%%%%%%

We develop an efficient framework for streaming reconstructions of turbulent velocity fluctuations from sparse measurements with the potential for real-time implementation. In this framework, we partition time-resolved data into an initial training period and a subsequent testing period of equal length. We enable efficient, inverse-free reconstructions by first computing linear estimators from the training data and later applying them to measurements from the testing data. Various estimators are considered to address scenarios with varying degrees of available training data. Specifically, we employ streaming (i) truncated (resolvent) response mode estimation (STRME), (ii) optimal resolvent-based estimation (SORBE), and (iii) truncated SPOD mode estimation (STSME) to address the cases where the training data consists of (i) no, (ii) limited, and (iii) full-field second-order statistics.

As a prototypical example of this framework, we reconstruct snapshots from a DNS of turbulent flow in a minimal channel at ${\rm Re}_\tau \approx 186$ using spatially isolated wall-parallel planes to simulate a multiplane PIV setup. We use the energetically dominant Fourier modes to efficiently represent the dynamics at each plane and evaluate their ability to reconstruct both the filtered and unfiltered snapshots. By operating in Fourier triplet space, we allow for parallel reconstructions of reduced size at each triplet. During training, we introduce blockwise inversion to efficiently and accurately compute the resolvent operator. This technique is interpretable through the governing equations in terms of input-output relationships and a modified projection. During testing, we enable efficient streaming reconstructions by recursively updating the triplet space measurements via incoming data using a temporal sliding discrete Fourier transform (SDFT).

The STRME estimators can be formed from only the governing equations and a mean profile, but rely on an unrealistic uncorrelated forcing assumption in the present formulation. They provide a simple and informative means for local reconstructions and streamwise statistical reconstructions. The SORBE estimators effectively incorporate limited training data with the governing equations to improve reconstruction accuracy. They allow for a tailored balance of the cost-accuracy tradeoffs associated with incorporating training statistics and testing measurements. The STSME estimators require full-field training statistics and do not consider the governing equations. They provide an efficient means of representing energetic flow structures on a modal basis.

With full-field flow statistics, SORBE and STSME converge to the generalized Wiener filter. 
This estimator produces the most accurate reconstructions in the present investigation and it has filtered and unfiltered errors of 41.5\% and 52.0\% (i.e., 17.2\% and 27.0\% of the TKE), respectively, when $N_{\rm plane} = 7$ measurement planes are used. \rev{However, these errors are larger than those of the formally optimal noncausal estimator since the sliding temporal window produces time-varying flow statistics.} For all methods, significant reductions in error are obtained by computing the errors over a truncated wall-normal domain based on proximity to the measurements. The spatial error profiles demonstrate that this is primarily because spatially localized streamwise and spanwise flow structures are relatively well captured in these regions. The errors grow less rapidly for the wall-normal fluctuations since they are less localized in the wall-normal direction.

One \rev{promising} feature of the present framework is its potential amenability to real-time reconstructions of turbulent flows in experimental settings. Even the relatively unoptimized code used in the present investigation can reconstruct each incoming snapshot in a fraction of a second ($\sim 250$ ms). \rev{This time delay suggests that the present technique can help enable relatively accurate real-time flow estimation in laboratory settings (e.g., using a water tunnel) when paired with a multiplane PIV measurement system and a real-time PIV scheme. The time delay can be further reduced, thereby alleviating experimental design constraints, by optimizing the implementation and computational hardware used for the reconstruction scheme.} 

Moving forward, experimental validation is a natural next step to validate the efficacy of the present methodology in real-time settings. The velocity fluctuation statistics used for the present estimators \rev{are within reach for modern experimental measurements}, and the resolvent operator relates them to the forcing statistics (as in SORBE), which are more difficult to measure directly. In estimating flow statistics, determining the required spatial resolution and optimal underlying models is important to determining the extent of training data required to produce transfer functions with sufficient fidelity. In experimental reconstructions, further considerations are required to address more complex boundary conditions and real-time processing of raw data.

Further, numerous methodological improvements are likely to yield more efficient and accurate flow reconstructions. For example, the recently developed causal ORBE framework \citep{Mar2022} is a logical choice to employ in real-time applications. Another particularly promising avenue would be to combine frameworks to improve reconstruction accuracy with minimal training data. For example, estimators could be initialized using the STRME framework and continuously refined using flow statistics estimated from streaming measurements. Moreover, whereas the present frameworks involve computing optimal linear estimators, better performance may be obtained by introducing nonlinearity into the reconstruction formulation, either explicitly in the estimator or through, e.g., an eddy-viscosity model. Further, whereas we heuristically select measurement locations, optimizing the sensor locations to best capture the relevant dynamics in the flow is likely to improve reconstruction accuracy. While we consider a fixed measurement configuration during testing, adaptive measurement schemes may provide a means to significantly reduce measurement costs by capturing flow features of interest as they evolve. Beyond estimation, adapting the streaming framework to accommodate real-time flow control is a promising avenue of investigation. Further developments are required to address the need for in-time actuation based on real-time measurements.

%%%%%%%%%%%%%%%%%%%%%%%%%%%%%%%%%%%%%%%%%%
\begin{acknowledgments}
%%%%%%%%%%%%%%%%%%%%%%%%%%%%%%%%%%%%%%%%%%

The support of the Office of Naval Research (ONR) under Grant No. N00014-17-1-3022 is gratefully acknowledged. R.A. was supported by the Department of Defense (DoD) through the National Defense Science \& Engineering Graduate (NDSEG) Fellowship Program.

%%%%%%%%%%%%%%%%%%%%%%%%%%%%%%%%%%%%%%%%%%
\end{acknowledgments}
%%%%%%%%%%%%%%%%%%%%%%%%%%%%%%%%%%%%%%%%%%

%%%%%%%%%%%%%%%%%%%%%%%%%%%%%%%%%%%%%%%%%%
%\newpage
%%%%%%%%%%%%%%%%%%%%%%%%%%%%%%%%%%%%%%%%%%

%%%%%%%%%%%%%%%%%%%%%%%%%%%%%%%%%%%%%%%%%%
\appendix
%%%%%%%%%%%%%%%%%%%%%%%%%%%%%%%%%%%%%%%%%%

%%%%%%%%%%%%%%%%%%%%%%%%%%%%%%%%%%%%%%%%%%
\section{Eddy-viscosity resolvent operator}\label{sec:app:ev}
%%%%%%%%%%%%%%%%%%%%%%%%%%%%%%%%%%%%%%%%%%

Following previous studies \citep{Ill2018,Ama2020}, the modified (discrete) basic linear dynamics are given by
\begin{equation}% this is the one we use
    \boldsymbol{L^t_B} = 
    \begin{bmatrix}
    \boldsymbol{L^t_c} & -{\rm i} k_x \boldsymbol{\partial_y \nu_t} + \boldsymbol{\partial_y U} & \boldsymbol{0} \\
    \boldsymbol{0} & \boldsymbol{L^t_e} - \boldsymbol{E_e} & \boldsymbol{0} \\
    \boldsymbol{0} & -{\rm i} k_z \boldsymbol{\partial_y \nu_t} & \boldsymbol{L^t_c}
    \end{bmatrix},
\end{equation}
where $\boldsymbol{\partial_y \nu_t}$ is defined in an analogous form to $\boldsymbol{\partial_y U}$. The modified diagonal subblocks are given by
\begin{equation}
    \boldsymbol{L^t_{c,e}}  = \boldsymbol{L_{c,e}} - \boldsymbol{E_{c,e}} - \frac{1}{{\rm Re}_\tau}\frac{\nu_t}{\nu}\boldsymbol{\hat{\nabla}_{c,e}}^2, \quad \boldsymbol{E_{c,e}} = \boldsymbol{\partial_y \nu_t \partial_y},
\end{equation}
where the discrete constituents of $\boldsymbol{E_{c,e}}$ are appropriately defined with respect to cell center and edge values.
Here, $\nu_t$ is the turbulent eddy viscosity, which is given by
\begin{equation}
    \frac{\nu_t}{\nu} = \frac{1}{2}\left\{ 1 + \frac{\kappa^2 {\rm Re}_\tau^2}{9}\left(1 - \eta^2 \right)^2 \left( 1 + 2 \eta^2 \right)^2 \left[ 1 - {\rm exp}\left( \frac{{\rm Re}_\tau}{A} \left( | \eta | - 1 \right) \right) \right]^2 \right\}^{1/2} - \frac{1}{2},
\end{equation}
where $\eta = y/h$, $\kappa = 0.426$, and $A = 25.4$ \citep{Puj2009}.
Inverting the basic linear dynamics as before, now with extra terms, we express the basic resolvent as
\begin{equation}
    \boldsymbol{R^{B,t}_u} = \left(\boldsymbol{L^t_B}\right)^{-1} = \begin{bmatrix}
    \left(\boldsymbol{L^t_c}\right)^{-1} & \left(\boldsymbol{L^t_c}\right)^{-1} \left( {\rm i}k_x \boldsymbol{\partial_y \nu_t} - \boldsymbol{\partial_y}\boldsymbol{U} \right) \left( \boldsymbol{L^t_e} - \boldsymbol{E_e} \right)^{-1} & \boldsymbol{0} \\
    \boldsymbol{0} & \left( \boldsymbol{L^t_e} - \boldsymbol{E_e} \right)^{-1} & \boldsymbol{0} \\
    \boldsymbol{0} & \left(\boldsymbol{L^t_c}\right)^{-1} \left({\rm i} k_z \boldsymbol{\partial_y \nu_t} \right) \left( \boldsymbol{L^t_e} - \boldsymbol{E_e} \right)^{-1} & \left( \boldsymbol{L^t_c} \right)^{-1}
    \end{bmatrix},
\end{equation}
which again acts between aligned quantities in $x$ and $z$ for convenience. Thus, just as two inverses, $\boldsymbol{L_c}^{-1}$ and $\boldsymbol{L_e}^{-1}$, are required (for staggered data) to compute the standard basic resolvent, two different inverses, $\left(\boldsymbol{L^t_c}\right)^{-1}$ and $\left( \boldsymbol{L^t_e} - \boldsymbol{E_e} \right)^{-1}$, are required to compute the eddy-viscosity basic resolvent. As before, using blockwise inversion to enforce continuity gives the full eddy-viscosity resolvent as
\begin{equation}
    \boldsymbol{R^t_u} = \boldsymbol{S}^H \left[ \boldsymbol{R^{B,t}_u} - \boldsymbol{R^{B,t}_u} \boldsymbol{\hat{\nabla}} \left( \boldsymbol{\hat{\nabla}}^T \boldsymbol{R^{B,t}_u} \boldsymbol{\hat{\nabla}} \right)^{-1} \boldsymbol{\hat{\nabla}}^T \boldsymbol{R^{B,t}_u} \right] \boldsymbol{S}.
\end{equation}
This form is analogous to the form derived for the standard resolvent operator, and therefore it may also be understood as a modified projection that acts to eliminate the pressure terms and enforce continuity.

%%%%%%%%%%%%%%%%%%%%%%%%%%%%%%%%%%%%%%%%%%
\section{Temporal Fourier transform accuracy}\label{sec:app:sdft}
%%%%%%%%%%%%%%%%%%%%%%%%%%%%%%%%%%%%%%%%%%

\begin{figure}[!htpb]
    \centering
    \includegraphics[trim=200 50 200 50,clip,width=\textwidth]{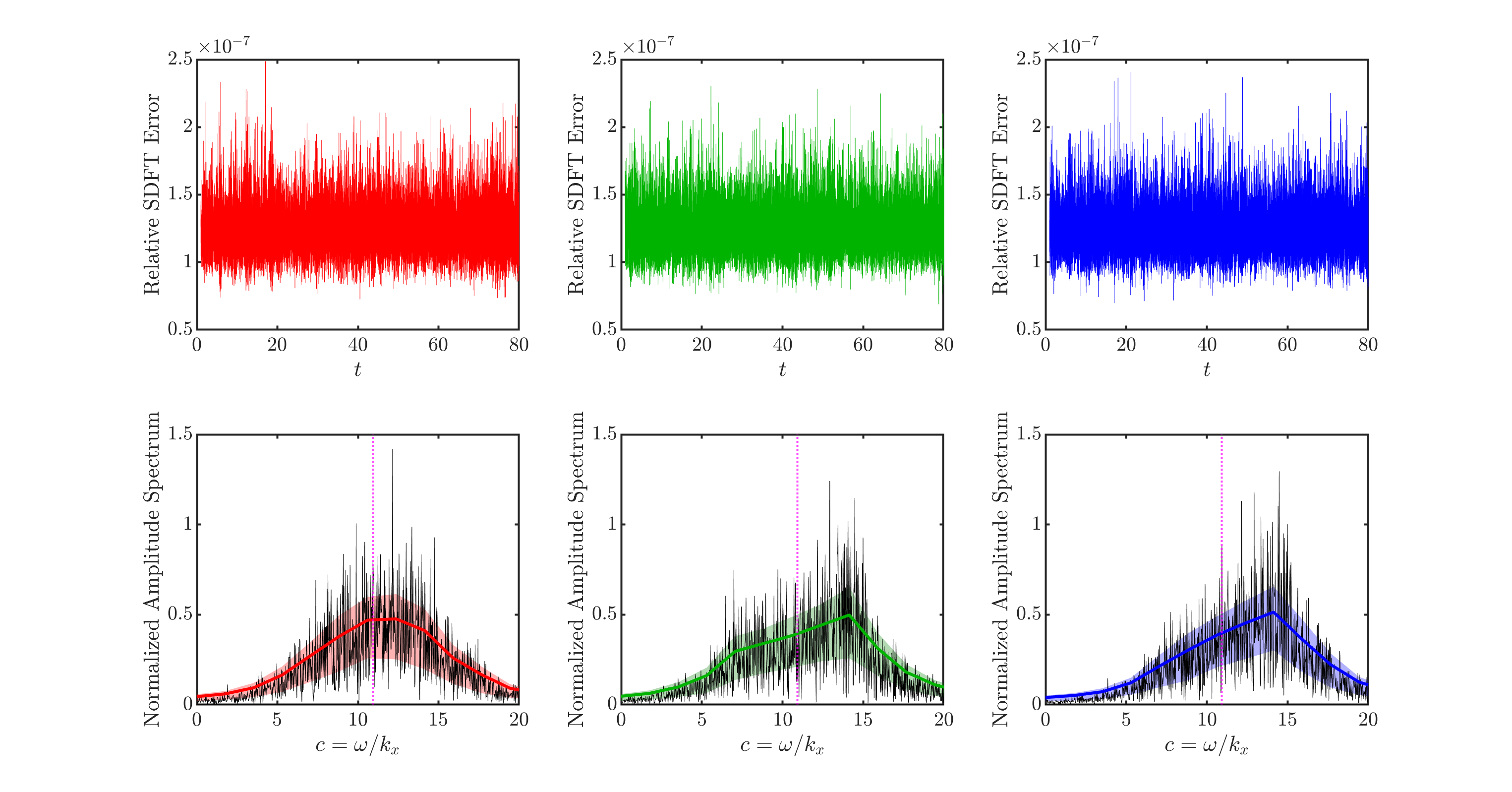}% new (normalized individual spectra)
    \caption{Single precision relative SDFT errors (top) and normalized temporal Fourier amplitude spectra (bottom) of $\hat{u}$ (left), $\hat{v}$ (middle), and $\hat{w}$ (right) computed for $(k_x, k_z) = (3.54, 7.08)$ and $y^+ \approx 14.7$ over the training period. The relative SDFT errors are evaluated with respect the $\texttt{fft}$ coefficients computed over the same temporal window ($\Delta T = 1$). The normalized spectra show the $\Delta T' = 80$ amplitudes (black curves) and the $\Delta T = 1$ amplitudes (colored curves: \rev{square root of mean variance}, shading: \rev{$20{\rm th} - 80{\rm th}$ percentiles}). The long-time and short-time spectra are normalized to reproduce unit signal energy, and in the latter case normalization is performed prior to computing statistics. The magenta dotted lines represent the local convection velocity, $U(y^+) \approx 10.9$.}
    \label{fig:sdft_ex}
\end{figure}

We implicitly absorb into the noise vector the errors associated with computing the temporal Fourier transforms of measurements (i) using a (recursive) sliding DFT and (ii) over a finite-length window. It is therefore important to at least qualitatively characterize the validity of these methods, the former in the context of real time or streaming reconstructions and the latter in the context of transfer functions designed for noncausal estimates. Figure \ref{fig:sdft_ex} depicts the errors associated with both of these methods for $(k_x, k_z) = (3.54, 7.08)$. The relative SDFT error plots show that the SDFT does not meaningfully contribute to the errors absorbed into the noise vector beyond those incurred by employing the \texttt{fft} over the same temporal window. The normalized temporal Fourier amplitude spectra address the validity of the finite-length temporal window. When properly normalized, the spectra computed over a temporal window of size $\Delta T = 1$ bear qualitative resemblance to those computed over $\Delta T' = 80$. Further, the variability in the amplitudes computed over subsequent smaller windows is in line with the amplitudes computed over the longer interval. This suggests that information regarding the noise incurred by truncating the integral associated with the temporal Fourier transform may be inferred to some degree. Nevertheless, to retain simplicity, we employ the simplified noise CSD model, $\boldsymbol{\tilde{S}_{nn}} = \epsilon\boldsymbol{I}$, when computing the transfer functions in the present investigation.

%%%%%%%%%%%%%%%%%%%%%%%%%%%%%%%%%%%%%%%%%%
\section{Coefficients with limited training data}\label{sec:app:fvsxi}
%%%%%%%%%%%%%%%%%%%%%%%%%%%%%%%%%%%%%%%%%%

\begin{figure}[!htpb]
    \centering
    \includegraphics[trim=0 0 0 0, clip, width=0.75\textwidth]{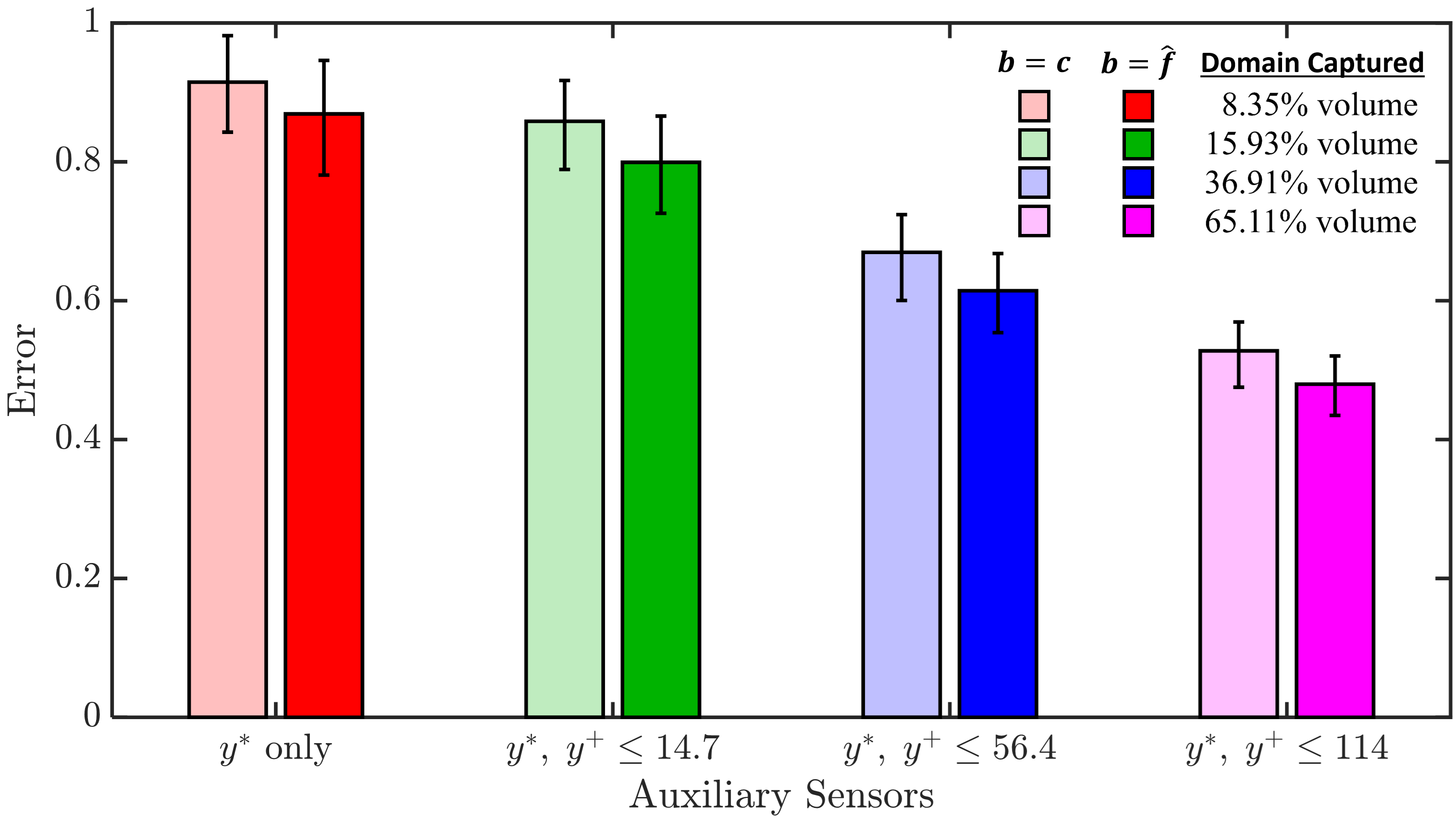}
    \caption{Comparisons of the full-field filtered errors, $\epsilon_{\rm filt}(t)$ (bars: \rev{square root of mean variance}, error bars: \rev{$20{\rm th} - 80{\rm th}$ percentiles}), for reconstructions using the data-driven extension of STRME (light bars) and using SORBE (dark bars). The errors are shown for estimators ($N_{\rm plane} = 7$) that incorporate various amounts of auxiliary training data to estimate coefficient statistics during training.}
    \label{fig:orbe_trme_compare}
\end{figure}

The optimal choice of coefficients for flow reconstruction remains an open question. Here, we empirically demonstrate that using the forcing (as in SORBE) produces more accurate reconstructions than using the mode weights (as in STRME) in the present investigation. Figure \ref{fig:orbe_trme_compare} compares the spatially integrated errors for estimators ($N_{\rm plane} = 7$) that incorporate forcing and mode weight statistics computed using various (limited) amounts of training data. In both cases, the coefficient statistics are estimated from the same auxiliary observations in a manner analogous to (\ref{eq:auxforc}). The results show that the forcing-based SORBE estimator produces consistently lower errors than the mode weight-based estimator representing a data-driven extension of STRME. While this validates our choice of SORBE to address the case where limited training data are available, further work is needed to find coefficients with statistics easily inferred from sparse training data.

%%%%%%%%%%%%%%%%%%%%%%%%%%%%%%%%%%%%%%%%%%
\section{Effect of truncated errors}\label{sec:app:tempsig}
%%%%%%%%%%%%%%%%%%%%%%%%%%%%%%%%%%%%%%%%%%

\begin{figure}[!htpb]
    \centering
    \includegraphics[trim=25 10 100 50, clip, width=\textwidth]{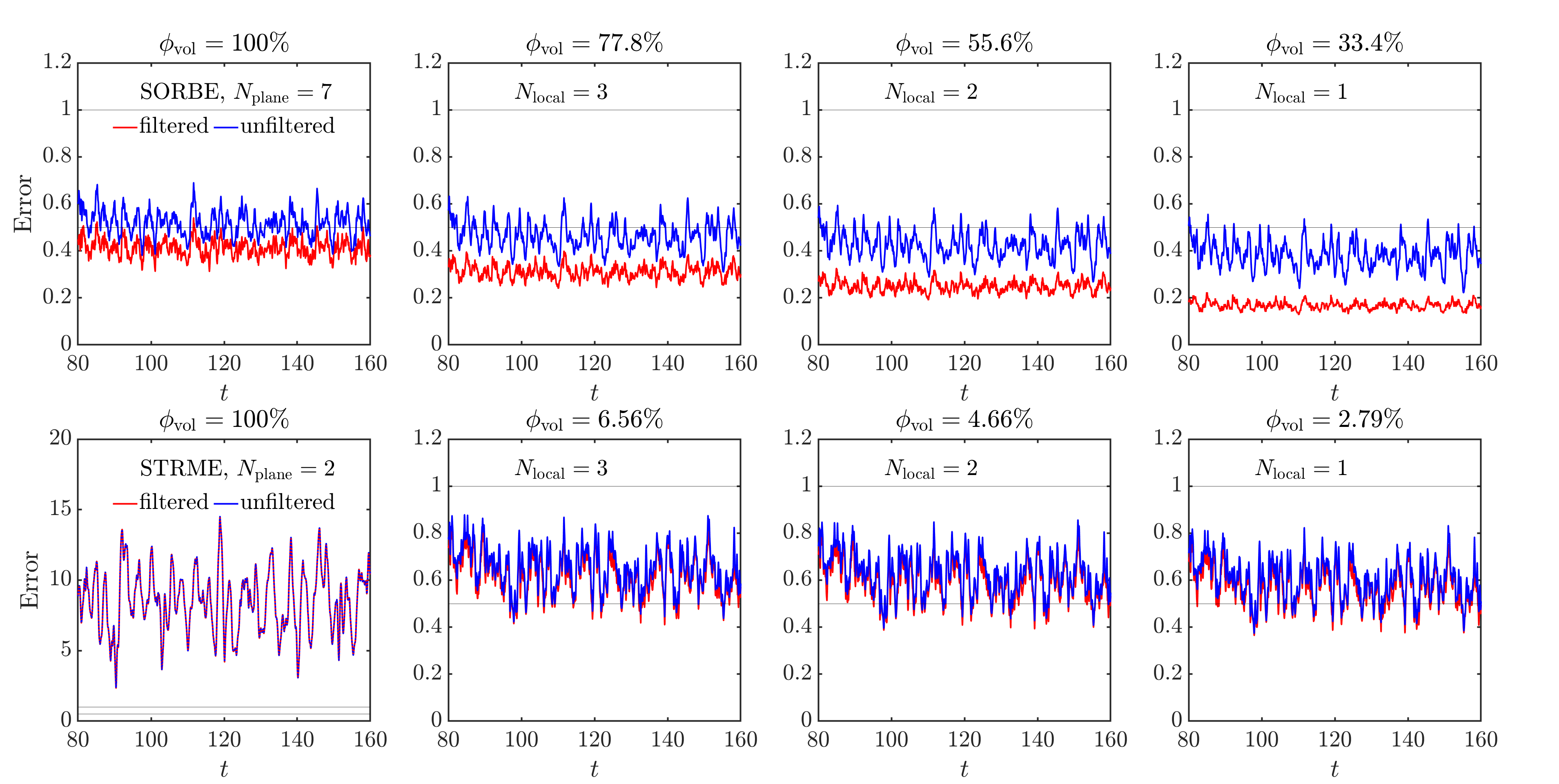}
    \caption{Temporal evolutions of the full-field errors, $\epsilon_{\rm filt}(t)$ and $\epsilon_{\rm full}(t)$, and the localized errors, $\epsilon^*_{\rm filt}(t)$ and $\epsilon^*_{\rm full}(t)$, for $N_{\rm local} = \;$ 3, 2, and 1. These cases are listed such that the fraction of the channel captured by each metric, $\phi_{\rm vol}$, decreases from left to right.}
    \label{fig:res:tempsig}
\end{figure}

To further probe the effect of truncating the wall-normal domain over which errors are computed, we consider truncations to within $N_{\rm local} = 1, 2$ and $3$ cells of the measurement planes. Figure \ref{fig:res:tempsig} compares the temporal evolutions of the full-channel and localized error metrics during the testing period for selected cases. For the $N_{\rm plane} = 7$ generalized Wiener filter, the localized reconstruction errors decrease with decreasing $N_{\rm local}$, especially with respect to the filtered fluctuations. Since seven measurement planes are used, the fraction of the domain captured when integrating the errors ($\phi_{\rm vol}$) remains relatively large for all cases shown. This indicates that truncating the errors allows for improved reconstruction accuracy while retaining a large (albeit limited) domain of applicability. For the $N_{\rm plane} = 2$ STRME estimator, the values of $\phi_{\rm vol}$ are considerably smaller since fewer measurement planes are considered. These localized errors are much smaller (by an order of magnitude) than the full-field error, but apply only to considerably smaller subdomains. However, these large error reductions suggest that the $N_{\rm plane} = 2$ STRME estimator may be suitable for localized reconstructions about each measurement plane. In this limited context, the uncorrelated forcing assumption has a relatively small impact on reconstruction accuracy.

%%%%%%%%%%%%%%%%%%%%%%%%%%%%%%%%%%%%%%%%%%
%\newpage
%%%%%%%%%%%%%%%%%%%%%%%%%%%%%%%%%%%%%%%%%%

% The \nocite command causes all entries in a bibliography to be printed out whether or not they are actually referenced in the text. This is appropriate
% for the sample file to show the different styles of references, but authors most likely will not want to use it.
% \nocite{*}

%%%%%%%%%%%%%%%%%%%%%%%%%%%%%%%%%%%%%%%%%%
\bibliography{references}% bibliography via BibTeX.
%%%%%%%%%%%%%%%%%%%%%%%%%%%%%%%%%%%%%%%%%%

%%%%%%%%%%%%%%%%%%%%%%%%%%%%%%%%%%%%%%%%%%
\end{document}